%%%%%%%%%%%%%%%%%%%%%%%%%%%%%%%%%%%%%%%%%%%%%%%%%%%
%%%%%%%%%%%%%%%%%%%%%%%%%%%%%%%%%%%%%%%%%%%%%%%%%%%
%%%%                                           %%%%
%%%%  (Plain-)TeX file for the revised paper:  %%%%
%%%%                                           %%%%
%%%%      Constituent Quarks from QCD          %%%%
%%%%                                           %%%%
%%%%                 by                        %%%%
%%%%                                           %%%%
%%%%    Martin Lavelle and David McMullan      %%%%
%%%%                                           %%%%
%%%%                                           %%%%
%%%%          which is published in            %%%%
%%%%                                           %%%%
%%%%       Physics Reports C279 (1997) 1       %%%%
%%%%                                           %%%%
%%%%%%%%%%%%%%%%%%%%%%%%%%%%%%%%%%%%%%%%%%%%%%%%%%%
%%%%%%%%%%%%%%%%%%%%%%%%%%%%%%%%%%%%%%%%%%%%%%%%%%%

%This paper used some of the AMS fonts,
%if you do not have these find the definitions of
%\Real and \Bold where alternatives are given

%figures macros
%\input epsfig.sty
%%%\def\fig#1#2#3{\vbox{\epsfig{file=#1,height=#2,width=#3}}}
%\def\fig#1#2{\vbox{\epsfig{file=#1,width=#2}}}
\input epsf
\def\fig#1#2{$$\epsfxsize=#2\epsfbox{#1}$$}

\font\bigbold=cmbx12
\font\eightrm=cmr8
\font\sixrm=cmr6
\font\fiverm=cmr5
\font\eightbf=cmbx8
\font\sixbf=cmbx6
\font\fivebf=cmbx5
\font\eighti=cmmi8  \skewchar\eighti='177
\font\sixi=cmmi6    \skewchar\sixi='177
\font\fivei=cmmi5
\font\eightsy=cmsy8 \skewchar\eightsy='60
\font\sixsy=cmsy6   \skewchar\sixsy='60
\font\fivesy=cmsy5
\font\eightit=cmti8
\font\eightsl=cmsl8
\font\eighttt=cmtt8
\font\tenfrak=eufm10
\font\sevenfrak=eufm7
\font\fivefrak=eufm5
\font\tenbb=msbm10
\font\sevenbb=msbm7
\font\fivebb=msbm5
\font\tensmc=cmcsc10
\font\tencmmib=cmmib10  \skewchar\tencmmib='177
%\font\sevencmmib=cmmib10 at 7pt \skewchar\sevencmmib='177
\font\sevencmmib=cmmib7 \skewchar\sevencmmib='177
%\font\fivecmmib=cmmib10 at 5pt \skewchar\fivecmmib='177
\font\fivecmmib=cmmib5 \skewchar\fivecmmib='177
%Some Families

\newfam\bbfam
\textfont\bbfam=\tenbb
\scriptfont\bbfam=\sevenbb
\scriptscriptfont\bbfam=\fivebb
\def\Bbb{\fam\bbfam}

\newfam\frakfam
\textfont\frakfam=\tenfrak
\scriptfont\frakfam=\sevenfrak
\scriptscriptfont\frakfam=\fivefrak

\newfam\cmmibfam
\textfont\cmmibfam=\tencmmib
\scriptfont\cmmibfam=\sevencmmib
\scriptscriptfont\cmmibfam=\fivecmmib
\def\bold#1{\fam\cmmibfam\relax#1}

%Definition of 8 point

\def\eightpoint{%
\textfont0=\eightrm   \scriptfont0=\sixrm
\scriptscriptfont0=\fiverm  \def\rm{\fam0\eightrm}%
\textfont1=\eighti   \scriptfont1=\sixi
\scriptscriptfont1=\fivei  \def\oldstyle{\fam1\eighti}%
\textfont2=\eightsy   \scriptfont2=\sixsy
\scriptscriptfont2=\fivesy
\textfont\itfam=\eightit  \def\it{\fam\itfam\eightit}%
\textfont\slfam=\eightsl  \def\sl{\fam\slfam\eightsl}%
\textfont\ttfam=\eighttt  \def\tt{\fam\ttfam\eighttt}%
\textfont\bffam=\eightbf   \scriptfont\bffam=\sixbf
\scriptscriptfont\bffam=\fivebf  \def\bf{\fam\bffam\eightbf}%
\abovedisplayskip=9pt plus 2pt minus 6pt
\belowdisplayskip=\abovedisplayskip
\abovedisplayshortskip=0pt plus 2pt
\belowdisplayshortskip=5pt plus2pt minus 3pt
\smallskipamount=2pt plus 1pt minus 1pt
\medskipamount=4pt plus 2pt minus 2pt
\bigskipamount=9pt plus4pt minus 4pt
\setbox\strutbox=\hbox{\vrule height 7pt depth 2pt width 0pt}%
\normalbaselineskip=9pt \normalbaselines
\rm}

%More general stuff

\def\pagewidth#1{\hsize= #1}
\def\pageheight#1{\vsize= #1}
\def\hcorrection#1{\advance\hoffset by #1}
\def\vcorrection#1{\advance\voffset by #1}

\newcount\notenumber  \notenumber=1              %Numbering does
\newif\iftitlepage   \titlepagetrue              %not start on title
\newtoks\titlepagefoot     \titlepagefoot={\hfil}%page
\newtoks\otherpagesfoot    \otherpagesfoot={\hfil\tenrm\folio\hfil}
\footline={\iftitlepage\the\titlepagefoot\global\titlepagefalse
           \else\the\otherpagesfoot\fi}

\def\abstract#1{{\parindent=30pt\narrower\noindent\eightpoint\openup
2pt #1\par}}
\def\smc{\tensmc}

%A nicer footnote

\def\note#1{\unskip\footnote{$^{\the\notenumber}$}
{\eightpoint\openup 1pt
#1}\global\advance\notenumber by 1}

%%% Macro to generate the equation #'s automatically.
%%% To use start each new section (eg 3) with the commands
%%% \secno=3 \meqno=1 :this will start the equations with (3.1)
%%% Then in place of \eqno(3.1) type \eqn\descriptivename . To refer
%%% back to the equation simply type (\descritivename)
%%% For the appendix set \secno=0, \appno=1\meqno=1 etc
%%% If there are no sections, then set \secno=0

\global\newcount\secno \global\secno=0
\global\newcount\meqno \global\meqno=1
\global\newcount\appno \global\appno=0
\newwrite\eqmac
\def\romappno{\ifcase\appno\or A\or B\or C\or D\or E\or F\or G\or H
\or I\or J\or K\or L\or M\or N\or O\or P\or Q\or R\or S\or T\or U\or
V\or W\or X\or Y\or Z\fi}
\def\eqn#1{
        \ifnum\secno>0
            \eqno(\the\secno.\the\meqno)\xdef#1{\the\secno.\the\meqno}
          \else\ifnum\appno>0
            \eqno({\rm\romappno}.\the\meqno)\xdef#1{{\rm\romappno}.\the\meqno}
          \else
            \eqno(\the\meqno)\xdef#1{\the\meqno}
          \fi
        \fi
\global\advance\meqno by1 }

\def\eqnn#1{
        \ifnum\secno>0
            (\the\secno.\the\meqno)\xdef#1{\the\secno.\the\meqno}
          \else\ifnum\appno>0
            \eqno({\rm\romappno}.\the\meqno)\xdef#1{{\rm\romappno}.\the\meqno}
          \else
            (\the\meqno)\xdef#1{\the\meqno}
          \fi
        \fi
\global\advance\meqno by1 }
%%% Macro to assist in the references
%%% At the begining of the paper list the references in the order
%%% that they appear by the command \refn
%%% So if the first reference is to be
%%%  D. McMullan and I. Tsutsui Nucl. Phys. B121 (1994) 12
%%% then type \refn\us{D. McMullan and I. Tsutsui\np{121}{94}{12}}
%%% In the text this is simply refered to by [\us].
%%% At the end of the text type \listrefs

\global\newcount\refno
\global\refno=1 \newwrite\reffile
\newwrite\refmac
\newlinechar=`\^^J
\def\ref#1#2{\the\refno\nref#1{#2}}
\def\nref#1#2{\xdef#1{\the\refno}
\ifnum\refno=1\immediate\openout\reffile=refs.tmp\fi
\immediate\write\reffile{
     \noexpand\item{[\noexpand#1]\ }#2\noexpand\nobreak.}
     \immediate\write\refmac{\def\noexpand#1{\the\refno}}
   \global\advance\refno by1}
\def\semi{;\hfil\noexpand\break ^^J}
\def\nl{\hfil\noexpand\break ^^J}
\def\refn#1#2{\nref#1{#2}}

\def\up#1{$^{[#1]}$}

\def\cmp#1#2#3{{\it Commun. Math. Phys.} {\bf {#1}} (19{#2}) #3}
\def\jmp#1#2#3{{\it J. Math. Phys.} {\bf {#1}} (19{#2}) #3}
\def\ijmp#1#2#3{{\it Int. J. Mod. Phys.} {\bf A{#1}} (19{#2}) #3}
\def\mplA#1#2#3{{\it Mod. Phys. Lett.} {\bf A{#1}} (19{#2}) #3}
\def\pl#1#2#3{{\it Phys. Lett.} {\bf {#1}B} (19{#2}) #3}
\def\np#1#2#3{{\it Nucl. Phys.} {\bf B{#1}} (19{#2}) #3}
\def\npps#1#2#3{{\it Nucl. Phys. (Proc. Suppl.\/)}
{\bf B{#1}} (19{#2}) #3}
\def\pr#1#2#3{{\it Phys. Rev.} {\bf {#1}} (19{#2}) #3}
\def\prB#1#2#3{{\it Phys. Rev.} {\bf B{#1}} (19{#2}) #3}
\def\prD#1#2#3{{\it Phys. Rev.} {\bf D{#1}} (19{#2}) #3}
\def\prl#1#2#3{{\it Phys. Rev. Lett.} {\bf #1} (19{#2}) #3}
\def\rmp#1#2#3{{\it Rev. Mod. Phys.} {\bf {#1}} (19{#2}) #3}
\def\ann#1#2#3{{\it Ann. Phys.} {\bf {#1}} (19{#2}) #3}
\def\prp#1#2#3{{\it Phys. Rep.} {\bf {#1}C} (19{#2}) #3}
\def\tmp#1#2#3{{\it Theor. Math. Phys.} {\bf {#1}} (19{#2}) #3}
\def\zpC#1#2#3{{\it Z. Phys.} {\bf C{#1}} (19{#2}) #3}
\def\zpA#1#2#3{{\it Z. Phys.} {\bf A{#1}} (19{#2}) #3}
\def\fortschr#1#2#3{{\it Fortschr. d. Phys.} {\bf {#1}} (19{#2}) #3}
\def\amjp#1#2#3{{\it Am. J. Phys.} {\bf{#1}} (19{#2}) #3}
\def\nc#1#2#3{{\it Nuovo Cimento} {\bf{#1}} (19{#2}) #3}
\def\hpa#1#2#3{{\it Helv. Phys. Acta.} {\bf{#1}} (19{#2}) #3}
\def\canjp#1#2#3{{\it Canad. J. Phys.} {\bf{#1}}(19{#2}) #3}
\def\empty#1#2#3{{\bf{#1}} (19{#2}) #3}

%%% Format for this paper

\pageheight{23cm}
\pagewidth{15.5cm}
\hcorrection{-2.5mm}
\magnification \magstep1
\baselineskip=16pt plus 1pt minus 1pt
\parskip=5pt plus 1pt minus 1pt

%%% Some macros specific to this paper

\def\frac#1#2{{#1\over#2}}
\def\dfrac#1#2{{\displaystyle{#1\over#2}}}
\def\tfrac#1#2{{\textstyle{#1\over#2}}}
\def\({\left(}
\def\){\right)}
\def\<{\langle}
\def\>{\rangle}
   %Partial derivatives
\def\2pd#1#2#3{\frac{\partial^2#1}{\partial#2\partial#3}}

\def\sqr#1#2{{\vcenter{\vbox{\hrule height.#2pt
        \hbox{\vrule width.#2pt height#1pt \kern#1pt
           \vrule width.#2pt}
        \hrule height.#2pt}}}}
\def\square{\mathop{\mathchoice\sqr64\sqr64\sqr{4.2}3\sqr33}}
\def\ni{\noindent}
\def\lqq{\lq\lq}
\def\rqq{\rq\rq}
\def\slash{\!\!\!\!/}
\def\ni{\noindent}
\def\Tr{{\rm Tr}\,}
\def\c{\gamma}
\def\d{\delta}
\def\a{\alpha}
\def\b{\beta}
\def\e{\epsilon}
\def\L{{\cal L}}
\def\ve{\varepsilon}
\def\pa{\partial}
\def\Au{A^{\scriptscriptstyle U}}
\def\hu{h^{\scriptscriptstyle U}}
\def\psiu{\psi^{\scriptscriptstyle U}}
\def\hc{h_{\rm c}}
\def\vc{v_{\rm c}}
\def\vv{v_{\rm v}}
\def\psic{\psi_{\rm c}}
\def\psiv{\psi_{\rm v}}
\def\psig{\psi_{_{\Gamma}}}
\def\phys{{\hbox{\sevenrm phys}}}
\def\Qm{Q_{\hbox{\sevenrm mat}}}
\def\bra#1{\<#1|}
\def\ket#1{|#1\>}
\def\v#1{v_{_{(#1)}}}
\def\A{{\cal A}}
\def\G{{\cal G}}
\def\O{{\cal O}}
\def\Real{{\Bbb R}}     %If you do not have this font use the defn below
\def\Z{{\Bbb Z}}
\def\quark{\langle\bar\psi\psi\rangle}
\def\Sr{S_{\hbox{\sevenrm R}}}

\def\vdenom#1#2{{\left(\dfrac{(#1_1-#2_1)^2}{1-v^2}+ (#1_2-#2_2)^2+
(#1_3-#2_3)^2\right)}}
\def\tr{{\hbox {\rm tr}}}
%%%%%%%%%%%%%%%%%%%%%%

%%% The references

{\eightpoint

\refn\AACHEN
{For a recent review of very many aspects  and the status of QCD see,
{\sl QCD --- 20 Years later}, ed.'s P.M.\ Zerwas and H.A.\ Kastrup
(World Scientific, two volumes, Singapore, 1993)}

\refn\GELLMANN
{M.\ Gell-Mann and Y.\ Ne'eman, {\sl The Eightfold Way},
(W.\ Benjamin, New York, 1964)}

\refn\KOKK
{J.J.J.\ Kokkedee, {\sl The Quark Model}, (W.\ Benjamin, New
York, 1969)}

\refn\NEWPARTICLES
{M.\ Gell-Mann, Caltech preprint CTSL--26, 1961, (unpublished)
reprinted in Ref.\ \GELLMANN}

\refn\DIS{J.I.\ Friedman, H.W.\ Kendall and R.E.\ Taylor,
in {\sl Nobel Lectures in Physics 1981--1990}, ed. G.\ Ekspong
(World Scientific, Singapore 1993)}

\refn\ASYMPTF
{D.\ Gross and F.\ Wilczek, \prl{26}{73}{1343}\semi
H.D.\ Politzer, \prl{26}{73}{1346}}

\refn\STROCCHI
{F.\ Strocchi and A.S.\ Wightman, \jmp{15}{74}{2198}}

\refn\LMQUARK
{M.\ Lavelle and D.\ McMullan, Dublin and Mainz preprint DIAS-STP-93-04
and MZ-TH/93-03 (unpublished); \pl{329}{94}{68}}

\refn\LMSBT
{M.\ Lavelle and D.\ McMullan, \pl{347}{95}{89}}

\refn\JAUCH
{J.M.\ Jauch and F.\ Rohrlich, {\sl The Theory of Photons and Electrons},
Second Expanded Edition, (Springer-Verlag, New York, 1980)}

\refn\DOLLARD
{J.D.\ Dollard, \jmp{5}{64}{729}}

\refn\KULISH
{P.\ Kulish and L.\ Faddeev, \tmp{4}{70}{745}}

\refn\LMSYM
{M.\ Lavelle and D.\ McMullan, \prl{71}{93}{3758}}

\refn\LMPROP
{M.\ Lavelle and D.\ McMullan, \pl{312}{93}{211}}

\refn\DIRACINCANADA
{P.A.M.\ Dirac, \canjp{33}{55}{650}}

\refn\DIRAC
{P.A.M.\ Dirac, {\sl The Principles of Quantum Mechanics}, (Oxford
University Press, 1958)}

\refn\WEINBERG
{S.\ Weinberg,  \prl{31}{73}{494}}

\refn\GROSS
{D.J.\ Gross and F.\ Wilczek, \prD{8}{73}{3633}}

\refn\FRITZSCH
{H.\ Fritzsch, M.\ Gell-Mann and H.\ Leutwyler, \pl{47}{73}{269}}

\refn\Xtra
{E.\ d'Emilio and M.\ Mintchev, \fortschr{32}{84}{473, 503}}

\refn\XXtra
{O.\ Steinmann, \ann{157}{84}{232}; \hpa{58}{85}{232}}

\refn\SHAB
{S.V.\ Shabanov, \mplA{6}{91}{909}; \pl{255}{91}{398};\nl
L.V.\ Prokhorov and S.V.\ Shabanov, \ijmp{7}{92}{7815}}

\refn\OTHERRUSSIAN
{F.A.\ Lunev, \mplA{9}{94}{2281}}

\refn\MACK
{G.\ Mack, \fortschr{29}{81}{135}}

\refn\KHVED
{A.\ Khvedelidze and V.\ Pervushin, \hpa{6}{94}{610}}

\refn\REST
{Y.\ Neeman and D.\ Sijacki, CERN-TH-7461-94}

\refn\KUGO
{T.\ Kugo and I.\ Ojima, {\it Prog. Theor. Phys. Suppl. No.\ {\bf 66}}
(1979) 1}

\refn\POVH
{B.\ Povh et al., {\sl Particles and Nuclei},
(Springer-Verlag, Berlin, Heidelberg, 1995)}

\refn\GASI
{S.\ Gasiorowicz and J.L.\ Rosner, \amjp{49}{81}{954}}

\refn\FLAM
{D.\ Flamm and F.\ Sch\"oberl, {\sl Introduction to the Quark
Model of Elementary Particles} Vol.\ 1, (Gordon and Breach, New York,
1982)}

\refn\GASSA
{J.\ Gasser and H.\ Leutwyler, \np{94}{75}{269}}

\refn\MONT{L.\ Montanet et al., {\sl Review of Particle Properties},
\prD{50}{94}{1173}}

\refn\COLOUR
{For the introduction of colour see M.\ Gell-Mann in Ref.\
\AACHEN}

\refn\JACKIW
{R.\ Jackiw, {\sl Topological  Investigations of Quantized Gauged
Theories,} in {\sl Current Algebra and Anomalies},
ed.'s S.B.\ Treiman et al.
(Princeton University Press, Princeton, 1985)}

\refn\NAKANISHI
{N.\ Nakanishi and I.\ Ojima, {\sl Covariant Operator Formalism of Gauge
Theories and Quantum Gravity} (World Scientific, Singapore, 1990)}

\refn\HENNEAUX
{M.\ Henneaux and C.\ Teitelboim, {\sl Quantization of Gauge Systems}
(Princeton University Press, Princeton, New Jersey, 1992)}

\refn\HAAG
{R.\ Haag, {\sl Local Quantum Physics}, (Springer-Verlag, Berlin,
Heidelberg, 1993)}

\refn\MAISZWAN{For further arguments in this direction see,
D.\ Maison and D.\ Zwanziger, \np{91}{75}{425} and references therein}

\refn\MOON
{R.\ Haag and D.\ Kastler, \jmp{5}{64}{848}}

\refn\ROEPSTORFF
{G.\ Roepstorff, \cmp{19}{70}{301}}

\refn\FROHLICH
{J.\ Fr\"ohlich, G.\ Morchio and F.\ Strocchi, \ann{119}{79}{241}}

\refn\GERVAIS
{J.L.\ Gervais and D.\ Zwanziger, \pl{94}{80}{389}}

\refn\BUCHHOLZ
{D.\ Buchholz, \pl{174}{86}{331}}

\refn\POLONYI
{J.\ Polonyi, \pl{213}{88}{340}}

\refn\MT
{D.\ McMullan and I.\ Tsutsui, \ann{237}{95}{269}}

\refn\SHABANOV
{S.V.\ Shabanov, Dubna preprint JINR-E2-92-136 (unpublished)}

\refn\LMRAD
{M.\ Lavelle and D.\ McMullan, \zpC{59}{93}{351}}

\refn\FADDEEV
{L.D.\ Faddeev,  in {\sl Methods in Field Theory,} ed.'s R.\ Balian
and J.\ Zinn-Justin (North-Holland/World Scientific, 1981)}

\refn\SINGER
{I.M.\ Singer, \cmp{60}{78}{7}}

\refn\MITTER
{P.K.\ Mitter and C.M.\ Viallet, \cmp{79}{81}{457}}

\refn\FUCHS
{J.\ Fuchs, M.G.\ Schmidt and C.\ Schweigert, \np{425}{94}{107}}

\refn\GRIBOV
{V.N.\ Gribov, \np{139}{78}{1}}

\refn\ENCYCLO
{{\sl Encyclopedic Dictionary of Mathematics\/}, ed.'s S.\ Iyanaga and
Y. Kawada (MIT Press, Cambridge, 1977)}

\refn\HATFIELD
{See, e.g., B.\ Hatfield, {\sl Quantum Field Theory of Point Particles
and Strings}, (Addison-Wesley, Redwood City, 1992)}

\refn\ZWANZIGER
{D.\ Zwanziger, \prD{11}{75}{3481}}

\refn\CHUNG
{V.\ Chung, \prB{140}{65}{1110} }

\refn\KIBBLE
{T.W.B.\ Kibble, \pr{173}{68}{1527}; \empty{174}{68}{1882};
empty{175}{68}{1624}}

\refn\JOHNS
{J.K.\ Storrow, \nc{54}{68}{15}; \empty{57}{68}{763}}

\refn\MUTA
{See, e.g., Chap.\ 6 of T.\ Muta, {\sl Foundations of Quantum
Chromodynamics}, (World Scientific, Singapore, 1987)}

\refn\SCHROER
{B.\ Schroer, \fortschr{173}{63}{1527}}

\refn\YENNIE
{H.M.\ Fried and D.R.\ Yennie, \pr{112}{58}{1391}}

\refn\BGS{D.J.\ Broadhurst, N.\ Gray and K.\ Schilcher,
\zpC{48}{91}{111}}

\refn\TOM
{J.C.\ Breckenridge, M.J.\ Lavelle and T.G.\ Steele, \zpC{65}{95}{155}}

\refn\JOHNSON
{K.\ Johnson, \ann{10}{60}{536}}

\refn\HAGEN
{R.\ Hagen, \pr{130}{63}{813}}

\refn\HECKATHORN
{D.\ Heckathorn, \np{156}{79}{328}}

\refn\ADKINS
{G.S.\ Adkins, \prD{27}{83}{1814}}

\refn\VIENNA
{For an overview see: {\sl Physical and Nonstandard Gauges}, ed.'s
P.\ Gaigg et
al, Springer Lecture Notes in Physics {\bf 361} (Springer,
Heidelberg, 1990)}

\refn\WILSONOPE
{K.G.\ Wilson, \pr{179}{69}{1499}}

\refn\SVZ
{M.\ Shifman, A.\ Vainshtein and V.\ Zakharov, \np{120}{77}{385;
448; 519}}

\refn\POLITZER
{H.D.\ Politzer, \np{117}{76}{397}}

\refn\BAGAN
{E.\ Bagan and T.G.\ Steele, \pl{226}{89}{142}}

\refn\PASCUAL
{P.\ Pascual and E.\ de Rafael, \zpC{12}{82}{12}}

\refn\MLMOREV
{M.\ Lavelle and M.\ Oleszczuk, \mplA{7}{92}{3617}}

\refn\COLLINS
{J.C.\ Collins, {\sl Renormalisation}, (Cambridge University Press,
Cambridge, 1984)}

\refn\MLMOSR
{M.\ Lavelle and M.\ Oleszczuk, \mplA{275}{92}{133}}

\refn\STINGL
{M.\ Stingl, M\"unster preprint MS-TPI-94-13;\nl
U.\ H\"abel et al, \zpA{336}{90}{423, 435}}

\refn\ELIAS
{V.\ Elias and M.D.\ Scadron, \prD{30}{84}{647}}

\refn\MSINVIENNA
{M.\ Lavelle and M.\ Schaden, in Ref. \VIENNA}

\refn\BROADH
{D.J.\ Broadhurst et al, \pl{329}{94}{103}}

\refn\WILSON
{K.G.\ Wilson, \prD{10}{74}{2445}}

\refn\FISCHLER
{W.\ Fischler, \np{129}{77}{157}}

\refn\FRADVILK{E.S.\ Fradkin and G.A.\ Vilkovisky, \pl{55}{75}{224}}

\refn\USAGAIN{M.\ Lavelle and D.\ McMullan, \ijmp{7}{92}{5245}}

\refn\BRWE
{L.S.\ Brown and W.I.\ Weisberger, \prD{20}{79}{3239}}

\refn\GROMES
{D.\ Gromes in \prp{200}{91}{127}}

\refn\ADM
{T.\ Appelquist, M.\ Dine and I.J.\ Muzinich, \pl{69}{77}{231};
\prD{17}{78}{2074}}

\refn\LATTPOT
{See, e.g., G.\ Cella et al, \npps{42}{95}{228}}

\refn\CAHILL
{K.\ Cahill and D.R.\ Stump, \prD{20}{79}{2096}}

\refn\LATTCOULP
{C.\ Parrinello et al, \pl{268}{91}{236}}

\refn\LATTCOULH
{M.W.\ Hecht et al, \prD{47}{92}{285}}

\refn\TASSOS
{See M.L.\ Paciello et al, \pl{341}{94}{187} and ref.'s
therein}

\refn\LATTART
{P.\ de Forcrand and J.E.\ Hetrick, \npps{42}{95}{861}}

\refn\CAST
{K.\ Cahill and D.R.\ Stump, \prD{20}{79}{540}}

\refn\LATTBAR
{See, e.g, J.\ Kamesberger et al, Few Body Systems, Suppl.
{\bf 2}, (1987) 529}

\refn\CGLUE
{M.\ Tanimoto, \pl{116}{82}{198}}

\refn\MORECGLUE
{S.\ Ishida et al, \prD{47}{92}{179}}

\refn\OHNUKI
{Y.\ Ohnuki, {\sl Unitary Representations of the Poincar\'e Group and
Relativistic Wave Equations} (World Scientific, Singapore, 1988)}

\refn\JACKIWOLD
{R.\ Jackiw, {\sl Field Theoretic Investigations in Current Algebra,}
in {\sl Current Algebra and Anomalies}, ed.'s S.B.\ Treiman et al.
(Princeton University Press, Princeton, 1985)}

\refn\BARUT
{A.O.\ Barut and R.\ Raczka, {\sl Theory of Group Representations and
Applications} (Polish Sci., Warsaw, 1977)}

\refn\JACKPAPER
{R.\ Jackiw, \prl{41}{78}{1635}}

\refn\JACKMAN
{R.\ Jackiw and N.S.\ Manton, \ann{127}{80}{257}}

\refn\COHERENT
{W.\ Zhang, D.H.\ Feng and R.\ Gilmore, \rmp{62}{90}{867}}
\refn\GODDARD
{P.\ Goddard and D.\ Olive, \ijmp{1}{86}{303}}

\refn\THOOFT
{G.\ 'tHooft, Cargese  Lectures 1979}

\refn\FMS
{J. Fr\"ohlich, G.\ Morchio and F.\ Strocchi, \pl{97}{80}{249};
\np{190}{81}{553}}

\refn\WETT
{B.\ Bergerhoff and C.\ Wetterich, \np{440}{95}{171};\nl
M.\ Reuter and C.\ Wetterich, \np{408}{93}{91}}

\refn\HISTORY
{C.M.\ Bender, T.\ Eguchi and H.\ Pagels, \prD{17}{78}{1086}}

\refn\HERSTORY
{R.D.\ Peccei, \prD{17}{78}{1097}}

\refn\VANB
{P.\ van Baal, \np{369}{92}{259}}

\refn\TWENTY
{D.\ Zwanziger, \np{345}{90}{461}, \nl
M.\ Schaden and D.\ Zwanziger, New York preprint NYU-ThPhSZ94-1}

\refn\HQET
{M.\ Neubert, \prp{245}{94}{259}}

\refn\POMMYEXPT
{R.\ Ball, private communication}

\refn\KISS{L.S.\ Kisslinger and Z.\ Li, \prl{74}{95}{2168}}

\refn\SLOW
{E.\ Bagan, M.\ Lavelle and D.\ McMullan, Barcelona/Plymouth 
preprint UAB-FT-379/PLY-MS-95-08, to appear in Physics Letters B}

\refn\FAST
{E.\ Bagan, M.\ Lavelle and D.\ McMullan, Barcelona/Plymouth 
preprint UAB-FT-384/PLY-MS-96-01, submitted for publication}

}
%%% frontpage

\null
%\leftline{\bigbold DRAFT}
\rightline{UAB-FT-369}
\rightline{PLY-MS-95-03}
\vfill
\centerline{\bigbold CONSTITUENT QUARKS FROM QCD}

\vskip 30pt
\centerline{\smc Martin Lavelle}
\vskip 5pt
{\baselineskip=13pt
\centerline{Grup de F\'\i sica Te\`orica and IFAE}
\centerline{Edificio Cn}
\centerline{Universitat Aut\'onoma de Barcelona}
\centerline{E-08193 Bellaterra (Barcelona)}
\centerline{Spain}
\centerline{(e-mail: lavelle@ifae.es)}
}
\vskip 10pt
\centerline
{\smc and}
\vskip 8pt
\centerline{\smc David McMullan}
\vskip 5pt
{\baselineskip=13pt
\centerline{School of Mathematics and Statistics}
\centerline{University of Plymouth}
\centerline{Drake Circus, Plymouth, Devon PL4 8AA}
\centerline{U.K.}
\centerline{(e-mail: d.mcmullan@plymouth.ac.uk)}
}
\vskip 60pt
\abstract{%
{\bf Abstract.}\quad
Starting from the observation that colour charge is only well
defined on gauge invariant states, we construct
perturbatively gauge invariant, dynamical dressings for individual
quarks.  Explicit calculations show that an infra-red
finite mass-shell
renormalisation of the gauge invariant, dressed propagator is possible
and, further, that operator product effects, which generate a running
mass, may be included in a gauge
invariant way in the propagator. We explain how these fields may
be combined to form hadrons and show how the interquark potential
can now be directly calculated.
The onset of confinement is identified with an
obstruction to building a non-perturbative dressing. We
propose several methods to extract the hadronic scale from the
interquark potential.
Various extensions are discussed.%
}
\bigskip
\centerline{{\it Physics Reports {\bf C279} (1997) 1}}

\vfill\eject

%%%%%%%%%%%%%%%%%%%
%%%             %%%
%%% Section1    %%%
%%%             %%%
%%%%%%%%%%%%%%%%%%%

\secno=1 \meqno=1
\ni
{\bf 1. Introduction}
\bigskip
\ni More than twenty years after the birth of Quantum Chromodynamics
(QCD) it seems fair to say that physicists are convinced that QCD is
the correct theory of the strong interactions\up{\AACHEN}.
The successes of Chromodynamics are many: the $Q^2$ dependence of
structure functions at high $Q^2$ shows the scaling violations that QCD
predicts and the study of the long expected quark and gluon  jets
now forms a sub-field of the physics of strong interactions. Moving
down to lower energies, the somewhat approximate approach of QCD sum
rules has predicted many hadronic parameters, such as masses, rather
well. Many phenomenological models have
also received some input from the insights spawned by QCD.
This theory is so far fully supported by experiment.

And yet despite the many successes of the theory much, if not
most, of low energy hadronic physics cannot be extracted directly
from QCD. The
most visible problem of the theory being the non-observation of the
fundamental QCD particles: the quarks and gluons. This has led to the
dogma of confinement which teaches that quarks and gluons are
permanently confined inside colour singlets --- the observed hadrons.
Another problem is our inability to calculate the structure functions
themselves (in contrast to their $Q^2$ dependence) and, more recently,
there has been controversy about how the spin of the proton is
produced\up{\AACHEN}.

Quarks were initially introduced as constituent particles so as to
introduce a semblance of order into the \lqq particle
zoo\rqq\up{\GELLMANN, \KOKK}.
Mesons are in this picture made up of a quark and an antiquark,
and baryons of three quarks. The hadrons can then be fitted into
multiplets and their spectrum understood. In this way new particles
were even predicted\up{\NEWPARTICLES}. Later a colour quantum number was
introduced to explain the statistics of the baryons. Deep inelastic
scattering experiments\up{\DIS} showed that there were seemingly
freely moving objects inside the nucleus and these \lq partons\rq\
were identified with the naive quarks of the quark model.
The discovery\up{\ASYMPTF} of
asymptotic freedom showed that in non-abelian gauge theories the
coupling grows small at short distances and the quarks of QCD were then
identified with the partons.

QCD is a non-abelian gauge theory modelled upon Quantum
Electrodynamics (QED). The interaction between the quarks
is carried by massless vector (gauge) bosons, the gluons. As is well
known a decisive difference between QED and QCD is that the gluons also
carry QCD (colour) charges and the quarks feel the colour charges of the
gluons. It is, however, not possible to directly identify the quarks
of the
QCD Lagrangian with the constituent quarks which explain hadronic
spectroscopy. The most obvious reason for this is the fact that the
masses of the QCD Lagrangian quarks are, generally
speaking, much smaller than typical hadronic mass scales. (Three times
the Lagrangian mass of the $u$ quark is only about 1\% of the mass of
the nucleon.) In addition, deep inelastic scattering experiments
show that
about half the momentum of the nucleon is carried by particles which,
unlike quarks,
do not have an electric charge. These are believed to be the
gluons. It is also seen that nucleons contain, as well as the three
valence quarks which account for quantum numbers such as baryon
number and electromagnetic charge, quark-antiquark pairs ---
the sea quarks. The interpretation of deep inelastic scattering, it
should be further stressed, relies upon the quarks having small masses.
We see
that although quark fields are now more solidly rooted, in the theory of
QCD, we have lost touch with the initial motivation for their
introduction, viz. hadronic spectroscopy.

The success of QCD together with that of the
spectroscopic description of hadrons in terms of
more massive, constituent quarks leads in fact to our modern picture of
hadrons, where the three valence quarks of the nucleon, say, are
dressed with gluons and quark-antiquark pairs in
such a way that they become much heavier.
Although the very language being used leads to an appealing picture
of this dressing as a swirling cloud of
glue and sea-quarks, the precise nature of the dressing is, to say
the least, hard to pin
down within the context of QCD.
This dressing, with colour
charged objects, has in the first instance to explain the successes of
the constituent quark model. As such it has to be constructed in
such a way that the quantum numbers
of the hadrons are still produced. Insofar as the constituent quark
picture applies, one expects quasiparticle type constituent quarks to
also have quantum numbers such as colour charges and spin.
In addition, this dressing must be such that it is compatible with the
confinement of coloured objects.

What is required is a clear physical reason for the need for such a
dressing in QCD. Our view is that if  QCD is  the correct description
of strong interactions then its structure should
restrict the dressing so as to conform
with the above observed properties. The main aim of this paper is to
investigate
the guiding principles behind the process of dressing such  charged
fields in a gauge
theory. In particular,
we will discuss why one in fact {\it must} dress
quark (and gluonic) fields and how far this dressing may be taken.
We will argue that
these results point to a systematic approach to dressing both quarks and
gluons  that make it at least plausible that a constituent
quark picture can emerge from QCD.

Our first motivation for dressing the Lagrangian  quarks is to note that
they are not
gauge invariant. The gauge symmetry shows that not all fields are
physically significant. To produce a gauge invariant quark field it is
necessary to dress the fermions with coloured gluons. In QED electrons
must
also be dressed
with photons, as will be discussed below, but a major difference is
that photons are not electrically charged, in contrast to the colour
charged gluons. We will, however, see that
the gluonic dressing is such that it actually gives a quark a well
defined
colour charge.
Since we want to  use the standard tools of theoretical
physics, our approach to dressing quarks will be perturbative. We
will give
explicit dressings to low orders in the coupling and then show how
they may be systematised.

One  views dressing as surrounding  the charged particle
with  a cloud
of gauge fields. We will see that, at least in this perturbative
description,
this cloud spreads out over the
whole space, resulting in  a highly non-local structure. This is not
unexpected in QED\up{\STROCCHI}, but in QCD we will see that there is
a non-perturbative obstruction to the construction of this
dressing\up{\LMQUARK}.
This breakdown of the quark description should be identified with
confinement. We cannot in the non-perturbative domain talk about quark
fields.  This is to be contrasted with the situation in
both  QED and spontaneously broken non-abelian gauge theories
(i.e., with Higgs' scalars) 
where such problems do not arise\up{\LMSBT}.

As well as gauge invariance,
dressing is necessary because use of, for example,
the free electron as an asymptotic field in QED leads to infra-red
divergences (see, for example, the summary in Supplement~4 of
Ref.\ \JAUCH).
Essentially the long range of the Coulomb
force, which only falls off as $1/R$, is responsible for
this\up{\DOLLARD,\KULISH}. In QCD the infra-red problems seem more
severe.
We will argue, and also show by explicit calculation, that the gauge
invariant dressing can remove the infra-red problems.

A physically quite reasonable, but at first sight surprising, aspect of
this dressing is that it
depends upon the velocity of the quark. In the context of QED this is
simply a reflection of the fact that the electric and magnetic fields
of a charged particle vary with its velocity. We will demonstrate that
this non-covariance is really unavoidable.
 A technical consequence of
this, though, will be that our formulation will be non-covariant.
In conjunction
with the non-locality of the dressing, this means that we will need to
carefully show that standard field theoretic methods can still be applied
to this description of constituent quarks.

Some parts of the work reported here, and certain closely related
topics, have already been reported, albeit in rather terser form,
elsewhere\up{\LMSYM, \LMPROP, \LMQUARK,\LMSBT}.
However, we have tried to make this paper
self-contained.
As we have stated above, the basic theme in this approach is to use  gauge
invariance as the guiding principle in the  construction of the dressings
appropriate
for  the  charged particles of the system and, in particular, for the
quarks
and gluons. Of course, it is almost a truism to insist upon the gauge
invariance of all physical quantities
in a  gauge
theory. What is different here is that we are requiring
gauge invariance for not just S-matrix elements, but for all the basic
constructions in the theory, i.e., also for the propagators and the states
of
the system.
The idea that gauge invariance should dictate how
electrons are dressed goes back to Dirac (see Ref.\ \DIRACINCANADA and 
Section~80 of Ref.\ \DIRAC).
There are understandable reasons for why
this suggestion seems to have  gone largely unnoticed. In the context of
QED
the Bloch-Nordsieck theorem allows one to, at least perturbatively,
circumvent the infra-red problems associated with the use of the (gauge
non-invariant)
Lagrangian fermion. Given that Dirac's expression for the electron 
is, as we will see, both non-local and
non-covariant, it is tempting to dismiss it as a purely  formal
construction that cannot be accommodated within the standard perturbative,
field
theoretic catechism. We argue, though, that in QCD such a pragmatic view
is
less defensible. Indeed, it has long been speculated that the
additional infra-red
problems  found there could lead to a  confinement mechanism  through some
form of
infra-red slavery\up{\WEINBERG-\FRITZSCH}. However, as this is usually
approached
within a conventional, local, covariant formulation of quantum field
theory, Dirac's non-local, non-covariant,
abelian construction has had little impact.

This is not to say, though, that
Dirac's arguments have been completely ignored. We have become aware of
several groups that have either been directly influenced by Dirac's
observations, or 
have arrived at similar points of view through their own
investigations\up{\Xtra,\XXtra}. In Ref.\ \SHAB\  the obstruction
to extending Dirac's treatment of the electron to non-abelian gauge
theories
has also been noted, and its relation to confinement discussed.
Gauge invariance is  also seen to  take a central
role in
Ref.'s\ \OTHERRUSSIAN--\REST\ albeit from somewhat different
perspectives.
We should also,
for completeness,  mention the work of Kugo and Ojima\up{\KUGO} where
confinement is related to gauge invariance (or more properly, BRST
invariance), although the connection of
that approach to the one presented here is far from clear to us.
None of these works, though, deals with the perturbative aspects
of the dressing. Hence they are unable to interpret the construction in
terms of constituent
quarks.

We will now outline the structure of this paper and highlight the key
results. After this introduction,
in Sect.~2, we summarize the successes of the constituent quark model.
This
short review will serve as a constant source of motivation for the more
technical arguments that will follow. Then, in Sect.~3, the basic
properties
of QCD will be discussed. As well as allowing us to fix our notation,
this
gives us an opportunity to present several important aspects of gauge
theories which will be central to our discussion. In particular, we will
discuss the definition of colour
charges, proving that the colour charge only has a well defined
meaning on gauge invariant states. It immediately follows that the
Lagrangian quarks and gluons do not have well defined colour charges.
We then recall that charged states
are always non-local and that their Lorentz transformations are not
the standard ones. Having developed sufficient background material,
in Sect.~4 we present explicit dressings of quarks to lowest order
in perturbation theory, and show that these dressed quarks have well
defined colour charges. This construction makes manifest the
non-locality and non-covariance expected from the dressing, and we
show that such dressed quarks may be combined to form colourless
hadrons in
the way commonly done in the constituent quark model. In Sect.\ 5 we
systematise the results of the preceding section, and  show how gluons
may also be dressed. An algorithm for dressing colour charges is given.
The central theme of this section, and one which underlies this paper,
is the close relationship between dressings and gauge fixings.
This allows us to then identify the
existence of the Gribov ambiguity with the breakdown of the
non-perturbative dressing of a quark. This unavoidable feature of QCD
means that it is impossible to construct asymptotic quark states, and
single quarks cannot be observed.

In Sect.\ 6 we study the practical use of the dressed quark. We see in
a one loop calculation that a dressed quark's propagator is gauge
invariant and infra-red finite in a  mass shell renormalisation scheme
where the on-shell quark has the velocity appropriate to its dressing.
This allows
us to identify Dirac's original dressing as that of a static particle.
We further see that it is
possible to incorporate quark condensate effects in a gauge invariant
fashion into the dressed quark propagator and that they introduce a
running mass. In Sect.~7 the interaction between  static quarks is
discussed. We show that the perturbative interquark static potential
may, with the help of the dressings, be obtained in a simpler fashion
than is usually the case. We make two proposals for how to extract the
confinement scale in such a heavy meson system.

The final part of this paper deals with how light quarks may be dressed.
In Sect.\ 8 the structure of Lorentz transformations in gauge theories
is investigated. We see that the Lorentz generators must be modified
when they act on charged states. These results are then brought to
bear on the construction of boosted charges in Sect.\ 9. Gauge invariant
dressings for moving charges are given. In Sect.\ 10 a summary of our
results is
presented, along with some speculations on future directions.
Finally,
there is an appendix where the dressing for a static quark to
order $g^3$ is explicitly derived.

\vfill\eject

%%%%%%%%%%%%%%%%%%%
%%%             %%%
%%% Section2    %%%
%%%             %%%
%%%%%%%%%%%%%%%%%%%

\secno=2 \meqno=1
\ni
{\bf 2. The Constituent Quark Model}
\bigskip
\ni The aim of this section is to rapidly review the successes of the
constituent quark model. More detailed discussions of this
model may be found in Ref.'s \POVH--\FLAM. These results show
that, despite
confinement, we can still use quarks in low energy
hadronic physics --- although these constituent quarks
are not the bare Lagrangian fermions.

Hadrons may be essentially divided into those containing heavy
$c$ and $b$ quarks\note{The $t$ quark decays too rapidly for
hadrons to form.}\ and those made up of light quarks. Heavy
quark-antiquark pairs, quarkonia, have masses which are roughly equal
to twice
those of the valence quarks. These mesons may be described in terms
of non-relativistic potential models which describe confinement in
terms of a phenomenological potential. The success of non-relativistic
models, and their
ability to describe the splitting of the energy levels in terms of a
chromomagnetic interaction, is strong evidence that it makes sense to
discuss these hadrons in terms of their constituent quarks.
The quarks here are dressed quarks
but their masses are believed to be close to those of the Lagrangian
masses.

It is much more surprising  that we can also describe
light hadrons in terms of constituent quarks. Recall that the
Lagrangian masses of the three light quarks may be
estimated to be: $m_u\approx4\,{\rm MeV}$, $m_d
\approx 7\, {\rm MeV}$ and $m_s\approx 150\,{\rm MeV}$ from
chiral symmetry arguments\up{\GASSA}.
These are more or less negligible
compared to the standard hadronic masses. The physical scale which
must generate the hadronic masses is the QCD scale,
$\Lambda_{\rm QCD}$, which
enters the theory through renormalisation.
It turns out, however, to be possible to
generate the spectrum of light hadrons remarkably well
by assuming they are composed of quark-antiquark pairs or three
quarks, which quarks now, however, have larger, so-called, constituent
masses. It is, of course, very well known\up{\GELLMANN,\KOKK}
that the hadrons can be fitted into multiplets such as the
vector octet. A corollary of this is that certain states are forbidden,
i.e., they cannot be constructed in the simple quark model out of three
spin one-half quarks or from a quark and an antiquark. The problems
with the
experimental signals for these
states is another confirmation of the applicability of the simple
(coloured) quark model.
At the simplest quantitative level the quark model
corresponds to noting that the mass of the
rho meson is very roughly equal to two thirds that of the
nucleon and the observation that the difference in the masses of a
meson formed from two $u$ or $d$ quarks and one constructed out of
a strange quark and a single $u$
or $d$ quark is roughly equal to the mass gap
between the latter and a meson solely built out of $s$ quarks.
However, much greater accuracy can be attained.

Let us now briefly recall the quantitative precision of the quark
model\up{\POVH, \GASI}. The
model predicts that the masses of the mesons are given by
$$
m_{\bar q'q}=m_{\bar q'} +m_q+\Delta^1_{{\sevenrm spin}}
\,,
\eqn\mmeson
$$
and those of the baryons should be described by
$$
m_{qq'q''}=m_{q}+m_{q'}+m_{q''}+\Delta^2_{{\sevenrm spin}}
\,,
\eqn\mbaryon
$$
where the $m_q$ are the {\it constituent} masses of the quarks and
the $\Delta^i_{{\sevenrm spin}}$ describe the hyperfine splitting.
This term emanates from a spin-spin
interaction produced by one gluon exchange, we shall not go into its
exact form here, suffice it to say that it
depends upon the masses of the quarks, their spins and
the square of the wave function of a quark pair for vanishing
separation\up{\GASI}. (For baryons this term entails a sum over all the
possible quark pairs.) The model has therefore three free parameters:
the mass of the constituent $u$ and $d$ quarks (assumed to
be equal), that of the $s$ quark and the square of the wave function.

The following two tables contain the results of fits for mesons and
baryons respectively. (The fits are taken from Ref.\ \GASI, and
experimental values\note{These
experimental values are slightly rounded
off and, where necessary, are averages over the various charge states
of the hadrons.}\ from Ref.\ \MONT.)
For the mesons one uses the following constituent quark masses:
$m_u=m_d=310\,{\rm MeV}$ and $m_s=483\,{\rm MeV}$
and obtains (in units of MeV):
\bigskip
\centerline{\vbox{
\halign{
\bf #\hfil\quad&&\hfil#\quad\cr
Meson & $\pi$ & $K$
& $\eta$ & $\rho$/$\omega$ & ${K^{*}}$
& $\phi$\cr
Mass (Expt.) & 137 & 496 & 548 & 769/782 &
894 & 1019 \cr
Mass (Theory) & 140 & 485 & 559 & 780 & 896 & 1032 \cr
}
}
}
\noindent while the baryons may be fitted with slightly different
parameters. From the masses $m_u=m_d=363\,{\rm MeV}$ and
$m_s=538\,{\rm MeV}$ one finds:
\bigskip
\centerline{\vbox{
\halign{
\bf #\hfil\quad&&\hfil#\quad\cr
Baryon & $N$ & $\Lambda$ &
$\Sigma$ & $\Xi$ & $\Delta$ &
${\Sigma^{*}}$
& ${\Xi^{*}}$
 & $\Omega$ \cr
Mass (Expt.) & 939 & 1116 & 1193 & 1318 & 1232 & 1385 & 1533 & 1673 \cr
Mass (Theory) & 939 & 1114 & 1179 & 1327 & 1239 & 1381 & 1529 & 1682 \cr
}
}
}
\noindent The goodness
of these fits is remarkable. They show that hadrons may be
very well described in terms of dressed, constituent quarks. The
constituent masses differ slightly for the quarks in mesons and those
in baryons. This is not unexpected since they are confined in quite
different
systems. However, the similarity of the masses is great and supports
the view that the dressing of the quarks is primarily
responsible for the
constituent mass which is only somewhat modified by the dynamics
of confinement.

Before looking at further quantitative predictions, we need to recall
that the statistics of the constituent quark picture was responsible
for the introduction of the colour quantum
number\up{\COLOUR}. The combination of
fermionic quarks could only produce a ground state particle like the
$\Delta^{++}$ (made from three $u$ quarks, with zero angular momentum
and parallel spins, i.e., its wave function is completely symmetric) if
colour was introduced and one postulated a totally antisymmetric
colour wave function, $\epsilon^{abc}$, for the constituent
quarks. Once we introduce colour we may construct general
baryonic wave functions with the necessary symmetry properties.

Using then the spin wave functions of three coloured
constituent quarks in the various baryons, we can also predict
the magnetic moments of the baryons\up{\GASI}. The
above baryons all have zero angular momentum and their magnetic moments
may all be assumed in the framework of the quark model to be a direct
sum of the magnetic moments of the constituent quarks. These last
have magnetic moments
$$
\mu_q=\frac{e_q}{2m_q}
\,,
\eqn\muq
$$
where $e_q$ is the electric charge of the quark. One may now, for
example, find the ratio of the magnetic moments of the proton and the
neutron from their wave functions without reference to the quark masses
(assuming only that $m_u=m_d$) and a good agreement is obtained. As far
as the absolute values are concerned, one obtains using the constituent
quark masses which fitted the masses of the baryons the following
predictions (in units of nuclear magnetons):

\bigskip
\centerline{\vbox{
\halign{
\bf #\hfil\quad&&\hfil#\quad\cr
Baryon & $p$ & $n$ & $\Lambda$ & $\Sigma^+$ &
$\Sigma^-$ &
$\Xi^0$ &
$\Xi^{-}$ \cr
Moment (Expt.) & 2.79 & -1.91 & -0.61 & 2.46 & -1.16 & -1.25 &
-0.65  \cr
Moment (Theory) &
2.79 & -1.86 & -0.58 & 2.68 & -1.05 & -1.40 & -0.47 \cr
}
}
}
\noindent This
accuracy is again very satisfactory, especially considering the
extraordinary simplicity of the model. We note also that these last
predictions test not only the size of the constituent masses but also
the symmetry properties of the baryonic wave functions. Further
predictions of the quark model may be found elsewhere\up{\POVH, \GASI}.

\bigskip
The above agreement between experimental data and the predictions of
the constituent quark model shows that it is useful to think of
hadrons as being made up of dressed quarks; this although the
confinement mechanism is conspicuous by its absence in the above
considerations. Additionally these are not the current quarks of
QCD: the masses are quite different. Furthermore, as we shall see in
the following section, the Lagrangian
quarks of QCD have ill-defined colour charges because of
their gauge dependence, a defect that is at odds with
the constituent quark model.

Of course the constituent quark model breaks down if high energy probes
are used. This is because such probes test the structure of the
constituent quarks and the naive model lacks this structure. The need
for structure is seen most simply in the existence of the need for two
different limiting masses --- the current and constituent masses. The
dressing of the constituent (valence) quarks must be made up of
gluons and sea quarks.
At short distances and very large momenta the
coupling is small and the dressing may be to a good approximation
neglected. At larger distances we know that the coupling
increases and that perturbation theory will collapse, at this stage
the quark picture we propose will also break down and we can only
talk of hadrons. This should also
be reflected in one's choice of effective models.

In the next section we will introduce the QCD concepts required for
this paper, in particular we will discuss colour charges and states. In
the light of this treatment we will be able to start dressing a
Lagrangian quark.
\vfil\eject

%%%%%%%%%%%%%%%%%%%
%%%             %%%
%%% Section3    %%%
%%%             %%%
%%%%%%%%%%%%%%%%%%%

\secno=3 \meqno=1
\ni
{\bf 3. Colour charges and states in QCD}
\bigskip
\ni
In this section we review the structure of QCD as a field theory, paying
particular attention to its gauge
symmetries and their generators.  We will show that colour
quantum numbers can only be defined on gauge invariant states.
This has direct consequences for any description of coloured fields,
i.e., quarks and gluons. Furthermore, we will see that gauge
invariant charged states are necessarily both non-local and
non-covariant. Hence any approach to constituent quarks necessitates
going beyond the traditional, local, covariant formulation of QCD.
The implications of this general account of charged states will all
be visible in the explicit solutions which we will start to develop
in the next section.

As discussed in the introduction, QCD is now widely accepted as the
correct field theoretic description
of the strong interaction. It is
a gauge theory  based upon the non-abelian structure group $SU(3)$ rather
than the abelian group $U(1)$ familiar from QED. This non-abelian
structure manifests itself
in the fact that the gauge fields
interact not just with the matter
fields  but also with themselves.
The non-linearities inherent in such
self interactions obstruct any simple extrapolation from our
experience with the structure of QED to QCD.

The basic fields in QCD are the vector fields,
$A_\mu^a$, and the spinor fields $\psi_n$. In these the index
$a$ ranges over the values
$1,\dots,8$
(the dimension of the $SU(3)$ group), while $n$ takes three distinct
values (the dimension of $SU(3)$'s fundamental representation). (We
will, on occasion, deal with more general gauge theories in which case
$SU(3)$ is replaced by an $SU(N_c)$ group.)
These fields are conventionally identified
with the gluons and the quarks. Although we shall see that this
identification is, at best, misleading, it is so ingrained into the
physics literature that we will also loosely refer to these fields in this
way. When a distinction is important, though, we shall refine this
terminology and refer to these fields as the Lagrangian quarks and gluons.

Let $T^a$  be a basis of the Lie algebra $su(3)$. We
take these to be anti-hermitian operators in some representation of
$SU(3)$ which satisfy the Lie algebra commutator
$$
[T^a,T^b]=f_{abc}T^c\,,\eqn\commtwo
$$
where the $f_{abc}$ are the structure constants of $su(3)$. For the
quark fields the appropriate representation is the fundamental
one where we take $T^a_{mn}=(\lambda^a/2i)_{mn}$, with $\lambda^a$ the
Gell-Mann matrices. For the gauge fields we use the adjoint
representation with \hbox{$T^a_{bc}=-f_{abc}$}.
The gauge fields can be viewed in a more compact way as a $su(3)$
valued field $A_\mu$, where\up{\JACKIW}
$$
A_\mu:=A_\mu^aT^a\,.\eqn\lietwo
$$
We can recover the components via $A^a_\mu=-c\Tr(A_\mu T^a)$, where $c$
is a
constant determined by the representation (in the fundamental
representation $c=2$).

The field strength is given by
$$
F_{\mu\nu}^a=\pa_\mu A^a_\nu-\pa_\nu A^a_\mu+gf_{abc}A^b_\mu
A^c_\nu\,,\eqn\no
$$
where $g$ is the coupling constant. The Yang-Mills Lagrangian is then
$$
\L_{_{\rm YM}}=-\tfrac14F^{\mu\nu\,a}F_{\mu\nu}^a\,.\eqn\no
$$
The covariant derivative acting on the matter fields is
$$
(D_\mu\psi)_n=\pa_\mu\psi_n+gA^a_\mu T^a_{nm}\psi_m\,,\eqn\no
$$
and the matter Lagrangian is
$$
\L_{_M}=\bar\psi(i\c^\mu D_\mu-m)\psi\,.\eqn\no
$$
The QCD Lagrangian is then the sum of the Yang-Mills and matter ones:
$$
\L_{_{\rm QCD}}=\L_{_{\rm YM}}+\L_{_{\rm M}}\,.
\eqn\qcdlagtwo
$$
The equations of motion which follow from this are
$$
(D_\mu F^{\mu\nu})^a=gJ^\nu_a\,,\eqn\eqmtwo
$$
where the covariant derivative is now that appropriate to the adjoint
representation, and we have the matter current
$$
J^\mu_a:=-i\bar\psi\gamma^\mu T^a\psi\,.\eqn\mcurrenttwo
$$
In addition to this current we can also define the conserved
singlet vector current
$$
J^\mu=\bar\psi\gamma^\mu\psi\,.\eqn\singlettwo
$$

In the free quark theory the matter current is also conserved
and it may be used to determine  the colour charge carried by
such quarks.
In this way combinations of quarks
can be constructed
with a prescribed colour and, in particular,
globally colourless hadrons
may be given a non point-like definition.
However, when we couple
the quarks to gluons in QCD, we see from (\eqmtwo) that this
 matter current
is  not conserved.
 As such, its role in characterizing coloured states
has been lost. This reflects the fact that gluons also possess colour
charges.
To find an appropriate conserved colour charge in QCD we  first need to
review the
structure of some of the  symmetries of the Lagrangian (\qcdlagtwo).

Among the many symmetries of the QCD Lagrangian, the gauge symmetry is
surely primus inter pares. We recall that, by construction, (\qcdlagtwo)
is invariant  under the transformation
$$
\eqalignno{
A_\mu\to\Au_\mu:=&\,U^{-1}A_\mu U+\frac1g U^{-1}\pa_\mu
U&\eqnn\atranstwo\cr
\noalign{\hbox{and}}
\psi\to\psiu:=&\,U^{-1}\psi\,,&\eqnn\ptranstwo\cr
}
$$
where $U(x)$ is an element of the group $SU(3)$ (in the
appropriate representation) for each point $x$ in spacetime.
These gauge transformations contain the rigid transformations where
$U(x)=U$ is a constant group element. For such rigid transformations
we have
$$
\eqalignno{
A_\mu&\to U^{-1}A_\mu U\,,&\eqnn\rigidatwo\cr
\psi&\to U^{-1}\psi\,.&\eqnn\rigidptwo\cr
}
$$
The transformations (\atranstwo--\rigidptwo) are symmetries of the
action, as such they have associated with them conserved Noether
currents. Writing $U(x)=e^{\theta(x)}$, where $\theta=\theta^aT^a$,
we find that the Noether current for gauge transformations is
$$
j^\mu_\theta=-\frac1gF^{\mu\nu\,a}(D_\nu\theta)^a+J^\mu_a\theta^a\,.
\eqn\noethertwo
$$
For the
rigid transformations the parameter $\theta^a$ is a constant which
we can discard to give the conserved current
$$
j^\mu_a=-f_{abc}F^{\mu\nu\,b}A^c_\nu+J^\mu_a\,.\eqn\noetherrigidtwo
$$
The conservation of this current is a consequence of Noether's theorem
and follows directly from (\eqmtwo) which can be written as $\pa_\mu
F^{\mu\nu\,a}=gj^\nu_a$.

{}From (\noethertwo) we expect to be able to identify the generator of
gauge transformations with the charge
$$
\eqalign{
G(\theta)&=\int d^3x\,j^0_\theta(x)\cr
&=\int d^3x\, G^a(x)\theta^a(x)\,,
}
\eqn\nchargetwo
$$
where
$$
G^a(x)=-\frac1g(D_iE_i)^a(x)+J^0_a(x)\eqn\gaussgentwo
$$
and $E^a_i(x)=-F^a_{0i}(x)$ is the  chromo-electric field.

Using
the basic, non-vanishing, equal time (anti)commutators
$$
\eqalignno{
[E_i^a(x),A^b_j(y)]&=i\d^{ab}\d_{ij}\d(x-y)&\eqnn\ccrtwo\cr
\noalign{\hbox{and}}
[\psi^\dagger(x),\psi(y)]_{_{+}}&=\d(x-y)\,,&\eqnn\acrtwo\cr
}
$$
it is straightforward to show that the Noether charge (\nchargetwo)
generates the transformations
$$
\eqalign{
\d_\theta A^a_i(x):=&\,[iG(\theta),A^a_i(x)]\cr
{}=&\,\frac1g(D_i\theta)^a
}
\eqn\smallatwo
$$
and
$$
\eqalign{
\d_\theta \psi(x):=&\,[iG(\theta),\psi(x)]\cr
{}=&-\theta\psi(x)\,.
}
\eqn\smallpsitwo
$$
These are the infinitesimal forms of the gauge transformations  on
the $A_i$ and $\psi$ fields, but we have {\it not} recovered the gauge
transformation on the time component $A_0$ of the gauge field. The
reason for this is that this field is really not a dynamical variable.
 In the QCD Lagrangian (\qcdlagtwo) it enters as a multiplier field
(this is most easily seen from the fact that its momentum $E_0^a$ can
only be of the form $F^{00\,a}$, which trivially vanishes). This means
that this gauge theory is a constrained system and care must be taken
in identifying physical degrees of freedom. Variation of the Lagrangian
(\qcdlagtwo) with respect to the multiplier field $A^a_0$ leads to the
further constraints
$$
G^a(x)=0\,,\eqn\gausstwo
$$
which are the non-abelian generalizations  of Gauss' law.

In the quantum theory we cannot take the constraint (\gausstwo) to hold
as an operator
equation since this is incompatible with the above commutation
relations, rather it is implemented as a restriction on the allowed
physical states of the system. So, heuristically, a state $\ket\psi$
is said to be physical if
$$
G^a(x)\ket\psi=0\,.\eqn\physstatestwo
$$
That this is at least consistent follows from the commutator
$$
[G^a(x),G^b(y)]=-if_{abc}G^c(x)\d(x-y)\,,\eqn\no
$$
which reflects the fact that Gauss' law is also the generator of the
gauge transformations.

On such physical states the expectation values of the
constraints  are zero: $\bra\psi G^a\ket\psi=0$. This we can
interpret as naturally
corresponding to the classical condition (\gausstwo). We note, however,
that we may not use this weaker condition to replace (\physstatestwo).
If we were to use this superficially
more appealing  property to actually
define the physical states then it is simple
to see that such a restriction on the expectation
values does not pick out a linear subspace of  states, i.e., if
$\bra{\psi_1}G^a\ket{\psi_1}=\bra{\psi_2}G^a\ket{\psi_2}=0$, then
$(\bra{\psi_1}+\bra{\psi_2})G^a(\ket{\psi_1}+\ket{\psi_2})=
2{\rm Re}\bra{\psi_1}G^a\ket{\psi_2}$ which does not necessarily
vanish.
Hence the weaker condition would imply that there would
be no superposition principle.

How are we to interpret the condition (\physstatestwo)?
This  seems to imply that
the physical states are gauge invariant, but even in QED this would
lead to difficulties. The problem is that the  gauge transformations
also include the rigid ones where $\theta$ is a constant. In this case
the
abelian charge becomes the electric charge, and
physical states annihilated by this can
only have charge zero. This is clearly something of an obstacle to our
aim
of constructing  charged states.
To avoid this we must have a
non-trivial action of the rigid transformations on the physical
states.
To determine then what (\physstatestwo) means in
terms of the gauge invariance of the system we will analyse how the
colour charge is to be defined in QCD.

{}From (\noetherrigidtwo) we would expect to identify the colour charge,
$Q=Q^aT^a$, with
$$
\eqalign{
Q^a&=\int d^3x\,j^0_a(x)\cr
&=\int d^3x\,\(J^0_a(x)-f_{abc}E^b_i(x)A^c_i(x)\)\,.
}\eqn\cchargetwo
$$
This is, by construction, conserved but it is {\it not\/} gauge invariant.
{}From our discussion in Sect.\ 2 we know that constituent quarks have a
well-defined colour and the question now naturally arises: how can we
reconcile this with the gauge structure of Chromodynamics? The answer
hinges on taking the gauge invariance of any physical state seriously.
Extending an argument due to Jackiw (see Exercise 2.5 in
Ref.~\JACKIW), we now show that
if this charge is acting
on a physical state, as defined by (\physstatestwo), then we have
$$
j^0_a=\frac1g\pa_iE^a_i\,,\eqn\no
$$
and we so find that
$$
\eqalign{
Q&=\frac1g\int d^3x\,\pa_iE_i\cr
&=\frac1g\lim_{R\to\infty}\int_{S^2_{_R}} d{\bold s}\cdot{\bold E}\,,
}\eqn\defcchargetwo
$$
where we have used Gauss' theorem to get an integral over the spatial
two-sphere $S^2_{_R}$ of radius $R$. Under a gauge transformation
we have $E_i\to U^{-1}E_iU$ so that the charge  $Q$ becomes
$Q^{\scriptscriptstyle U}$, where
$$
Q^{\scriptscriptstyle U}=\frac1g\lim_{R\to\infty}\int_{S^2_{_R}}
d{\bold s}\cdot U^{-1}{\bold E}U\,.\eqn\chrge
$$
In order to be able to extract  the group element from this integral
we must assume that it tends to a constant $U_\infty$ in a direction
independent way. Then
$$
Q^{\scriptscriptstyle U}=U^{-1}_\infty QU_\infty\,,
\eqn\QOK
$$
and we see that the colour charge acting on physical states is gauge
invariant under those gauge transformation that at spatial infinity
tend to a constant $U_\infty$ which lies in the centre of SU(3).
We postpone for the moment a detailed analysis of the structure  of
these gauge transformation, and content ourselves with the observation
that if $U_\infty=1$, the identity element of $SU(3)$, then the colour
charge (\cchargetwo) is invariant. 

We now define the group of
{\it local gauge transformations} to be those gauge transformations
that become the identity at spatial infinity. In addition we have the 
rigid (global) gauge transformations which are those that are 
constant in space-time. Note the important distinction between local 
and rigid gauge transformations: to define colour charge we are forced 
to restrict local gauge transformations to those which reduce to unity 
at spatial infinity, this means that the rigid transformations are {\it 
not} a special class of the local ones in this description. 
We thus see that in QCD we have to 
consider those gauge transformations which are a combination of such 
local and rigid transformations. 

We know from the instanton
structure of QCD, (see also  Sect.~5), that the local gauge
group is disconnected\up{\JACKIW}.
This means that it is only the part of the group
connected to the identity
that can be generated from the infinitesimal gauge transformations and
hence Gauss' law.
Thus, in order for the
colour charge to be well defined, the physical states, as defined by
(\physstatestwo), are invariant under those local gauge transformations
that belong to the identity component of the group of all gauge
transformations.
In a moment we will see that the BRST charge gives a more
succinct characterization of these states.

In summary, we have seen that the conserved colour charge (\cchargetwo)
 is only well defined acting on locally gauge invariant states. Under
 rigid transformations this charge transforms in the same way as the
matter charge constructed out of the matter current (\mcurrenttwo).
 From its construction, it is clear that we should expect both the
quarks and the gluons to carry colour charge. This is in marked
contrast to the situation in QED where it is only the electron that
has an electric charge, not the photon.

The gauge sector of QCD was initially described by a vector field
$A_\mu^a$ with $4\times8$ degrees of freedom (per space time point).
We have  seen that a kinematical consequence of gauge invariance was
that these are reduced to $2\times8$ degrees of freedom. (There are
$2\times8$ constraints coming from the primary constraint $E^a_0=0$
and Gauss' law.) This is analogous to the reduction in QED from the
four-potential to the two transverse polarizations for the photon.
This discussion, though, has been at the expense of manifest Lorentz
invariance. Now quantum field theory, to a large extent, owes its
success to the realization that the  ultra-violet structure of the
theory is best handled in a manifestly covariant formulation. That
is, in a perturbative description, the programme of  regularization
and then renormalization
has only been successfully carried out in local, manifestly covariant
theories.

To circumvent the non-covariance inherent in the dynamical content of the
QCD Lagrangian (\qcdlagtwo) a Lorentz invariant gauge fixing term may be
introduced. A convenient choice is the Lorentz class of gauges
$$
\L_{_\xi}=-\frac1{2\xi}(\pa_\mu A^{\mu\,a})^2\,,\eqn\covtwo
$$
which can be viewed as a generalization of the Lorentz gauge, $\pa_\mu
A^{\mu\,a}=0$. The addition of this gauge fixing term has the desired
effect of destroying the gauge invariance of the QCD Lagrangian,
resulting in all the components of the gauge field being  dynamical.
This democracy in the dynamics means, though, that we are
not allowed to just blithely add such a gauge fixing term
to the Lagrangian, since it has evidently changed
the physical content of the theory.
On top of this, maintaining covariance implies that the
state space will have an indefinite metric. In QED, in for example
the Feynman gauge (where $\xi=1$ and  $\square\pa_\mu A^\mu=0$),
one can project out the unphysical, negative norm states using the
Gupta-Bleuler positive frequency condition $(\pa_\mu A^\mu)^+\ket\psi=0$.
In QCD, though, this is not possible and additional ghost degrees of
freedom must be introduced. Heuristically, as ghosts are fermionic
scalars, each one contributes minus one degrees of freedom. So adding
eight ghost fields $c^a$ and eight anti-ghost fields $\bar c^a$ to
the gauged fixed QCD Lagrangian will result in a theory with the
expected $4\times8-2\times8$ degrees of freedom. The actual proof
of this claim is far from transparent. The key step in the argument
is to note that the ghosts contribute a term
$$
\L_{_{\rm ghost}}=-\bar c^a\pa^\mu D_\mu c^a\,,\eqn\no
$$
to the full QCD effective Lagrangian
$$
\L_{_{\rm eff}}=\L_{_{\rm QCD}}+\L_{_\xi}+\L_{_{\rm ghost}}\,.
\eqn\efftwo
$$
Although gauge invariance has been lost, the action constructed out
of this Lagrangian is  invariant under  BRST transformations. Acting
on the gluons and quarks these are:
$$
\eqalign{
\d A^a_\mu&=\frac1g (D_\mu c)^a\cr
\d\psi&=-c\psi\,.
}
\eqn\brsttwo
$$
Comparing these with (\smallatwo) and (\smallpsitwo), it is clear that
there is a close connection
between these global (but odd) transformations (\brsttwo) and the
original gauge
transformations  of the system. This simple observation lies at the
heart of the success of the BRST method.

This symmetry is generated by a BRST charge $Q_{_{\rm BRST}}$ which is
nilpotent, $Q_{_{\rm BRST}}^2=0$. The physical states are now defined as
being BRST invariant; a condition  which can be shown\up{\NAKANISHI,
\HENNEAUX} to recover the
states (\physstatestwo).
The requirement that the BRST invariant states
have a well defined colour charge now translates into the condition that
the ghosts tend to zero at spatial infinity. So, for example, a ghost
field configuration that only depends on time, $c^a=c^a(t)$, is not
allowed. From (\smallatwo), (\smallpsitwo) and (\brsttwo) we see that
for states built out of the gauge and matter fields, BRST invariance
is a precise and rather economical way of imposing the condition that
these states are invariant under those local gauge transformations that
are in the identity component of the group of gauge transformations.

In the rest of this paper we will work almost exclusively with the
Lorentz invariant effective QCD Lagrangian (\efftwo). However, we
will also persist, rather perversely,  in occasionally describing
the physical states as being gauge invariant. This must always be
translated into the statement that the states are BRST invariant.
Our justification for such an `economy with the truth' is that
there is little intuition in the wider physics community for BRST
invariance: gauge invariance is much more appealing physically. As
long as one is aware of all the caveats needed when applying gauge
invariance, this should not lead to any confusion.

The above discussion shows that there exists a local, covariant
formulation of QCD. In this description the ultra-violet structure
is manageable --- renormalization can be proven. What is more, the
colour charge is an observable when restricted to BRST invariant
states.

Starting from the vacuum, we can build up physical states by
applying gauge invariant operators to it. For example, in QCD
sum rules the  singlet current $J_\mu(x)$ may be used to study
certain colourless states. To actually construct
hadrons, though, we know from our discussion in the introduction
that such `point like' operators are inappropriate. This leaves
 two options: we can either attach a Wilson line between the two
fermions to maintain gauge invariance but at the cost of then solely
having charge
zero states, or we can look for a gauge
invariant generalization of the Lagrangian fermion, $\psi$. It is this
second route that we will follow in the rest of this paper: our
aim being to show to what extent such  charged states can be
constructed in QCD, and to analyse their dynamical consequences.
That there should be obstructions to fully carrying out this programme
is, we will argue, simply a restatement of confinement. The extent to
which such coloured charged states can be defined, though, will give a
range of validity for the constituent quark model in QCD.

Before embarking on a detailed account of how such charged states are
constructed, we wish to end this section with some general remarks on
the expected structure of such states. We will argue that,
with only a very few assumptions, a picture emerges
of the charged sectors that is far removed from the local, covariant
stand taken in the standard approaches to quantum field theories.
Indeed, many of the unfamiliar techniques that we will be forced to
introduce in our later construction of charged states will have natural
counterparts in this initial general discussion.

In our description  of the colour
charge (see the discussion around  Eq.~\defcchargetwo) we
assume that such an operator
exists and is non-zero on some states.
In terms of the field strength, we expect it to fall off like $1/R^2$
at spatial infinity, a fall off rate dictated by finite energy
considerations. This  allows us to identify the \lqq asymptotic field
strength\rqq\
$$
F^{{\rm as}\,a}_{\mu\nu}(t,\theta,\phi)=\lim_{R\to\infty}R^2
F^a_{\mu\nu}(t,R,\theta,\phi)\,.\eqn\asyfield
$$
We now assume that the physical states are such that the expectation
value of the field strength converges to a well defined distribution
(for a more precise account of this see, for example,
Ref.\thinspace\HAAG)
$$
f^a_{\mu\nu}:=\bra\psi F^{{\rm as}\,a}_{\mu\nu}\ket\psi
=\lim_{R\to\infty}R^2\bra\psi F^a_{\mu\nu}\ket\psi\,.
\eqn\no
$$
In fact, for what follows, we only need the limit
$$
\Phi^a({\bold n})=\lim_{R\to \infty}R^2\bra\psi{\bold n}\cdot{\bold E}^a
(R{\bold n})\ket\psi
\,,\eqn\cetwo
$$
which measures the asymptotic chromo-electric flux distribution in
the direction ${\bold n}$.

We note here that we cannot deduce from this alone that the potentials
should fall off like $1/R$ at spatial infinity, since there could be a
residual pure gauge term. However, we have seen that in order for the
colour charge to be well defined, such a pure gauge must, in fact, be
zero.

Given a physical state $\ket\psi$, its colour charge from
(\defcchargetwo) and (\cetwo) is
$$
\bra\psi Q^a\ket\psi=\int_{S^2}d\Omega\, \Phi^a({\bold n})\,.
\eqn\statectwo
$$
That is, as a consequence of Gauss' law, its charge is given by
the total chromo-electric flux through the sphere at spatial infinity.
In QED we expect to be able to construct charged states, so the abelian
version of the limit (\cetwo) is expected to hold. For QCD isolated
coloured charges have not been observed, so the existence of a non-zero
limit in (\cetwo) might be considered questionable. However, to probe
the extent to which
constituent quarks can be constructed, we now proceed to analyse the
consequences  of (\statectwo) in QCD.

States can always be written as $O\ket0$, where $O$ is a gauge invariant
operator and $\ket0$ is the vacuum\note{In QCD we would expect the
vacuum to incorporate some non-perturbative structure,
so $\ket0$ should not be thought of as just the perturbative vacuum.
We will return to this point in Sect.~6.}. In quantum mechanics the
(Hermitian)
operators correspond to the observables of the system. In gauge theories
being Hermitian and gauge invariant is not enough to be able to qualify
as an observable: in addition causality dictates that the operator must
be local --- that is, having support in some compact space-like
region\up{\HAAG}. In QED the field
strength $F_{\mu\nu}$ and current $J^\mu$ (both suitably smeared)
are observables. In QCD we have the singlet current $J^\mu$ and the
 colour charge
$Q^a$, for localized sources, as observables. On top of these one can
construct  observables
from localized Wilson loops and Wilson lines between fermions.
Recalling our
discussion of the charge operator around equation (\chrge), we see that
for any such observable $O$ we have\up{\STROCCHI}
$$
[Q^a,O]=\frac1g\lim_{R\to\infty}\int_{S^2_R}d{\bold s}\cdot[
{\bold E}^a,O]=
0\,,\eqn\no
$$
since $O$ has, by definition, compact support and hence vanishes outside
of some sphere of
finite radius. This implies that any state
created from the vacuum by such an observable will have charge zero.
Hence
{\it charged states are intrinsically non-local}.

This argument shows that, even in QED, if one wishes to maintain
locality for all fields then  charged states cannot exist in
nature.
Put another way: charged states  can only be described
by unobservable fields
(since they are non-local)\up{\MAISZWAN}.
A possible  response to this
is to maintain a zero charge by always including a balance between the
charges in the theory (the so called \lq\lq particle behind the
moon\rq\rq\
argument of Haag and Kastler\up{\MOON}). This is neither very appealing
nor is it economical. Indeed,
in the context of inflationary cosmology, the charge of our accessible
universe is not expected to be zero. This  implies that if we want to
correctly model the real world then charged states are needed and
hence some form of non-locality must be expected. We stress, this does
not mean that observables can now be
non-local, that would violate causality; rather we
must accept into the theory non-local fields that take us
from the vacuum to the charged sectors. Although these fields cannot
be directly identified with observables due to their non-locality,
they will have observable consequences since they create charged states.

We now wish to conclude this section with an account of the surprising
impact gauge invariance has on that stalwart symmetry --- the
Lorentz invariance of the theory.
In QED it has been noted\up{\ROEPSTORFF-\BUCHHOLZ, \HAAG} that the well
known problems
with the infrared structure\up{\JAUCH} obstruct the unitary
implementation
of the Lorentz group: the Lorentz symmetry is said to be
spontaneously broken (see, for example,
page 281 of Ref.~\HAAG).
Buchholz's analysis\up{\BUCHHOLZ} makes it clear that this obstruction
is  related to the gauge invariance of the system. In Sect.~8 we will
present an alternative account and interpretation
of this important result that directly uses the gauge invariance of
the system. Then, in Sect.~9 we will see how to use these results to
construct non-static charged states.
For now, though, let us briefly
summarize how this result is usually arrived at.

It is a general phenomena in both quantum mechanics and quantum field
theory that the state space breaks up into physically distinct
sectors --- the so
called superselection sectors of the theory\up{\HAAG}.
These are closed subspaces  of the state space which are invariant
under the action of the observables of the theory. In QCD the
$\theta$-parameter\up{\JACKIW}, which labels the $\theta$-vacua,
thereby labels
different superselection sectors. Another example in QCD are the
sectors labelled by triality\up{\POLONYI}. Both of these types of sectors
reflect  global properties of the Yang-Mills
configuration space: the $\theta$ parameter emerges because the gauge
group is disconnected; triality arises because $SU(3)$ has a non-trivial
centre. In general we expect many other such sectors arising from the
topological structure of the system\up{\MT}.
Furthermore, it is evident that, since
it is not possible for a
local operator to alter the flux (\cetwo), the asymptotic
field strength also labels different superselection sectors.  A
consequence of this is that the charge of the system is a label for
different superselection sectors\up{\HAAG}.

The fact that the asymptotic fields (\cetwo) label superselection
sectors now has a direct consequence for the action of the Lorentz group
on the system. Since
the asymptotic field of a charged particle can be
changed by applying a Lorentz transformation, such a transformation must
change the  superselection
sector. But this contradicts the expectation that the Lorentz generators
are  observables of the system.
This leaves us with two choices: either the Lorentz generators are not
gauge invariant, and hence not observables (which is physically
unappealing
as they will then
have no clear action
on the true degrees of freedom); or we take their gauge invariance as
fundamental and live with the fact that they cannot
furnish a representation of the Lorentz group on all the fields
(which is, to say the least, an unfamiliar stand in quantum field
theory). In Sect.~8 we will show explicitly how these general
considerations arise in practice, and how to deal with them. There we will
see that on the (local) physical observables and states the Lorentz
group does act in a physically acceptable way. However, on the gauge
invariant charged states  their action changes in quite a
dramatic way.

In conclusion, we have seen in this section that gauge theories
contain both physical and un-physical degrees of freedom, as well as
observable and un-observable gauge invariant operators.
Physics dictates much of
the mathematical structure associated with the physical observables of
the theory, i.e., their locality and Lorentz covariance properties.
The conventional (and highly successful) approach to these
theories requires that even the un-observable fields share these
mathematical structures. We have seen, though, that this
utilitarianism fails when charged sectors are allowed and we argue that
it is misleading to think of individual quarks or gluons as local,
covariant fields. Hence in any
approach to constituent quarks we must allow for non-locality and the
lack of manifest Lorentz invariance.
One aim of this paper is to show that
one can live with this in quantum field theory.

\vfil\eject

%%%%%%%%%%%%%%%%%%%
%%%             %%%
%%% Section4    %%%
%%%             %%%
%%%%%%%%%%%%%%%%%%%

\secno=4 \meqno=1
\ni
{\bf 4. Dressing the Quarks}
\bigskip
\ni
In the previous section we  derived some general results about charged
states in gauge theories. In particular we saw that any description
of  a charged state must be gauge invariant, non-local and
non-covariant. This
clearly rules out the Lagrangian fermion $\psi$ -- which, as we argued
earlier, does not correspond to a real field of the sort
seen in experiments and  has only a very limited asymptotic
validity corresponding to the case of the free theory, $g=0$.
The aim of this section  is to go beyond this and to obtain explicit
solutions   for  charged   states   in  QCD  to  low  orders   in
perturbation theory. These solutions will then be shown to have
well-defined colour charges, which means that they can be combined
just as in the naive quark model to form colourless hadrons.

A gauge invariant extension of a fermion may be formed
by attaching a string to it. In QED, where similar problems exist
with the construction of charged states (although their effects are,
as we will see, not so dramatic) such a  \lq string\rq\ electron is:
$$
\psig(x)\equiv\exp\left(
ie\int^{\bold x}_{-\infty}\!d{\bold z}_i A_i\right)\psi(x)\,,
\eqn\stringelec
$$
where the integral runs over some contour, $\Gamma$. (We assume that
this contour is local in time; a detailed critique of temporal
non-localities in any description of a physical field will be presented
in Sect.\ 5.) By attaching the string to the point at spatial infinity,
and
using the boundary condition that the gauge transformations reduce to
the identity
there, the gauge invariance of $\psig$ is easily seen.
The non-physical aspect of this ansatz is, however, immediately
apparent: the contour, $\Gamma$, is arbitrary and the dependence upon
it is non-physical. This unnatural dependence may, however, be
factorised out. We split the $A_i$ fields into the longitudinal and
transverse components in the standard way --- $A_i \equiv A_i^T+A_i^L$,
where $A_i^L\equiv \dfrac{\pa_i\pa_jA_j}{\nabla^2}$.
Recall that the non-local term
$$
\frac{f}{\nabla^2}(x)\equiv
-\frac1{4\pi}   \int\!d^3y  \frac{f(x_0, {\bold  y})}{\vert{\bold x}
-{\bold
y}\vert}\,,
\eqn\nabdef
$$
is (given our assumptions about the spatial fall-off of the potentials)
the inverse Laplacian  acting on $f$, i.e., $\nabla^2/\nabla^2=1$.
It is then obvious that
(\stringelec) may be rewritten as
$$
\psig(x)=N_\Gamma
\exp\left(
ie\frac{\partial_i A_i}{\nabla^2}(x)\right)\psi(x)\,,
\eqn\gamfactored
$$
where the dependence on $\Gamma$ is now only present in the first term
$$
N_\Gamma(x)\equiv\exp\left(
ie\int^{\bold x}_{-\infty}\!d{\bold z}_iA^T_i\right)\,.
\eqn\ngamm
$$
Dropping this unphysical first factor, we are forced to use what is
left to describe the electron.

In fact we recognise the second term in (\gamfactored) as just the
set of fields corresponding to electrons and positrons which
was proposed many years ago by Dirac\up{\DIRAC}. He suggested  using
$$
\psic(x)\equiv \exp\left(
ie\frac{\partial_i A_i}{\nabla^2}(x)\right)\psi(x)\,,
\eqn\coulelec
$$
where  the  $A_i(x)$  are  the  QED  gauge  fields (which are real)
and $e$  is the
electromagnetic coupling constant.
It is easily seen that these fields fulfill our minimal expectations from
Sect.~3,   i.e.,   they  are  gauge  invariant, and (manifestly)
non-local   and
non-covariant.  The notation $\psic$ is used to signify that this
field  dresses the charged field $\psi$ with the Coulombic electric
field. As such it is describing a static charge in QED.
This interpretation can be seen formally as follows:
let $\ket\ve$ be an eigenstate of the electric field, i.e., we
have $E_i(x)\ket\ve=\ve_i(x)\ket\ve$.
Then consider the (equal time, $x_0=y_0$)
state $\psic(y)\ket\ve$. It is also an eigenstate of
the electric field since,
$$
\eqalign{
E_i(x)\psic(y)\ket\ve&=[E_i(x),\psic(y)]\ket\ve+\psic(y)E_i(x)
\ket\ve\cr
&=\frac{e}{4\pi}\frac{{\bold x}_i-{\bold y}_i}{|{\bold x}-{\bold
y}|^3}\psic(y)\ket\ve+
\psic(y)\ve_i(x)\ket\ve\cr
&=\(\ve_i(x)+ \frac{e}{4\pi}\frac{{\bold x}_i-{\bold y}_i}{|{\bold
x}-{\bold y}|^3}\)
\psic(y)\ket\ve\,.
}\eqn\reFFed
$$
{}From this we see that introducing $\psic(y)$
has changed the electric field by a term corresponding to
that from a static charge at the
point ${\bold y}$. This argument is rather appealing, but its
reliance on the canonical commutation relations (\ccrtwo)
implies that it is only really valid for free theories
(or the bare fields). For
interacting theories an infinite  renormalization of the
field is needed which could greatly complicate the above argument.
In the full theory it is more useful to focus  on the
Green's functions, and we will see in Sect.~6 that the
same interpretation of
$\psic$ as a static charged field emerges directly
from the renormalization of its two-point function.

Note, however, that the string ansatz dismissed above only possesses
an electric field along the string contour --- which is a further
indication of its essentially unphysical nature. We also
mention here that Shabanov\up{\SHABANOV}
has shown that if we take a static, non-local fermion-antifermion
pair joined by a string, so as to ensure gauge invariance, as an
initial state and then study its time development, one finds that it
radiates and eventually approaches a pair of the Dirac physical
fields from (\coulelec).

We now want to find a description  of quarks similar to that
of (\coulelec). We no longer have Dirac's guideline of the Coulombic
electric field, but we will let ourselves be led by gauge invariance
and the lack of any dependence upon an arbitrary path such as $\Gamma$
above.  It is not
enough to just replace the coupling $e$ with $g$, and the
potential $iA_i(x)$ by $A_i^a(x)T^a$ (recall our
anti-hermitian conventions for the non-abelian potentials),  since gauge
invariance  is then lost as may be easily seen.  However,
we note that at lowest  order in
perturbation  theory this direct extension of Dirac's description
of the electron to the quark does go through and we can define:
$$
\psic^{g^1}(x)=
\left(1+g\frac{\partial_iA_i^a}{\nabla^2}(x) T^a\right) \psi(x)
\,,
\eqn\psione
$$
which we may suggestively rewrite as
$$
\psic^{g^1}(x)=
\exp\left(
g\frac{\partial_iA_i^a}{\nabla^2}(x) T^a\right)\psi(x)
+{O}(g^2)
\,.
\eqn\psionea
$$
To see that this is indeed  gauge invariant  to lowest  order  in
the coupling recall that under a local gauge transformation,
$U(x)=\exp(g\theta^aT^a)$, we have in perturbation theory:
$$
\psi(x)\to \left(1-g \theta(x) \right)\psi(x) +{ O}(g^2)\,,
\eqn\gonetr
$$
and
$$
A_i(x)\to A_i(x)+ \partial_i\theta(x)
 +{ O}(g^1)\,,
\eqn\gzerotr
$$
where we  note that due to the factor of $1/g$ in (\atranstwo)
it is convenient in such a perturbative expansion in $g$ to include
a factor
of the coupling in $U$. The factor of $g$ in the exponential  of
$(\coulelec)$   also means   that   we  only   require   the    gauge
transformation on $A_i$ to zeroeth order in $g$.
{}From these we have
$$
\eqalign{
\psic^{g^1}(x)\to &
\left(1+g\frac{\partial_iA_i^a}{\nabla^2}(x) T^a +g \theta(x)
\right) \left( 1- g \theta(x)\right) \psi(x)\cr
= &
\left(1+g\frac{\partial_iA_i^a}{\nabla^2}(x) T^a\right) \psi(x)
+{O}(g^2)
\,,}
\eqn\trans
$$
which shows the desired  invariance.

Although  (\psione)  is  not  gauge  invariant  at order  $g^2$  it is
possible,  with a certain amount of effort, to obtain the correction
terms which make it so\up{\LMQUARK}. One finds
$$
\eqalign{
\psic^{g^2}(x)=\Biggl(\Biggr.1
&+g\frac{\partial_iA_i^a}{\nabla^2}(x) T^a
+\tfrac12g^2\left(\frac{\partial_iA_i^a}{\nabla^2}(x) T^a \right)^2
\cr
&-\tfrac12g^2f^{abc}\frac1{\nabla^2}\left(A_j^b\frac{\partial_j
\partial_iA_i^c}{\nabla^2}\right)\!\!(x)T^a\cr
&-\tfrac12g^2f^{abc}
\frac{\partial_j}{\nabla^2}\left(A_j^b\frac{\partial_iA_i^c}{\nabla^2}
\right)\!\!(x)T^a\Biggl.\Biggr)\psi(x)\,.
}
\eqn\psitwo
$$
At this stage we should comment on why we still interpret this field as
describing a static quark. Firstly this description reduces in the
abelian case ($f^{abc}\rightarrow0$) to Dirac's description which has
the electric field of a static charge. Secondly this is a minimal
description: we could imagine adding further separately gauge invariant
gluonic field combinations, which would not destroy the gauge
invariance of the dressing, but this would unnecessarily raise the
energy of the quark.

The gluonic counterpart  of this description and a systematic way
of obtaining  such  expressions  will  be presented  in the  next
section.   We now
want to see how this quark  behaves  under  the
action of the charge operator  (\cchargetwo) with the eventual  aim of
seeing how to build colourless  objects out of these fields.  For
simplicity  all  explicit  calculations in the remainder of this
section will  be  in  the framework of an $SU(2)$ gauge theory, a
more general treatment will be postponed until the next section.

Recall firstly that in the free quark theory (i.e., no gluons)
we only have a rigid gauge symmetry and the
charge  operator  reduces  to the matter charge
$$
\Qm^a= \int\! d^3x J^0_a(x)
\,.
\eqn\freeQ
$$
For the free $SU(2)$ theory we expect
the quarks to diagonalize the charge $\Qm^3$ (or more generally the
 Cartan sub-algebra) where now the antihermitian $T^a$ are related to
the Pauli matrices by $T^a=\sigma^a/2i$.
The addition of a free quark to (say) a charge zero state,
$\ket\Omega$,
alters the eigenvalue  of this charge operator by
$$
\eqalign{
\Qm^3
\psi_{1}(x) \ket{\Omega} =&
\frac i2\psi_{1}(x)\ket{\Omega} \cr
\Qm^3
\psi_{2}(x) \ket{\Omega} =&
-\frac i2 \psi_{2}(x)\ket{\Omega} \,,
}
\eqn\Qonqu
$$
and  a  meson  with zero net colour charge may  be
described  (as in the  constituent  quark  model)  by
$$
\bar\psi_n(x)\psi_n(y) \,.
\eqn\meson
$$
Under rigid gauge transformations  \lq red\rq\  may become
\lq blue\rq\  etc., but such a meson remains colourless  overall,
rather as an isospin singlet remains a singlet under isospin
rotations,  and a \lq  meson\rq\ does not change the charge $\Qm^3$
$$
\Qm^3 \bar\psi_n(x)\Gamma \psi_n(y)
\ket\Omega = 0\,,
\eqn\mesinv
$$
where $\Gamma$ denotes any potential Lorentz structure.
The analogous description of a baryon goes through directly.
So much regarding the free quark theory.

In QCD, however, we have a local gauge invariance  and it is naively
apparent that a configuration
of two or more Lagrangian quarks at different points would not stay
colourless under local gauge transformations. In fact, as we saw
in Sect.~3,  we can only associate  a charge  to gauge  invariant
fields and although we could introduce, say, strings connecting  the
fermions,
this would be foreign to our philosophy of trying to talk
about quarks.
Even more significantly perhaps, the charge operator in QCD
(unlike in QED) is not the same in the free and the interacting
theories, and although  we have dressed the quarks above in such a way
that  they are perturbatively  gauge  invariant,  we have dressed
them with gluons which themselves  are coloured  objects.   It is
therefore  not self-evident that we can straightforwardly
combine the dressed quarks
in the usual way to form hadrons.

To see that this is in fact possible  consider  the lowest  order
description  of an $SU(2)$ dressed
quark from $(\psione)$.  The commutator of the full
charge with the dressed quark receives
two terms (one from the gluons and one
from the fermions) and, although it is not immediately obvious,
they can
be shown by explicit calculation to combine to give
$$
\eqalign{
Q^3
{\psic}^{g^1}_1(x) \ket{\Omega} =&
\frac i2{\psic}^{g^1}_1(x)\ket{\Omega} \cr
Q^3
{\psic}^{g^1}_2(x) \ket{\Omega} =&
-\frac i2 {\psic}^{g^1}_2(x)\ket{\Omega} \,.
}
\eqn\Qonqphys
$$
This is just what we  obtained in the free (glue-less) theory.
We so learn that the
physical,  dressed  quark fields  have a well defined  colour
charge which has the properties of the charge
 of the  fermion  in  the free theory.  This means that
we can indeed combine the dressed quarks so as to form, e.g.,
(perturbatively) colourless, mesonic objects:
$$
\bar\psi_n(x)\psi_n(y)\to \bar{\psic}^{g^1}_n(x)
{\psic}^{g^1}_n(y)\,,
\eqn\physmeson
$$
and analogous baryonic structures. Recalling the distinction we were 
earlier forced to make between local and rigid gauge transformations 
(see just after Eq.\ \QOK), we see that this explicit 
calculation may be understood
rather simply: although the dressed quarks are invariant under local
gauge transformations, they transform just like the Lagrangian quark
under rigid transformations. The colour charge is, however, the charge
associated with the Noether current for such rigid transformations and
so we would indeed expect that {\it perturbatively
dressed quarks have colour charges
like those of the bare ones in the free theory}. This property will be
also be visible in all of the extensions (higher orders in $g$,
non-static) of
the dressed quark (\psitwo)
which will be presented below.

This is a very satisfying result which explains one of the more
surprising parts of the constituent quark model: it means that we can
combine coloured
dressed quarks to form mesons and baryons
just as we would naively do in the model. This, for example, allows
us to  write
the wave function of a baryon as a product of a colour wave
function for the constituent, dressed
quarks
%($\sim \epsilon_{mnr}$ in QCD)
with the remaining wave function
(i.e., the
spatial, spin and flavour parts) in accord with the Pauli principle.
This aspect of hadronic spectroscopy
was of course the initial motivation for the introduction of the
colour quantum number. A further implication of this result is that
we have not yet seen colour screening. The colour charges will be
screened by renormalisation effects and by interactions between the
quarks when we bring them together in bound states. We will present
our initial studies of the interactions between such quarks in
Sect.\ 7.

Before we can, however, start to combine the dressed quarks to form
hadrons, we are now
confronted by several immediate
questions. Firstly, can we systematise
the derivation of these dressed quark fields and are we able to
dress quarks (and gluons) non-perturbatively? Secondly, these fields
are both non-local and non-covariant: can one actually work with them in
a quantum field theory such as QCD?
In particular, can we carry though a programme of regularisation and
renormalisation for these fields and their
Green's functions?
We will address
these important issues (in the above order) in the following two sections.

\vfil\eject

%%%%%%%%%%%%%%%%%%%
%%%             %%%
%%% Section5    %%%
%%%             %%%
%%%%%%%%%%%%%%%%%%%

\secno=5 \meqno=1
\ni
{\bf 5. The Breakdown of the Quark Description}
\bigskip
\ni
In this section we explain  how the construction of the
perturbative quarks can be generalized and extended both to all orders
in perturbation theory and to the gluons. The key step in the argument
is the recognition of the intimate link between gauge fixing and dressing
any charged object. As well as providing a simple way of generating the
perturbative dressing, and showing that the colour charges of dressed
quarks are those of the constituent quark model to any order in
perturbation theory, this link will let us prove that in QCD there
is an obstruction to extending these results beyond perturbation
theory. This lack of any gauge invariant description of a single
quark or gluon underlies colour confinement.

We have seen that it is possible to construct perturbatively physical
quarks, at least to low orders in the coupling.
The solution (\psitwo) had the
characteristic non-locality and non-covariance expected of such a
charged object.
A natural question to ask is  what expression is (\psitwo) the low
order approximation to? In addition to this, for completeness, we
would  like to see a similar description of the gluons.

In order to dress the quark $\psi$ and make it gauge invariant we need
a gauge field dependent element of $SU(3)$ that transforms under the
gauge transformation (\atranstwo)  in the same way as $\psi$, i.e.,
 we need to find $h\in SU(3)$ such that under local gauge
transformations we have
$$
h\to \hu=U^{-1}h
\,.\eqn\htransfour
$$
Then a physical quark can be identified with the BRST invariant
combination
$$
\psi_\phys=h^{-1}\psi\,.\eqn\psiphysfour
$$
{}From (\psitwo) we see that our expression for the static quark
corresponds to taking $\psic^{g^2}=\hc^{-1}\psi$ with $\hc=e^{-\vc}
+O(g^3)$ and, to this order,
$$
\vc=g\frac{\pa_iA_i}{\nabla^2}-\tfrac12g^2\frac1{\nabla^2}\left[A_j,
\frac{\pa_j\pa_iA_i}{\nabla^2}\right]-\tfrac12g^2\frac{\pa_j}{\nabla^2}
\left[A_j,\frac{\pa_iA_i}{\nabla^2}\right]\,.\eqn\vcfour
$$
In this expression we have reverted to the compact expression (\lietwo)
for Lie-algebra valued fields and hence the commutators are Lie-algebra
commutators. We note that since $\pa_i$ and $1/\nabla^2$ commute, this
expression can also be written as a total divergence (see the
appendix):
$$
\vc=\frac{\pa_j}{\nabla^2}\(gA_j+g^2\left[\frac{\pa_iA_i}{\nabla^2},A_j
\right]+\tfrac12g^2\left[\frac{\pa_j\pa_iA_i}{\nabla^2},\frac{\pa_kA_k}{
\nabla^2}\right]\)\,.\eqn\no
$$
Given a dressing $h$ satisfying (\htransfour) then the BRST invariant
gluonic field is identified with the transformed field $A_i^h$,
i.e., with
$$
(A_{\phys})_i:=h^{-1}A_ih+\frac1gh^{-1}\pa_i h\,.\eqn\aphysfour
$$
To see that this is also gauge invariant note that under the gauge
transformation $A\to \Au$ we have
$$
\eqalign{
(A_{\phys})_i&\to\(\hu\)^{-1}\Au_i\hu+\frac1g\(\hu\)^{-1}\pa_i\hu\cr
&=h^{-1}U\Au_i U^{-1}h+\frac1gh^{-1}(U\pa_iU^{-1})h+\frac1gh^{-1}\pa_i
h\cr
&=h^{-1}\bigl(A_i+\frac1g\pa_iU\,U^{-1}\bigr)h
-\frac1gh^{-1}(\pa_iU\,U^{-1})h+\frac1gh^{-1}\pa_ih=(A_{\phys})_i\,.
}
\eqn\no
$$
Substituting expression (\vcfour) into (\aphysfour) yields the physical
gluonic field $(A_{{\rm c}})_i:=A_i^{\hc}$ which is to $O(g^1)$
$$
(A_{{\rm c}}^{g^1})_i=\(\d_{ij}-\frac{\pa_i\pa_j}{\nabla^2}\)
\(A_j+g\left[\frac{\pa_kA_k}{\nabla^2}, A_j\right]+\tfrac12g\left[
\frac{\pa_j\pa_kA_k}{\nabla^2},\frac{\pa_lA_l}{\nabla^2}
\right]\)\,.\eqn\agfour
$$
We have thus reduced the problem of finding gauge invariant, dressed
quarks and gluons to that of constructing such a field $h$. In other
words, once we have found $h$ we can directly construct dressed quarks
and gluons which may then be combined to form hadronic bound states.
Different $h$ dressing functions will be seen to correspond to both
static and non-static quarks.

Having seen the utility of being able to construct the field $h$
transforming as in (\htransfour), we now want to demonstrate that
the
existence of such a dressing field is
equivalent to finding a gauge fixing condition. This relation between
gauge fixing and the dressings will allow us to both study the
non-perturbative aspects of quarks and gluons and will give us a
highly efficient algorithm for extending the perturbative results
of the last section. That there is a
connection between dressing and gauge fixing should not come as too
much of a surprise. Indeed, from the specific static example (\vcfour)
we see that, at least to this low order in $g$, the Coulomb gauge
$\pa_iA_i=0$ removes the dressing ($\vc=0$). In order to show that
this is true to all orders for the static quark, and that an analogous
result holds for any
dressing, we need to  first define  what is meant by a gauge fixing
condition in this context.

We have argued that the QCD effective Lagrangian (\efftwo) must contain
a gauge fixing term. Recall that such a term is generically a function
of the potentials, $\chi^a(A)$, where the Fadeev-Popov determinant,
$\det|\d\chi(\Au)/\d U'|$, does not vanish\up{\FADDEEV}. This gauge
fixing is generally taken to be
covariant in order to simplify the ultra-violet structure of the theory.
Dressing the quark will involve a further gauge fixing which is
{\it not\/} this Lagrangian gauge fixing, and we will henceforth refer to
it as the {\it dressing gauge fixing}.
What we will see is that the dressing gauge is the gauge choice  that
removes the explicit dressing from the quarks, resulting in the
identification of the Lagrangian fermion with the physical quark
field.

In QED we see from (\coulelec) that in the Coulomb gauge $\psic=\psi$,
and hence then the Lagrangian fermion $\psi$
is a physical charged field.
This appears to be at odds with our general result that such fields are
intrinsically non-local. We recall, though, that in the Coulomb gauge,
which is clearly non-covariant, the vector fields have space-time
commutator (see p.\ 484 of Ref.\ \JAUCH)
$$
[A_i(x),A_j(y)]=i\(\d_{ij}-\frac{\pa_i\pa_j}{\nabla^2}\)
\triangle(x-y)\,,\eqn\no
$$
where $\triangle(x)$ is, for the free field, the Jordan-Pauli function
$\triangle(x)=-\frac1{2\pi}\varepsilon(x_0)\d(x^2)$. Locality would
imply that this commutator vanishes for space like separations. But
the term
$$
\eqalign{
\frac1{\nabla^2}\triangle(x)&=\frac1{8\pi|\underline x|}
\bigl(\bigl|t+|\underline x|\bigr| -\bigl|t-|\underline x|\bigr|\bigr)\cr
&=\frac1{4\pi}\varepsilon(t)\theta(x^2)+\frac{t}{4\pi|\underline x|}
\theta(-x^2)
}
\eqn\no
$$
clearly does not vanish for $x^2<0$.
In this sense, the non-locality in the charged sector
is no longer manifest in this gauge --- but it is definitely there!
We note in this example that we are still free to impose conditions
on the temporal
component of the potential. So the radiation gauge would also remove
the dressing in this case (recall that this is the Coulomb gauge
and Amp\`ere's law combined and hence involves two gauge fixing
conditions). The implementation of this type of gauge fixing in the
path integral formalism is, though, quite involved\up{\LMRAD}.

Although the Coulomb gauge dressing will play a central role in this
paper, it is not the only dressing available to us. Indeed in
Sect.~9 a whole class of dressings will be presented
describing boosted quarks. Before we get involved in analysing
precisely how gauge fixing and dressings go together, though, we
need to address a rather natural question:  is it possible  to find
a dressing that is removed in a covariant gauge? Clearly, in QED, we
can formally construct the gauge invariant and apparently covariant field
$$
\psi_{\hbox{\sevenrm cov}}(x)=\exp\(ig\frac{\pa_\mu A^\mu}{\square}\)
\psi(x)\,.\eqn\badfour
$$
Now the Lorentz gauge $\pa_\mu A^\mu=0$ is both covariant and local.
So in this gauge the dressing described in (\badfour) would vanish and
we have
a local covariant charged field $\psi$\thinspace! Now this really is at
odds with our general discussion, and there is a temptation to keep
(\badfour) and conclude that the general results were  just idle
mathematical speculation! A less impulsive response, though,
is needed here. Both (\coulelec) and (\badfour) are no more than
symbolic expressions---we must give an, if not rigorous then at least
operational, prescription for how they are
to be defined. In the
context of perturbative field theory we want to construct the propagator
for these fields which will involve their time ordered products:
i.e., $\bra0T\bar{\psi}_{\hbox{\sevenrm cov}}(x)\psi_{\hbox{\sevenrm cov}}(y)
\ket0$. Now
(\coulelec) is only spatially non-local and, as we shall see in great
detail in the next section, its propagator can be constructed using
reasonably standard field theory techniques. However, the covariant
expression (\badfour) is, additionally, non-local in time.
Again this might seem to present no real obstacle as the overall time
ordering should sort out the (admittedly complicated) ordering in the
fields. But we must remember that these fields are  interacting
fields --- we are in the Heisenberg picture. In order to develop a
perturbative description we need something akin to
the Gell-Mann and Low description of the propagator in terms of
time ordered products of free fields. To do this we have to assume
that in
the distant past and future  these Heisenberg fields (and their
vacuum) \lq\lq agree\rq\rq\  with the free fields\up{\HAAG}. But given
that these fields are non-local in time such an identification is
impossible, and hence a perturbative description of (\badfour) is
not possible.
These problems can be avoided in
the Euclidean regime if we  only deal with the  Wightman functions
associated with the field (\badfour). The only physical interpretation
of the construction is then that these correspond to two-point functions
for the high temperature regime for some  five dimensional gauge theory.
Hence the fields (\badfour) cannot be given a meaning in a four
dimensional gauge theory and, we argue,  do not correspond to any
physical dressing of the charged particles.   The conclusion we reach
from this discussion is that
{\it the dressing  gauge fixing is never covariant}, i.e., the
nonlocality is only in the spatial directions.

It is useful here to restate this important
distinction between the gauge fixings:
the Lagrangian gauge fixing that we employ will be chosen to be covariant
so that
standard perturbative
techniques can be used; the dressing gauge fixing will, in contrast,
necessarily be non-covariant so that
charged sectors can be constructed.

We now want to make one further restriction on the dressing before
starting our general analysis. In contrast to the discussion above,
this new restriction is {\it not\/} essential to our formalism, but
will significantly simplify the presentation.  The problem that we
want to side-step is  that in the effective Lagrangian (\efftwo) the
generators for the gauge transformations on the gauge fields split
into the Gauss law part (\gaussgentwo) and the primary constaint, $E_0$,
which is the momentum conjugate to $A_0$. We have seen that Gauss' law
generates the local gauge transformations on the spatial components of
the gauge field. Being simply a momentum, though, the primary constraint
only generates a translation of the temporal components, and as such it
does not appear to generate the expected gauge transformations on $A_0$.
However, we are neglecting the ghost fields in this, and in a more
careful analysis\up{\HENNEAUX} one finds that the momentum conjugate
to the anti-ghosts
smears the primary constraint. When the momentum variables are then
eliminated the expected BRST transformation on the temporal component of
the  gauge field is  recovered. To avoid this level of complexity,
though,  we will make the further restriction that the dressing field
be constructed only out of the spatial components of the gauge field.
This we will see covers all the dynamical configurations of the dressing
needed for this paper (i.e., all boosted states for a quark).
The extension to a more general dressing is  an interesting
topic, but one that we will not concern ourselves with in this paper.

Given these restrictions on the dressing, it is now quite easy to show
that a dressing field transforming as in
(\htransfour) implies the existence of a gauge fixing condition.
Recall from our discussion in Sect.\ 3
that it is Gauss' law that generates the gauge transformations.
Hence $\hu=e^{iG(\theta)}he^{-iG(\theta)}$, where $U=e^\theta$. Thus,
to lowest order in $\theta(x)=\theta^a(x)T^a$, we have from
(\htransfour)
$$
i[G_a(x),h(y)]=-T^ah(x)\d(x-y)\,.\eqn\no
$$
Writing $h(y)=e^{-v(y)}$, with $v(y)=v_b(y)T^b$, this gives
$i[G_a(x),v_b(y)]=\d_{ab}\d(x-y)+$terms which vanish when
$v$ is set equal
to zero. That is, using the idea of weak equivalence familiar
from constrained dynamics\up{\HENNEAUX}, we have that the
infinitesimal version of (\htransfour) is the (weak) equation
$$
i[G_a(x),v_b(y)]\approx \d_{ab}\d(x-y)\,.\eqn\gaugefixing
$$
The matrix
$i[G_a(x),v_b(y)]$ is precisely the Fadeev-Popov matrix one would
construct for functions $v_a$ if we were viewing them as gauge
fixing conditions\up{\FADDEEV}. Indeed, the condition (\gaugefixing)
tells us that the
Fadeev-Popov determinant associated with this gauge fixing is non-zero,
and hence it is a valid gauge fixing condition.

Our aim now is twofold: firstly we want to show that
the converse to this holds, i.e., that a gauge fixing
condition $\chi^a(A)=0$ can be used to construct the dressing $h$.
Secondly we will demonstrate that (\gaugefixing) does {\it not} have a
general solution in a non-abelian gauge theory, i.e., there is no
complete nonperturbative description of an asymptotic quark or gluon
field and we should not expect to see such fields.
Before starting the general discussion, let us see how the first point
is realised in QED and the Coulomb gauge. As we have said before, the
electromagnetic potential $A$ should really be thought of
as belonging to an orbit of gauge related potentials which,
in this abelian case, are described by $\O_A=\{A_i+\pa_i\Lambda\}$,
where $\Lambda$ is a scalar. The Coulomb gauge condition picks out
a unique representative from each such orbit. Thus, for any $A\in
\O_A$ there must be a potential dependent scalar $v$ such that
$\pa_i(A_i+\pa_iv)=0$. Clearly we have $v=-\pa_iA_i/\nabla^2$.
By construction, for another gauge related potential $A_i+\pa_i\Lambda$
the scalar $v$ transforms to $v-\Lambda$. Thus the gauge invariant
(Coulomb gauge satisfying) potential is $A^h_i=A_i-\frac{i}{e}
h^{-1}\pa_ih$, with $h=e^{iev}$. Having constructed the dressing
field $h$ from the Coulomb gauge fixing condition on the potentials,
we can directly write down the (static) electron using the general
formula (\psiphysfour).

We now want to extend this argument both to more general gauge
fixing conditions  and to non-abelian gauge theories. As well
as providing an efficient algorithm for perturbatively constructing
the dressing (see the appendix) this discussion will also show why
this construction {\it cannot\/} be extended in the non-abelian
theory to a non-perturbative dressing. However, since what
follows is of a slightly higher  mathematical sophistication than
the rest of this work, let us summarise how gauge fixing is used
in the construction of the dressing field $h$.

The strategy in the general situation is the same as in the simple
example above: we use gauge fixing in the Yang-Mills sector to
construct $h$, then use (\psiphysfour) to dress the quark. In
the Yang-Mills configuration space we denote the  orbit of the
gauge group that goes through the potential $A$ by
$$
\O_A:=\{\Au:U\in\G\}\,,\eqn\no
$$
where $\G$ is the group of all gauge transformations. Again, a
gauge fixing condition $\chi=0$ will \lq slice\rq\ the orbits,
picking a representative from each one. Given a potential $A$,
we now define $h$ to be the ($A$-dependent) element of $\G$ that
takes the potential $A$ to the point in the orbit $\O_A$ where the
gauge fixing condition holds, i.e., so that $\chi(A^h)=0$.
Pictorially this construction is summarized in Fig.~5.1.

\topinsert
%\vskip 4cm
%\fig{qfig5.eps}{4cm}{4cm}
\fig{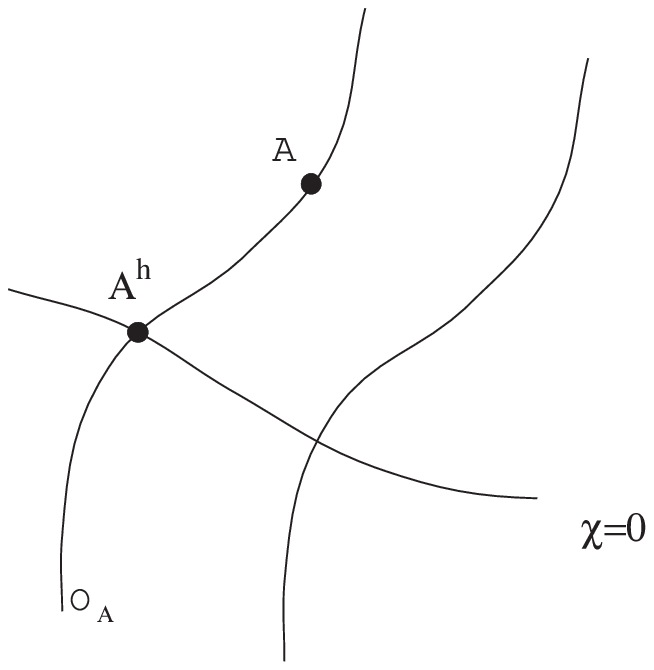}{4cm}

{\eightpoint\narrower
\noindent{\bf Figure 5.1} A diagram showing how gauge fixing is used
to define the dressing.\par}
\endinsert

To show that this construction gives a dressing with the
correct transformation properties, and that the Gribov ambiguity
implies that it can
only be defined perturbatively, we need  to briefly review some of
the more geometric aspects of
gauge theories and gauge fixing. We will conclude this section with a
brief discussion of the physical implications of this result.
%\bigskip
%\fig{qfig5.eps}{5cm}{4cm}
%\bigskip
%\noindent{\bf Figure 5.1} A diagram showing how gauge fixing is used
%to define the dressing.
%\bigskip

Some care is needed in the precise identification of the Yang-Mills
configuration space and we can only hope here to give an honest and,
hopefully, intuitive account of the rich structures to be found there.
For more details we refer to Ref.'s \SINGER--\FUCHS.
The first thing that we need to do is
worry a little about the asymptotic form of the fields.
Finiteness of the energy implies  that the gauge fields must at least
tend to  a pure gauge at  spatial infinity. However, we have seen that
in order to construct a colour charge, any gauge transformation that
is not rigid must tend to
the identity at spatial infinity. This allows us to identify the
Yang-Mills configuration space $\A$ with the set of all potentials:
$\A=\{A_\mu^a(x)\}$ that tend to zero at least as fast as $1/R$.
One can show that this can be made into an infinite dimensional
smooth manifold (modelled now on a Hilbert space). We denote the
group of local gauge transformations  by $\G$, and this can also
be viewed as an infinite dimensional Lie group. The upshot of this is
that the spaces of interest in Yang-Mills theory are analytically
quite complicated due to the fact that they are  infinite dimensional,
but their qualititive structures (such as their topology) have
much in common
with more familiar spaces such as Lie groups
and, as we shall see, coset spaces.

The group $\G$ has a (right) action on the space $\A$ given by
$A_\mu\to \Au_\mu$, where $U\in\G$. We call this  a right action
since $(A^{\scriptscriptstyle U_1})^{\scriptscriptstyle U_2}=
A^{\scriptscriptstyle U_1U_2}$. This $\G$ action is essentially
free\note{This is one of the places where things can get a little
involved. As we understand it\up{\FUCHS},  those connections which have a
non-trivial little group in $\G$ are deemed reducible. By lumping
gauge fields together with the same little group the Yang-Mills
configuration space breaks up into different strata. On each such
strata, though, the spaces are much the same. Hence this complication
does not alter the main conclusions we wish to reach about
gauge fixing.}\ so the orbit of $\G$ through a point $A\in \A$ is a
copy of the group $\G$, i.e.,
 being free implies that $\O_A\approx\G$.

The physical Yang-Mills configuration space is identified with the
space of orbits $\A/\G$, this is again a smooth infinite dimensional
manifold.  The mapping $\pi:\A\to\A/\G$, given by $\pi(A)=\O_A$ is a
projection onto the physical configuration space.
A gauge fixing condition allows us to construct a mapping (cross-section)
going the other way, i.e., gives us a mapping $\sigma:\A/\G\to\A$,
which will let us  view the physical configurations as special elements
of the extended configuration space $\A$. This in its turn will allow us
to disentangle the true degrees of freedom from the others in $\A$

To see this let us assume that $\chi^a(A)$ is a smooth gauge fixing
condition: the non-vanishing of the Faddeev-Popov determinant then implies
that the surface $\chi^a=0$ intersects  each orbit once.
This allows us to define a cross-section $\sigma$ by
$$
\sigma(\O_A)=\{\hbox{{\rm the unique element in the orbit where the
gauge fixing holds}}\}\,.\eqn\sectionfour
$$
We can now use this cross-section to separate out the true dynamics
from the gauge directions in $\A$. Consider the mapping $i$ from the
product space $\A/\G\times\G$ to $\A$ given by
$$
i(\O_A,U)=\sigma(\O_A)^{\scriptscriptstyle U}\,,\eqn\no
$$
that is, take the point in the orbit where the gauge fixing holds and
then apply the gauge transformation $U$ to it.
Since the  $\G$ action on $\A$ is assumed to be free, this mapping is
both one-to-one and onto.

On the product space the gauge action takes a very simple form
$$
(\O_A,U_1)\to(\O_A,U_1)^{\scriptscriptstyle U_2}:= (\O_A,U_1U_2)\,,
\eqn\prodgaugefour
$$
which allows us to clearly isolate the gauge invariant, and hence
physical, structures.
The  mapping $i$ makes clear the role of gauge fixing in isolating
the true degrees of freedom, i.e.,
to make manifest the physical content of gauge theories we exploit
gauge fixing to go from the  product space $\A/\G\times\G$, where
the physical
content is manifest to
the jumbled-up space $\A$, where there is a
highly non-trivial mixing of the physical and un-physical degrees of
freedom.
Of course, we really want to go the other way around and for this we
need the inverse mapping to $i$. This is easy enough to construct,
define the  mapping $j:\A\to\A/\G\times\G$ by
$$
j(A)=(\pi(A), h^{-1}(A))\,,\eqn\jfour
$$
where $h^{-1}(A)$ is here defined to be that unique element of $\G$ that
takes
$\sigma(\O_A)$, the gauge fixed element of the orbit through $A$, back
to $A$.
That is,
$$
\chi^a(A^{\scriptscriptstyle h})=0\,.\eqn\no
$$
That  $j$ is the inverse to the mapping $i$ follows easily from the
definitions.  For example we have
$$
\eqalign{
i\circ j(A)&=i\bigl(\pi(A),h^{-1}(A)\bigr)\cr
&=\bigl(A^{\scriptscriptstyle h}\bigr)^{\scriptscriptstyle h^{-1}}\cr
&=A\,.
}
$$
Now if we apply a gauge transformation to both sides of (\jfour)
we find that
$j(\Au)=(\pi(\Au),h^{-1}(\Au))$ and $
(\pi(A),h^{-1}(A))^{\scriptscriptstyle U}
=(\pi(A),h^{-1}(A)U)$,
using (\prodgaugefour). Thus we must have $h^{-1}(\Au)=h^{-1}(A)U$,
and we
have recovered the dressing transformation (\htransfour).

In summary, we have seen that in order to build a dressing field for
the quarks we can use a gauge fixing condition in the gauge sector to
construct $h$. The dressing of the quark is then given by (\psiphysfour).
That this process leads to an efficient algorithm for finding the
dressing field to higher order in $g$ is demonstrated in the appendix
where the dressing for the static quark to order $g^3$ is derived
from the Coulomb (dressing) gauge condition.

As an immediate application of this description of how the dressing
field is constructed we can now give an elementary  proof of the
important result that the dressed quarks transform as irreducible
representations of SU($N_c$)
under rigid transformations. (Recall that in (\Qonqphys) we showed
this for $N_c=2$ only to order $g$.)

We are required to show that under the rigid transformation (\rigidatwo)
the dressing transforms as $h\to\tilde h=U^{-1}hU$, so that the dressed
quark (\psiphysfour) transforms in the same way as the free quarks;
$\psi_\phys\to\tilde\psi_\phys=U^{-1}\psi_\phys$; under rigid gauge
transformations (recall the distinction between local and rigid gauge 
transformations made after Eq.\ \QOK). 
We know that $A^h$
is a potential, so it transforms under a rigid transformation as
$A^h\to\widetilde{A^h}=U^{-1}A^hU$. We now need to see
that $\widetilde{A^h}$
is in the same orbit as $\tilde A=U^{-1}AU$: this may be restated as
there being a field dependent $\tilde h\in \G$ such that
$\tilde A^{\tilde h}=\widetilde{A^h}$. It is now  simple to show
that $\tilde h=U^{-1}hU$ has the required properties (recall that $U$
is rigid).

The final topic we want to address in this section is the obstruction
to extending this construction of the dressing beyond perturbation
theory. From our discussion we have seen that a dressing for the quark
is equivalent to finding a gauge fixing. So the existence of a dressing
is inextricably connected with the existence of a   
gauge fixing condition. 
Now Gribov\up{\GRIBOV} has explicitly
demonstrated that the Coulomb gauge in Yang-Mills
theory does not exist outside of perturbation theory. What he showed was
that the condition
$$
\pa_i(A_i^{\scriptscriptstyle h})=0\,,\eqn\gribcoulfour
$$
does {\it not\/} uniquely fix $h$. Hence there is not a unique element
in the gauge orbit $\O_A$ satisfying the Coulomb gauge condition and
thus the section (\sectionfour) cannot be
constructed. This is {\it not\/}
an artifact of the Coulomb gauge condition and, as shown by
Singer\up{\SINGER},
occurs for all gauge fixing conditions defined on $\A$. This has
become known as the Gribov ambiguity.  This existence of this
ambiguity is
central to our account of confinement in QCD, hence it is helpful to
briefly explain how it arises.

We have identified
the group of all local gauge transformations $\G$ with the
mappings from space time $\Real^4$ to the structure group $SU(3)$ which
tend to the constant $1$ at spatial infinity. Our restriction to
dressings only made out of the spatial components of the gauge fields
implies that we are not interested in the time dependence of these
local gauge transformation, thus we can view the gauge group $\G$ as
mapping from space $\Real^3$ to $SU(3)$ which tend to 1 at infinity.
This gauge group has lots of interesting topological structures that
it inherits
from the group $SU(3)$. This follows from the fact that the homotopy
groups
 of $\G$ are simply related to those of $SU(3)$ by the relation
$$
\pi_n(\G)=\pi_{n+3}(SU(3))\,.\eqn\no
$$
The homotopy groups of $SU(3)$ are well known\up{\ENCYCLO},
and we note that $\pi_3(SU(3))=\Z$ and  $\pi_5(SU(3))=\Z$. The first
of these implies that
$$
\pi_0(\G)=\Z\,.\eqn\zerofour
$$
To understand what this means we recall that the zeroth homotopy group
measures the number of disconnected components a space has. Thus
(\zerofour) implies that the gauge group is disconnected, having
an infinite number of components. This is responsible for the instanton
structure of QCD. One can see this disconnectedness explicitly by
noting that the function on $\G$ given by
$$
\omega(U)=\frac1{24\pi^2}\int d^3x\,\epsilon^{ijk}\Tr\bigl(U^{-1}
\pa_iUU^{-1}
\pa_jUU^{-1}\pa_kU\bigr)
$$
is invariant under the right action $U\to Ue^\theta$, i.e., it
is invariant under gauge transformations in the identity component of
$\G$. In particular we see that on the identity component
$\omega(e^\theta)=0$.
However, it is possible to find gauge group elements $U$ such that
$\omega(U)$ is not zero\up{\JACKIW}, so these elements cannot be in
the identity
component of $\G$ and $\G$ is thus disconnected.

So how does this topology in $\G$ relate to gauge fixing? Well, if
$\G$ is disconnected then so is the product space $V\times\G$ for
any space $V$. In particular $\A/\G\times\G$ must be  disconnected.
If a gauge fixing exists then we have seen that $\A/\G\times\G$ is
diffeomorphic to the Yang-Mills configuration space $\A$. Hence $\A$
must be disconnected. But this is {\it not\/} true, for if $A_1$ and
$A_2$ are {\it any\/} elements of $\A$ then
$$
A(t):=tA_2+(1-t)A_1\,,\eqn\atrivialfour
$$
for $t\in[0,1]$,
is a line of Yang-Mills potentials in $\A$ from $A_1$ to $A_2$ --- the
space $\A$ is thus connected! From this contradiction we must conclude
that the spaces $\A/G\times\G$ and $\A$ are not diffeomorphic and hence
there can be no gauge fixing.

Now it might be argued that the fact that the group $\G$ is disconnected
really means that we should be just investigating the reduction from
$\A$
to the coset space of $\A$ with the identity component of $\G$. Indeed
the $\theta$-parameter in QCD measures the extent to which the states
are invariant under the full group $\G$ and  only for $\theta=0$ are
they fully invariant. However, even the identity component of $\G$
has a rich topological structure that obstructs the dressing gauge
fixing
in much the same way as this $\pi_0$ result. Indeed the fact that
$\pi_5(SU(3))=\Z$ implies that
$$
\pi_2(\G)=\Z\,.\eqn\no
$$
This means that there are non-contractable two-spheres in $\G$, and
the existence of a perfect
gauge fixing would imply the same for $\A$. This again is not the case
since (\atrivialfour) implies that $\A$ is topologically trivial.

In conclusion, we have seen in this section that it is possible to
construct dressed quarks and gluons, at least perturbatively, and
that such a dressing is closely related to a non-covariant gauge
fixing condition. As a consequence of this, we have seen that a
non-perturbative dressing would amount to the existence of a
globally well defined gauge fixing condition, which is not possible
due to the Gribov ambiguity. We have sketched a proof of this deep
result --- one of the very few rigorously known facts about the
non-perturbative structure of QCD.
Given that physical quarks would need to be dressed, this shows that
it is not possible to construct a
non-perturbative asymptotic quark field. This, we propose, is a
direct proof of quark confinement.
We recall, though, that the experimental results from  deep-inelastic
scattering are understood in terms of freely moving quarks
inside the nucleon and that the naive quark model also shows
that hadrons may be described to a good extent in terms of
constituent quarks. In other words, although free quarks
are not experimentally seen, many aspect of hadronic physics
can be described in terms of one or other sort of quark. We
now want to address the important question of whether the
perturbatively dressed quarks developed here can be used
to describe constituent quarks and the fields seen in deep-inelastic
scattering. The field theoretic
side of this identification will be the topic of the next section.

\vfil\eject

%%%%%%%%%%%%%%%%%%%
%%%             %%%
%%% Section6    %%%
%%%             %%%
%%%%%%%%%%%%%%%%%%%

\secno=6 \meqno=1
\ni
{\bf 6. The Dynamics of Dressed Quarks}
\bigskip
\ni We now want to see how
far we can get using dressed quark fields
in practical calculations. We stress again that the standard approach
to quantum field theory makes much use of locality and covariance in
order to push through the renormalisation programme. Since the dressed
quarks (\psitwo)
and gluons (\agfour) are neither local nor covariant we will in this
section investigate the practicability of working with these fields by
performing one-loop studies of the quark propagator both in
perturbation theory and in the framework of the operator product
expansion (OPE). To retain as many as possible of the advantages of
covariance in such calculations, and also
to  demonstrate the gauge invariant nature of the results, we
will always work in a general Lorentz gauge. We will see that the
one-loop perturbative propagator of the static quark can be
renormalised on-shell in a gauge invariant manner without any
trace of an infra-red divergence. The second half of this section
shows how the QCD vacuum structure can be incorporated through the
OPE and that this yields a gauge invariant, running mass for the
dressed quarks.

\smallskip
\ni {\underbar{i) Perturbation Theory of the  Constituent Quark
Propagator}}
\smallskip
\ni  Before we proceed to the
details of the calculation we wish to first recall some details of the
usual one-loop propagator (i.e., diagram Fig.\thinspace 6.1a) in the
Lorentz gauges.
%\bigskip

\midinsert
%\vskip 3cm
%\centereps{12cm}{3cm}{C:/martin/paper/qqfig61.eps}
%\fig {qqfig61.eps}{3cm}{12cm}
\fig {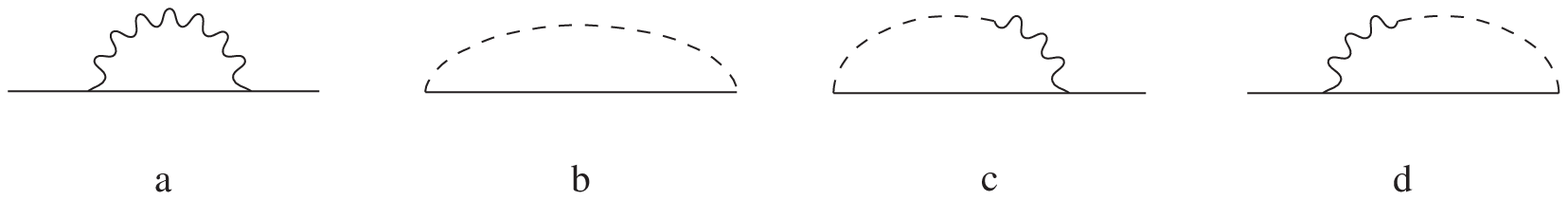}{12cm}

\smallskip
{\eightpoint \narrower
\noindent{\bf Figure 6.1} The one loop contributions to the physical
propagator. Solid lines correspond to the fermion propagator, wavy ones
to the vector boson propagator and dashed lines represent the
projection of the vector boson propagator that comes from the dressing
function.\par}
\endinsert

%\bigskip

This diagram is, up to a factor, the same in both QED and QCD. Its
contribution to the propagator
has a logarithmic ultra-violet divergence. This can be dealt with in
the standard way (see, for example, Ref.\ \HATFIELD). Most important for
 our
purposes is the fact that if one attempts to
perform a mass-shell renormalisation one finds an infra-red divergence.
This means that we cannot straightforwardly choose a subtraction point
such that the external fermions are on shell, i.e., this latter
divergence is
telling us that we are not able to
write a renormalised propagator which
represents a free on-shell fermion. It is well-known\up{\DOLLARD,\JAUCH}
that
the Coulomb force of QED falls off so slowly at
infinity that we cannot ignore it and use free fermions as our
asymptotic fields and this is the origin of the infra-red
divergence. Various people have considered redividing the QED
Hamiltonian into a new interaction term and a term which would lead to
different asymptotic fields\up{\KULISH, \ZWANZIGER}. Essentially the
problem may also be understood as a consequence of the photon being
massless. An infinite number of soft photons may and must
be added to, say, an
electron state and one is thus naturally led to using coherent states
to describe charged particles\up{\CHUNG-\JOHNS}.
The photon propagator does not suffer from such divergences
since it is chargeless and does not directly
interact with the Coulombic fields
produced by (and associated with) any charged particles.

In QCD
we do not have a description of the long distance nature of the force
between bare charges since quarks and gluons are confined. This is
supposed to be due to an
increase of the strong force at larger distances and so it would seem
only logical that we also
cannot employ asymptotic Lagrangian quark or gluon fields
in QCD. Indeed the QCD Green's functions display infra-red divergences
worse\up{\MUTA} than those of QED and it has previously been suggested
that they
are at the root of confinement\up{\WEINBERG-\FRITZSCH}.

The standard methods\up{\JAUCH, \HATFIELD, \MUTA} of circumventing the
infra-red divergence
are to give the photon (gluon)
a tiny mass or to not perform a mass shell renormalisation
scheme. This last option implying that we do not
having an asymptotically on-shell particle. What we want to see
below is {\it will the dressing of the fermion suggested above remove
the infra-red divergence and permit us to put it asymptotically
on-shell?\/}

Before starting the detailed perturbative calculation, it is worth
recalling that it is known\up{\FROHLICH, \BUCHHOLZ} that these
infra-red divergences
are  a manifestation of the fact that the mass operator does not have a
discrete set of {\it normalizable\/} eigenstates in the charged sector of
the theory. As such, the infra-particle ideas of Schroer\up{\SCHROER}
are usually used to discuss these states. Our approach is in
marked contrast to this. The states created by the dressed fields are
non-normalizable states and it is through this process that infra-red
finite results can emerge. In fact the states created have
much in common with coherent states, and it would
be interesting to see in a more precise
way how dressing the charged states relates to the standard\note{We also
note that the good infra-red properties of the Fried--Yennie
gauge\up{\YENNIE} can be
understood through the  gaugeon formalism (see,
for example, page 125 of Ref.\ \NAKANISHI). It is not clear, though, how
this interesting construction relates to
dressing the charged fields due to its
covariant nature.}\  use of coherent states in the infra-red sector.

Finally we further recall that it is known that the quark
pole mass is gauge parameter independent, which may be shown to be
a consequence of generalised Ward (Nielsen) identities\up{\TOM},
and that the
$Z_2$ fermion wave function renormalisation constant in an on-shell
scheme has also been shown to be gauge invariant up to two loops by
explicit calculation\up{\BGS}. Thus armed we can move on to the
field-theoretic description of the dressed propagator.

\bigskip
The
propagator of the static dressed quark (\psitwo) receives at one loop
contributions from the four diagrams of Fig.\thinspace 6.1. These
correspond respectively to the usual interaction vertices  of the free
fermion propagator (6.1a),
gluonic corrections which emerge purely from the dressing (6.1b) and
corrections which receive contributions from both the dressing and the
normal interaction vertices (6.1c \& d). Note that the dressing
projects out certain components of the full vector boson propagator.
Since we wish to preserve BRST invariance we will always use dimensional
regularisation which means that tadpole diagrams,
where we take both of the $g^2$
corrections from one of the dressing functions alone,
vanish. Note also that diagrams such as Fig.\ 6.1b need to be
considered at the level of the propagator.

Rather than calculate
the diagrams one at a time we choose to
first combine them all at integrand level. Note that
since the diagrams are, up
to colour factors, the same for QED and QCD we drop colour indices
henceforth. The first
step is to check if the propagator of the gauge invariant dressed
fermion is itself gauge invariant. Each of the four diagrams in Fig.\
6.1 has a piece linear in the gauge parameter, $\xi$. A little algebra
shows that their sum reduces to tadpoles which we recall
vanish in dimensional
regularisation. Any possible gauge dependence thus cancels\up{\LMPROP}.

Working in Feynman
gauge one now attaches where necessary
external factors of $(p\,\slash-m)
(p\,\slash-m)^{-1}$ in such a way that a free propagator is present
left and right of each of the terms from the four diagrams. This allows
us to better combine them.
Dropping some tadpoles one finds after a certain
amount of algebra that the integral may be reduced to that
which we would have from the usual fermion propagator (Fig.\ 6.1a)
evaluated in Coulomb gauge, i.e., we have
$$
S'(p)=g^2S_0(p)
\int\!\frac{d^{2\omega}k}{(2\pi)^{2\omega}} \gamma_\mu S_0(p-k)
\gamma_\nu D^{\mu\nu}_{\rm c}(k) S_0(p)
\,,
\eqn\Coulint
$$
where $D_{\rm c}^{\mu\nu}$ is the vector boson
propagator in Coulomb gauge.
This is of course what we would expect --- the exponential factors in
the dressing reduce to unity in Coulomb gauge ($\partial_i A_i=0$).

A reasonable amount is known about perturbation theory for
electrodynamics in Coulomb
gauge\up{\JOHNSON-\ADKINS}. We
will most closely follow
Ref.\ \ADKINS, although our notation will deviate from his in several
places.
The regulated propagator
$$
S'(p)=\frac1{p\,\slash -m-\Sigma(p)}
\,,
\eqn\Sreg
$$
is related to the renormalised propagator, $\Sr(p)$, by
$$
S'(p)=Z_2 \Sr(p)
\,.
\eqn\ztwo
$$
Writing the inverse regularised propagator as
$$
p\,\slash-m-\Sigma(p)=ap\,\slash+bm+c{p\,\slash}_0
\,,
\eqn\abcdef
$$
where ${p\,\slash}_0=p_0\gamma_0$ and $a$, $b$ and $c$ are as yet
unknown functions
depending upon $p^2$ and $p_0^2$, we see that
$$
S'(p)=\frac{a p\,\slash +c{p\,\slash}_0 -m b }{m^2\Delta^2}
\,,
\eqn\Sregtwo
$$
where
$$
\Delta^2(p^2,p_0^2)=a^2p^2+c^2p_0^2+2 a c p^2_0-b^2m^2
\,.
\eqn\Deltadef
$$
The physical mass is determined by
$$
\Delta^2(m^2,p_0)=0
\,.
\eqn\Deltazero
$$
This determines the mass renormalisation. We stress again that this is
known to be gauge parameter independent in the class of covariant
gauges.

Writing the mass as a sum of
the bare mass and the mass shift, $m=m_0+\delta m$, one finds the
following forms for the three functions in (\abcdef) to one
loop in $2\omega =(1,2\omega-1)$ dimensions\up{\ADKINS}
$$
\eqalign{
a(p^2,p_0)=&1+\frac\alpha{4\pi{\hat\epsilon}} +
\frac\alpha{4\pi}\big(\big.
\frac{19}6 -\int_0^1\! \frac{dx}{\sqrt{x}} (1-x)
\log {\cal X} -2\int_0^1\!dx (1-x)\log{\cal Y} \cr
&
\quad\quad\quad\quad +
2 \int_0^1\!dx \sqrt{x}\int_0^1\!du\log{\cal Z}\big.\big) \,,
\cr
b(p^2,p_0)=& -1-\frac\alpha{\pi}\left(\frac1{\hat\epsilon}+1\right)
+\frac{\delta m}m +
\frac\alpha{4\pi}\left( \int_0^1\!\frac{dx}{\sqrt{x}}\log{\cal X}
+2\int_0^1\!dx\log{\cal Y}\right)\,,
\cr
c(p^2,p_0)=& \frac\alpha{4\pi} \left( -\frac83+ \int_0^1\!
\frac{dx}{\sqrt{x}} (1-x)\log{\cal X}-2\int_0^1\!dx\sqrt{x}\int_0^1\!du
\log{\cal Z}\right)
\,,}
\eqn\oneloopfns
$$
where
$$
\frac1{\hat\epsilon}= \frac1{2-\omega} -\gamma_{\sevenrm E} +
\log(4\pi)\,,
\eqn\betterthanD
$$
and
$$
\eqalign{
{\cal X}=& 1+\frac{{\bold p}^2}{m^2}(1-x)\,,
\cr
{\cal Y}=& 1-\frac{p^2}{m^2}(1-x)\,,
\cr
{\cal Z}=& 1-\frac{p_0^2}{m^2}(1-u)
+\frac{{\bold p}^2}{m^2}(1-xu)-i\epsilon
\,.}
\eqn\XYZ
$$
The work of Ref.\ \ADKINS\ shows
that the reality properties of the full propagator (it
aquires an imaginary part for $p^2>m^2$) make physical sense.
Note further
that although some non-covariant integrals must be performed
to obtain these results, the integrals are well-defined
and the situation is therefore in marked contrast
to axial gauges where an integration prescription needs to be
introduced\up{\VIENNA, \LMRAD}.

Armed with these formulae we can study the mass shift, $\delta m$.
Since it is gauge invariant we would not expect it to depend upon $p_0$
and indeed explicit calculation shows that
$$
\frac{\partial \delta m}{\partial p_0}=0\,,
$$
which was assumed in Ref.\ \ADKINS. The calculation of $\delta m$
is most easily
performed for $p_0=m$, i.e., the static mass shell since there we find
that (\Deltazero) simplifies to\note{A factor of $\frac{\xi^2}{m^2}$ in
Eq.\ 34 of Ref.\ \ADKINS\ is superfluous.}
$$
\tilde a+\tilde b+\tilde c=0
\,,
\eqn\aandbandc
$$
where $\tilde a$
denotes that the function is taken at its static mass shell (i.e.,
$\tilde a=a(p^2=m^2,p_0=m)$, etc.). Since this is linear in $\alpha$ it
may be straightforwardly solved while at other (non-static) mass shell
momenta this is not the case and we must expand in the coupling only
retaining terms up to order $\alpha$. One finds\up{\ADKINS}
$$
\delta m=\frac{\alpha m}{4\pi}\left(\frac3{\hat\epsilon}+4\right)
\,,
$$
which is the covariant gauge result. This demonstrates that the gauge
invariant mass of the usual fermion propagator is just the mass of the
dressed fermion.

To fully renormalise the dressed fermion propagator we now consider
$Z_2$. We first consider the static mass shell. In other words we
choose our
renormalisation point to be at $p=(m,0,0,0)$, which we stress is what
our earlier considerations, i.e., the fact that the
electric field associated to the dressed fermion is that
of a static charge, would suggest. The renormalisation
condition is then that as $p\to (m,0,0,0)$ we have
$$
\Sr(p)\to \frac{m\gamma_0+m}{p^2-m^2}
\,,
\eqn\rencond
$$
which corresponds to a static asymptotic field. (This renormalisation
condition was used in Ref.\ \ADKINS, the motivation there was that this
was a reasonable renormalisation point for an electron in positronium
where one might expect the momenta to be small. Below we will see that
this choice is in fact forced upon us by the theory.)

The full propagator in this limit is
$$
S'(p)\to
\frac{
m\tilde a\gamma_0+m\tilde c\gamma_0 -
m \tilde b}{
m^2(p^2-m^2)\left[\partial/\partial p^2\Delta(p^2,p_0)\right]
|_{p^2=m^2,p_0=m}
}
\,,
\eqn\limitstat
$$
which from (\aandbandc) may be recast in the form
$$
S'(p)\to\frac{\tilde b(m\gamma_0+m)}{
(p^2-m^2) \left[2m^2\tilde b({\tilde a}'+{\tilde b}' +
{\tilde c}')-{\tilde a}^2 \right]}
\,,
\eqn\statlimit
$$
where the obvious notation ${\tilde a}'=\partial/\partial p^2
a(p^2,p_0) |_{p^2=m^2,\,p_0=m}$ applies. Comparing (\statlimit) and
(\rencond) we see that the multiplicative renormalisation condition
(\ztwo) holds. Substituting one finds that
$$
Z_2=1-\frac\alpha{4\pi}\frac1{\hat \epsilon}
\,.
\eqn\ztwostatic
$$
This is a very attractive result. We see that $Z_2$ is not plagued by
an infra-red divergence --- in contrast
to the covariant gauge propagator.
This implies that one can use this field as an
asymptotic one (at least in the static two-point function). This is
strong confirmation that the dressing we use is physical even at the
quantum level.

This mass shell
renormalisation of the dressed fermion propagator at one loop is
now complete. We have seen that the mass shift is gauge invariant and
equal to that of the usual fermion in covariant gauges. The wave
function renormalisation constant, $Z_2$, is, however, very different
to that in covariant gauges --- it is infra-red finite and could be
multiplicatively renormalised. (This last is in contrast to axial
gauges, where multiplicative renormalisation is often
impossible\up{\VIENNA}.)
However, we wish to now mention a further difference
between the renormalisation of the physical fermion and the usual
covariant approach. In the renormalised
propagator of, say, $\phi^4$ theory we find after a mass shell
renormalisation that the renormalised propagator approaches a free
field propagator if the particle is put on shell; since the propagator
only depends upon $p^2$ it makes no difference in which way it is put
on shell.  The renormalised
propagator which we have obtained above depends upon both $p^2$
and $p_0^2$ and only looks like a free propagator on the static mass
shell. The interpretation that the propagator describes a free field on
shell is not open to us for other, non-static, mass shell momenta. This
is a reflection of the breakdown of the Lorentz boost symmetry and we
wish to now study it further.

We have argued that the dressing of the fermion is correct
for a static charge. Let us consider what happens if we
use a non-static mass shell condition. To be
explicit we consider the moving mass shell point $p=m(5/4,0,0,3/4)= m
\eta^{*}$, i.e., $\eta^*$ is a (not purely temporal)
unit vector. We
have already seen above that this will have the same mass shift,
$\delta m$, and this may also be explicitly verified from the
condition here. Explicitly, $\delta m$ may be found from
$$
{\tilde a}_1+{\tilde b}_1+\frac{25}{16}{\tilde c}_1=0
\,,
\eqn\Movshell
$$
where the subscripts denote that only terms of order $\alpha$ in $a$,
$b$ and $c$ are to be retained (e.g., $a=1+\alpha a_1+\cdots$)
and the tildes now of course refer to
evaluating the functions at the moving pole. The usual result emerges
from (\Movshell).
A problem, however,
rears its head when we attempt to evaluate
$Z_2$. The on-shell renormalisation condition
(\rencond) now reads
$$
\Sr(p)\to
\frac{\frac54 m\gamma_0-\frac34m\gamma_3+m}{
p^2-m^2}=\frac{m\eta^*\!\!\slash+m}{p^2-m^2}
 \,,\quad\quad {\rm as}\quad p\to m\eta^{*}\,.
$$
Compare this with the full propagator $S'(p)$ which near the moving
pole is
$$
S'(p)\to
 \frac{\frac54 m (\tilde a+\tilde c)\gamma_0 -\frac34
m\tilde a\gamma_3-\tilde b m}{
(p^2-m^2)m^2\left[\partial/\partial p^2\Delta|_{p^2=m^2,\,p_0=
\frac{25}{16}m^2}
 \right]}\,,
\eqn\movingfullprop
$$
where the derivative in the denominator is evaluated at the non-static
pole position. To first order in $\alpha$ this may be rewritten as the
sum of two terms
$$
\eqalign{
S'(p)\to  &
 \frac{m\eta^*\!\!\slash+m}{
(p^2-m^2)m^2\left[\partial/\partial p^2\Delta|_{p^2=m^2,p_0=
\frac{25}{16}m^2}
 \right]}\cr
&
+\alpha\frac{\frac54m(\tilde a_1+\tilde c_1)\gamma_0 -
\frac34m\tilde a_1\gamma_3-\tilde b_1m}
{p^2-m^2}
\,,}
\eqn\Sprimeone
$$
We remark that $\Delta^2$ is to be expanded in the coupling.

It is apparent from the second term in this last equation
that the renormalisation is not now
multiplicative and explicit calculations bear this out. The
non-multiplicative term may be calculated and is seen to be
non-vanishing but both
ultra-violet and infra-red finite. This is in accord with the work of
Heckathorn\up{\HECKATHORN}
who showed that the UV poles were multiplicatively renormalisable.

Consider the first term on the RHS of (\Sprimeone). It is
visible that the UV poles are as in the static mass shell scheme. A
short calculation shows, however, that there is now a new infra-red
divergence of the form
$$
\int_0^1\! dx\frac{\sqrt{x}}{25-9x}\int^1_0\!\frac{du}u
\,.
$$
The interpretation of this divergence is clear. We have calculated the
one-loop, dressed propagator corresponding to a static charge. If we
use a static renormalisation point we can perform a multiplicative,
infra-red finite, mass shell renormalisation. If we choose any
non-static mass shell renormalisation point, even for small
velocities, we acquire not only non-multiplicative terms but also
an infra-red problem since the dressing we employ is only suitable for
a static charge. This is an explicit manifestation of the non-covariant
nature of the dressing, which we claimed in Sect.\ 3 and we will return to
this in Sect.'s 8 and 9.

The results of this subsection show the great sensitivity
of the theory to the type of dressing employed.
In Sect.~9 we will study how to boost a charge and how the
dressing must then be altered. We will in particular see that Coulomb
gauge does not then remove the dressing, which indicates why it is not
possible to carry out a non-static
mass shell renormalisation scheme for static charges. We conclude this
subsection by repeating that no problems were encountered in
renormalising the physical fields as long as the correct
renormalisation scheme was used.

\goodbreak
\bigskip
\ni {\underbar{ii) Operator Product Expansion of the Constituent Quark
Propagator}}
\smallskip
\ni Notwithstanding the success of the above perturbative description
of the dressed quarks, it is natural to want more. Low energy hadronic
physics cannot be understood solely in terms of perturbation theory
and,
although we may hope to describe constituent quarks in hadrons as (at
short distances) weakly interacting quasi-particles, it seems likely
that the dressed quarks ought to receive some non-perturbative input
which should, e.g., generate a rather larger mass than the current
one. The rest of this section will concern the interaction of the
perturbatively dressed quarks with the non-perturbative QCD vacuum.

A widely applied method of incorporating some non-perturbative
QCD physics is the operator product expansion (OPE) as used in QCD sum
rules. The OPE\up{\WILSONOPE} for the
product of two local operators
takes the general form of a weak equality valid between states:
$$
A(x)B(0)\approx\Sigma_n C_n{\cal O}_n\,, \quad {\rm as}\,\,x_\mu\to 0
\,,
\eqn \OPEdef
$$
and it is used to
separate short distance (the $C_n$)  and long distance effects (the
${\cal O}_n$). The former coefficients may
be calculated perturbatively while the
latter are non-perturbative operators which,
at the current level of our
understanding of QCD, are best estimated from phenomenology.

In the form of QCD
sum rules\up{\SVZ}
this has developed into a very productive way of predicting
hadron masses and investigating other hadronic properties. In the sum
rules the short distance effects are the
perturbatively calculable coefficients of various
universal, non-perturbative condensates in correlation functions of
certain gauge-invariant currents. These are then compared with the
results of models (resonance saturation) for the same correlation
functions and the model parameters (e.g., the mass of the relevant
hadron) may be obtained from a fit. The condensates (whose values are
themselves obtained from fits) are the vacuum expectation values of
certain operators. They are essentially used to
parameterise the non-perturbative nature
of the QCD vacuum and are to be thought of as unknown functions of the
QCD scale rather than new parameters. A simple
example is the quark condensate, $\langle \bar \psi
\psi\rangle$, whose value may be extracted from the PCAC relation,
$(m_u+m_d)\langle \bar \psi \psi\rangle=-m^2_{\pi}f^2_\pi$, which
relates the light quark masses, the pion form factor and the pion mass
to a condensate whose value is a measure of
the extent of spontaneous chiral symmetry breaking. The OPE corrections
generally
fall off as powers of the momenta flowing through the Green's function
or correlation function under consideration. So for large momenta they
are negligible, while
for very small momenta the OPE breaks down but in some
middle region they are hoped to be both important and reliably
calculable.

We now want to see how the
leading order effects of the non-perturbative vacuum
influence the propagator of the dressed quark. It is not obvious that
the OPE can be directly applied to the non-local dressed fields.
However, the central fermion fields are local and so we may hope that
at least the coefficient of the quark condensate may be found.
Specifically we want to
check that this lowest order term from the OPE
yields a gauge-invariant correction to the
propagator and to discover its exact form.

We recall in this connection that Politzer suggested\up{\POLITZER}
many years ago that the OPE
of the Lagrangian quark propagator, and in particular the
contribution from the quark condensate, $\quark$, could generate the
constituent quark mass. It has later been realised that
his result was highly gauge dependent. What we will see, however,
 is that this
idea finds a natural application in the dressed quarks and that it
there leads to a gauge-independent, non-perturbative correction to
the physical quark propagator. We now begin
by summarizing the OPE of the usual propagator\note{There are different
ways of calculating the OPE corrections. Ref.\BAGAN\ provides a
recent discussion.}.

\topinsert
%\vskip 3cm
%\centereps{12cm}{3cm}{C:/martin/paper/qqfig62.eps}
%\fig {qqfig62.eps}{3cm}{12cm}
\fig {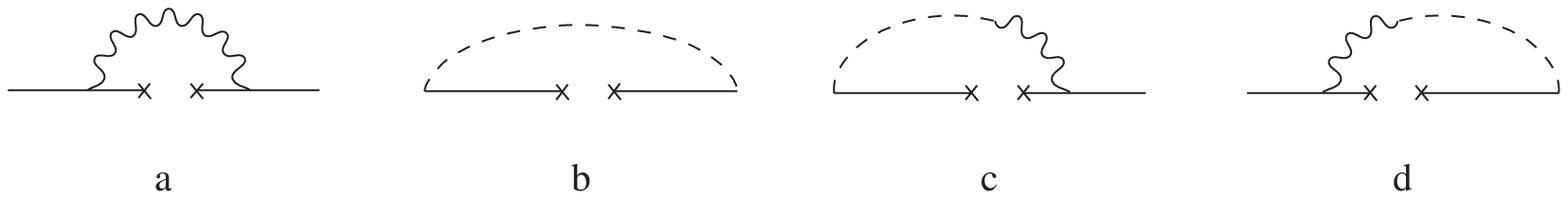}{12cm}

{\eightpoint \narrower
\noindent{\bf Figure 6.2} Insertion of quark condensates into the
dressed propagator. The crosses denote quark vacuum insertions. Other
notation is as in Fig.\ 6.1.\par}
\endinsert

The first non-perturbative
term in the OPE of the usual quark propagator comes from the quark
condensate, $\quark$. It arises from Fig.\thinspace 6.2a. Politzer
worked at lowest order in the quark mass and obtained\note{A factor was
corrected in Ref.\PASCUAL.}\
in Landau
gauge the following contribution to the  quark self energy
$$
\Sigma(p^2)=\frac{3(N_c^2-1)\pi\alpha_s\langle
\bar\psi\psi\rangle}{2N_c^2p^2}
\,.
\eqn\selflow
$$
It was then
argued that this could be understood as an effective, running
mass and such ideas have enjoyed some popularity in
phenomenology. The
mass falls off rapidly at high momenta and the mass scale introduced by
(\selflow) for initially chiral quarks is
(using $\quark=(-270{\rm MeV})^3$) of a size which it is
natural to want to
identify with the constituent quark mass of the light $u$ and $d$
quarks.
Unfortunately a more exact analysis shows, as we now discuss,
that this mass term in the standard
propagator does not in fact
lend itself to any straightforward interpretation.

Generally in $D$ dimensions for an $SU(N_c)$  gauge  group
and to first order in the quark mass one finds\up{\PASCUAL}
in an arbitrary Lorentz gauge from Fig.\ 6.2a:
$$
\Sigma(p^2)=\frac{(N_c^2-1)\pi\alpha_s\langle
m\bar\psi\psi\rangle}{2N_c^2p^2}    \left[   \frac{2\xi   (2-D)}D
\frac{p\,\slash}{p^2} +(D-1+\xi)\frac1m\right]
\,.
\eqn\selfhigh
$$
This shows that this correction to the propagator is gauge dependent
and hence cannot be associated with any physical meaning.

In fact the OPE of a gauge dependent quantity such as this propagator
is even more subtle than this (for a review see Ref.\thinspace
\MLMOREV). Although $\quark$ is the only gauge and Lorentz invariant
dimension three condensate, there are several such quark (and gluon)
condensates of dimension four and (\selfhigh) is actually too simple.
To obtain $(\selfhigh)$ use was made of the equation of motion for the
quark condensate, $\langle\bar\psi(D\slash-m)\psi\rangle=0$.
In QCD sum rules, where
one works with correlation functions of gauge (BRST) invariant currents
this is valid since it is known\up{\COLLINS}
that the only operators which can occur there
are gauge invariant ones (such as the $\quark$ or
$\langle G^2\rangle$ condensates),
equation of motion operators such as the above
(which can be set to zero in physical
states) and BRST variations (whose expectation values vanish). This
means that we are entitled  in the sum rules
to use the equation of motion to
replace $\langle\bar\psi\partial\,\slash\psi\rangle$
by $\langle m\bar\psi\psi
\rangle$ since the $\langle\bar\psi gA\,\slash\psi\rangle$ terms will
take care of themselves. (For an explicit demonstration of this
in the OPE of the two-point correlation
function of the singlet vector current see Ref.\thinspace \MLMOSR.) Since
the quark propagator is not gauge invariant this use of the equation of
motion is not so useful. This fundamental difference between the
QCD propagators
and the gauge invariant sum rules means that the
condensates do not have to appear in particular combinations in the
propagators and so it is
necessary to calculate the coefficients of {\it all} the
 condensates of a particular dimension.
The OPE of the
propagator at first order in the quark mass and leading order in
$\alpha_s$ is found\up{\MLMOSR}, from the diagrams
of Fig.'s 6.2a and other additional diagrams, to be such that
the condensates enter the OPE of the propagator
in gauge dependent combinations and with gauge dependent coefficients.
No direct physical meaning may be read off from this OPE of the usual
propagator, although
these results may be used to constrain solutions of the Schwinger-Dyson
equations\up{\STINGL}. In what follows we will restrict ourselves to
considering the dimension three condensate, $\quark$.

We have seen above that the pole mass of the usual propagator is gauge
invariant in perturbation theory and that this is understood
theoretically. It has been argued that the OPE correction to the
the pole mass is also gauge invariant\up{\ELIAS}.
This may, e.g., be seen from
(\selfhigh) by setting the mass to the pole mass and it follows that
at this pole the explicit gauge parameter dependence cancels. We
point out, however, that,
since the quark condensate is negative, the pole mass
which would follow from (\selfhigh) would be negative. (For Euclidean
momenta the effective mass is, although gauge dependent,
indeed positive.)  These sign problems notwithstanding,
we do not see a gauge independent pole if all dimension four quark
condensates are included\up{\MLMOSR}. It may be that summing
higher condensates and repeatedly using the equation of motion to
replace $\langle\bar\psi(\partial\,\slash)^n\psi\rangle$ by
$m^n\langle\bar\psi\psi\rangle+{O}(g)$ produces such a
cancellation. However, the OPE is surely
not valid at the scale of the $u$
or $d$ constituent masses and, even if were we to accept such
calculations as
order of magnitude estimates, what we would really like is a {\it
physically
meaningful, gauge invariant  running mass for the dressed quarks}.
The constituent masses obtained from hadronic spectroscopy will then be
formed from some average over the effective masses of the quarks inside
the hadrons. The rest of this section describes a first step in this
direction.

\bigskip
The leading order OPE correction to the physical propagator comes from
calculating the coefficient of $\quark$ at order $m^0$ and $\alpha_s$.
This may, in close analogy to the perturbative calculation of the first
half of this section, be found from the diagrams of Fig.\thinspace 6.2.
We work in a general Lorentz gauge. It
is natural to divide the calculation into two steps: firstly we
calculate the terms which depend upon the gauge parameter in the
diagrams and, after we have satisfied ourselves that their sum
vanishes, we calculate the gauge-independent OPE correction.

Since we are calculating the entire propagator at lowest order in the
current quark mass, we have to expand the external propagators (in
Fig.'s 6.2a,
c and d) in the mass. Higher order condensates, of the form
$\langle\bar\psi(\partial\,\slash)^n\psi\rangle$,
can be related via the
equation of motion to $m^n\langle\bar\psi\psi\rangle+{O}(g)$ and
should be introduced
in order to investigate the gauge independence of these
corrections. We will not study such terms here. The lowest
order calculation is completely
standard and the perturbative gluon propagator
in the three diagrams which receive a contribution from the gluon cloud
around the quark is just as in the first part of this section.

Each of the diagrams in Fig.\thinspace 6.2 includes a term linear in
the gauge parameter $\xi$.
To lowest order in $m$ we find that the magnitudes
of all four terms are
identical and the sum in the dressed propagator vanishes. This shows
that the leading OPE correction is, as we had hoped, gauge-invariant.

The remainder may be found by evaluating each of the diagrams in
Feynman gauge. The result may be readily checked to be:
$$
S_F(p^2)= \frac 1{p\,\slash}  + \frac{(N_c^2-1)\pi\alpha_s\langle
\bar\psi\psi\rangle}{2N_c^2p^4}       \left[      \left(      D-1
-\frac{p_0^2}{{\bold p}^2} \right) \right] \,,
\eqn\dressope
$$
which result has been previously obtained\up{\MSINVIENNA}
as the leading OPE
correction to the Coulomb gauge quark propagator\note{A factor of 3 is
misprinted in that reference.}. In the instantaneous approximation this
is just the standard  Landau gauge result.

Summing up this Coulomb gauge result we have:
$$
S_F(p^2)
 =
\frac1{p\,\slash-\Sigma(p_0^2, {\bold p}^2)}
\,,
\eqn\sumope
$$
where $\Sigma$ may be read off from (\dressope).

We now have to interpret this result. Although (\sumope) seems to
vanish for a static quark, which would be a further signal of
confinement, we should recall that the OPE is only valid in the
deep-Euclidean region, i.e., for large ${\bold p}^2$.
What we can, however, say
from (\dressope) and (\sumope) is that OPE effects from the quark
condensate introduce gauge-invariant, non-perturbative corrections in
the dressed quark propagator which have a power-like fall off and
introduce a running mass scale. We see also from $\Sigma$ in (\sumope)
that at Euclidean momenta,
where we may trust the OPE, the sign of the self energy is that of a
(running) positive mass and that the scale which the non-perturbative
effects have introduced is of the order of the light quark
constituent masses. Even if we do not want to extrapolate (\sumope)
down to small momenta, we find the appearance of this constituent mass
scale (and its gauge invariance) highly suggestive.

We should perhaps conclude this section by remarking\up{\BROADH} that
for heavy
quarks the fermionic condensate is small, $\langle\bar QQ\rangle\approx
-\frac1{12}\langle\frac{\a_s}\pi G^2\rangle/m_Q$, where
$\langle\frac{\a_s}\pi G^2\rangle$ is the gluon condensate ($\approx
(360{\rm MeV})^4$. This means that any change in the heavy quark masses
from the condensate will be very small. Higher dimensional condensates
will be power suppressed and typical condensate scales are very small
compared to the heavy quark masses. This implies that the heavy quark
current and constituent masses are very similar.

The calculations of this section raise several immediate questions.
Clearly it is interesting to see how far we can take this OPE approach.
In addition one would like to see to what extent the incorporation of
such non-perturbative effects as instantons is feasible. As we proved in
Sect.\ 5, a full non-perturbative description of the dressing is
impossible. However, although studying the breakdown of our construction
for a single quark is important, it is perhaps more physical to study
the limitations of the dressing in quark-antiquark and three quark
colour singlet systems --- since these states have a well-defined
hadronic description. The next section will therefore be devoted
to the simplest such system: a meson made up of two ultra-heavy quarks.

\vfil\eject

%%%%%%%%%%%%%%%%%%%
%%%             %%%
%%% Section 7   %%%
%%%             %%%
%%%%%%%%%%%%%%%%%%%

\secno=7 \meqno=1
\ni
{\bf 7. The Potential Between Two Heavy Quarks}
\bigskip
\ni
Wilson\up{\WILSON} has
suggested that studying a non-local operator whose behaviour in a certain
limit yields the static fermion-antifermion potential will give the
confinement scale. This approach to the potential has been widely
studied: in perturbation theory this Wilson loop has been
calculated\up{\FISCHLER} to $O(g^6)$ and on the lattice it has been
extensively investigated. We will reconsider this approach in the
light of our construction in the latter part of this section and
argue that a very similar operator may be constructed out of dressed
fields. This will give us one relatively familiar route to the
confinement scale (signalled by the, as we have shown, inevitable
breakdown of the dressing).
We now, however, wish to directly calculate the potential between
two dressed fermions in perturbation theory. For QED this will give
the Coulomb potential. In QCD we will see that our approach involves,
to any given order in the coupling, calculations which are  one loop
less than the corresponding Wilson loop approach. This further shows
the utility of the dressed quarks presented in this paper and opens
the door to higher order calculations of the potential than have yet
been performed.

Recalling Dirac's original motivation for introducing the dressing,
as discussed in Sect.~4, we can write the electric field associated
with the static fermion as
$$
E_i^a(x)\ket{\psic(y)}= [E_i^a(x), \psic(y)]\ket0
\,,
\eqn\eforone
$$
where $\ket{\psic(y)}=\psic(y)\ket0$. Here we assume that
$E_i^a(x)\ket0=0$, which is valid in this static situation.
For two charges the electric field may be analogously shown to be
$$
E_i^a(x)\ket{\bar\psic(y)\psic(y')}=
\bigl([E_i^a(x), \bar\psic(y)]\psic(y')
+\bar\psic(y)[E_i^a(x), \bar\psic(y')]\bigr)\ket0
\,,
\eqn\efortwo
$$
where now $\ket{\bar\psic(y)\psic(y')}= \bar\psic(y)\psic(y')\ket0$.
Given that both of these dressings correspond to static (ultra-heavy)
charges, which have no magnetic fields, the appropriate measure of the
energy of this configuration is determined by the Hamiltonian\note{The
Hamiltonian appropriate to Lorentz gauges can be written as $H=
\frac12\!\int \!d^3x(E^2+B^2)+[Q_{\rm  BRST}, \Psi]$, where
$Q_{\rm BRST}$ is the BRST charge and $\Psi$ is a ghost number
minus one function which contains all the information about the
gauge fixing\up{\FRADVILK}. The vacuum and the two particle dressed
states needed here are constructed so as to be BRST invariant, hence the
commutator term in the Hamiltonian does not contribute to the
potential  --- effectively we can use $H=\frac12\int d^3x (E^2+B^2)$.
Now our static dressing does not generate a magnetic field, so the
static interfermion potential is determined by the electric field
part of this Hamiltonian. Had we chosen to work in Coulomb gauge,
the dressing would have vanished, but the above argument for the
form of the effective Hamiltonian does not go through\up{\USAGAIN}.}
$$
H=\frac12\int d^3x E^a_i(x) E^a_i(x)\,.
\eqn\heff
$$
The energy can be divided into a self-energy part and the intercharge
potential. This last part depends upon the separation of the charges,
we now wish to calculate this. For simplicity we start with the QED case.

In the abelian theory we find using (\reFFed) that
$$
E_i(x)\ket{\bar\psic(y)\psic(y')} =
\frac e{4\pi} \left(\pa_i\frac1{\vert {\bold y}-{\bold x}\vert} -
\pa_i\frac1{\vert {\bold y}'-{\bold x}\vert}\right)\ket{\bar\psic(y)
\psic(y')}\,.
\eqn\abbie
$$
Thus we obtain the expected result for the potential between such
heavy fermions
$$
\bra{\bar\psic(y)\psic(y')}H\ket{\bar\psic(y)\psic(y')}=
-\frac{e^2}{4\pi}\frac1{\vert{\bold y}-{\bold y}'\vert}\,,
\eqn\potty
$$
where we have dropped the divergent self-energy terms.

In QCD the order $g^2$ calculation mirrors the previous argument.
Note, however, that with our conventions for the $T^a$ matrices the
trace  $\tr T^aT^b$ is negative definite. One thus obtains the QED
result once more, but with the correct colour prefactor. This simple
calculation has, in contrast to the more familiar Wilson loop
technique, {\it not\/} involved the vector boson propagator!
It seems that the dressing has directly incorporated these
contributions. We will now sketch the argument that such simplifications
are present at all orders in this approach to the potential.

The explicit static dressing found above is generically of the form
$$
h\sim 1+ g A + g^2 A^2 + g^3 A^3 +\cdots + g^n A^n + \cdots
\,,
\eqn\genus
$$
where by $A^r$ we simply denote the power of the vector potential
in any expression. Thus, following our above line of argument, we
see that such a dressing contributes an electric field of the form
$$
E\sim g + g^2 A + g^3 A^2 +\cdots
\,.
\eqn\elecgenus
$$
The energy of a static quark-antiquark system will then be of the
form
$$
\bra{\bar\psic\psic} E^2\ket{\bar\psic\psic}\sim g^2 \bra01\ket0 +
g^4\bra0 {A^2}\ket0
+g^6\bra0 A^4\ket0 +\cdots
\,.
\eqn\elecener
$$
The first term in this schematic series is just our above calculation
of the Coulombic potential. The next order in the potential involves
the calculation of $\bra0A^2\ket0$ {\it to order $g^0$}, i.e., the
free propagator. The free propagator of a general Lorentz gauge may
be employed due to the gauge invariance arguments mustered above.
The form of the dressing dictates the various projections of the
propagator that enter into the calculation of the $g^4$ potential.
This term in the potential can
in the Wilson loop scheme be found from a one loop perturbative
calculation which is then nested inside the contour of the
loop\note{The level of complexity of this perturbative calculation is
such that it is the most widespread diagnostic tool for prescriptions
for dealing with the singularity in the axial gauge
propagator\up{\VIENNA}.}. We see that the dressing transforms a
one loop calculation into an algebraic one involving only the
free propagator.

This is not to say that the still higher order calculation of the
potential remains purely algebraic in this scheme. At order $g^6$,
which is the highest order term extracted from the Wilson loop
method till now, we have two contributions: one from the $g^6 A^4$
term in (\elecener),
which involves a tree level four point function, and a second
from the $g^4 A^2$ term where now, rather than the free propagator,
the $g^2$ term in the vector propagator appears. Again this involves
a perturbative calculation one loop lower than its Wilson loop
counterpart.

We thus see that the approach to the interquark potential detailed
here involves significantly less complicated integrals than the
conventional one. That the projections due to the dressing can be
algebraically involved is by comparison of little importance. We
believe the $O(g^8)$ term in the static potential is now quite
feasible --- the most complicated term being projections of the
two loop, Lorentz gauge propagator.

Our philosophy in this paper has been to use dressed quarks as
fundamental entities insofar as this is possible. We have seen
above that we can discuss the perturbative interquark potential
in this fashion. Cahill and Stump\up{\CAST} considered in an
interesting paper the potential from a different perspective.
These authors introduced a cloud of glue to dress a non-local
quark-antiquark pair corresponding to an ultra-heavy meson.
This glue was such that the resulting system was gauge invariant,
i.e., they considered
$$
\bar\psic(y)K(y,y')\psic(y')\,,
\eqn\specialk
$$
where, to ensure gauge invariance, the field dependent ansatz for
$K$ had to transform under a gauge transformation as $K\to U^{-1}
(y)KU(y')$. These authors argued that a Wilson line was too singular
to be a good ansatz for the physical cloud of glue. Using energy
minimisation arguments they then  arrived at a formula for $K$
which yielded the usual potential to order $g^4$. Upon examining
their formulae we can recognise that the gluonic cloud used in
their paper is a simple product of our static quark dressing
$$
K(y,y')=h_{\rm c}(y) h^{-1}_{\rm c}(y')\,,
\eqn\factoredk
$$
which as we have seen gives a gauge invariant description of such a
meson. Combining their analysis with ours we arrive at another
characterization of confinement: we can no longer talk about
quarks at the scale from which the mesonic glue {\it does not factor}
into a product of two, individual constituent glue factors.

Given the widespread use of the Wilson loop method, it is natural to
try to reformulate it in terms of the dressed quark picture. We
therefore now recap the essential details of that approach.
Consider a pair of heavy fermions connected by a gluonic
string along some path, $\Gamma$. We will assume that the path $\Gamma$
lies in the time slice of the fermions but is otherwise arbitrary.
We now define the state
$$
\ket{\Gamma,t}\equiv \bar\psi({\bold x},t) P \exp \left( i
\int^{\bold y}_{\bold x} \!
d{z}_i A^i({\bold z})\right)\psi({\bold y},t)\ket0
\,,
\eqn\pathstate
$$
where the integral is along $\Gamma$ and $P$ implies a path ordered
exponential. The overlap of two such states at time $t=0$ and $t=T$
yields for large (Euclidean) $T$
$$
\eqalign{
\lim_{T\rightarrow\infty}
M(T, \Gamma)\equiv & \bra{\Gamma, 0}\Gamma, T\rangle\cr
=& \bra{\Gamma, 0} e^{-HT}\ket{\Gamma, 0}\cr
\sim & e^{-E_0T}
\,,}
\eqn\overlap
$$
where we have inserted a complete set of energy eigenstates. We see
that only the lowest energy eigenstate, with energy
$E_0$, contributes for very large $T$. Since we will
restrict our attention to {\it static} quarks, (\pathstate)
should have non-vanishing overlap with such mesonic states where the two
fermions are situated at ${\bold x}$ and ${\bold y}$, in
consequence we can use (\overlap) to extract
the dependence of the mesonic ground state energy upon the
quark-antiquark separation, $R$. This should be independent of the
particular path, $\Gamma$, chosen to link the quarks (although a string
whose path, in units of typical hadronic scales, extended over a very
large distance away from the quarks would presumably have a miniscule
overlap with such a meson).
We may now apply standard approaches to rewrite
(\overlap) as a Wilson loop.

Consider the following evolution amplitude
$$
\eqalign{
M(T,\Gamma)\equiv \bra0\bar\psi({\bold x},0) P
\exp &\left( i \int^{\bold y}_{\bold x} \!
dz_i A^i({\bold z},0)\right)\psi({\bold y},0) \times \cr &
\bar\psi({\bold x},T) P
\exp \left( i \int^{\bold y}_{\bold x} \!
dz_i A^i({\bold z},T)\right)\psi({\bold y},T)
\ket0
\,.}
\eqn\timeevolve
$$
The {\it static Wilson loop} is now obtained in three steps: firstly
the fermion fields (or at least the flavour involved in Eq.\thinspace
\timeevolve) are integrated out, this replaces the fermions by
propagators in a background field and a determinant which is unity if
we ignore closed fermion loops (quenched approximation). Some
of these contributions
correspond to pair creation and annihilation diagrams which we ignore
since we want to study the potential. The next step is to consider the
propagators in the static approximation as was first done by Brown and
Weisberger\up{\BRWE} who made the gauge invariant approximation that the
quark two-point Green's function obeys
$$
(\gamma_0D^0-m)S_0(x,y)=\delta^{(4)}(x-y)
\,,
\eqn\staticgf
$$
where $S_0$ is the static propagator in a background field. (The
solution to this equation contains a factor of $\delta^{(3)}({\bf
x}-{\bold y})$.) Details of
all these steps may be found in a lucid review by Gromes\up{\GROMES}.
Since the quarks do not move they have potential but no kinetic energy
and this is what the overlap (\overlap) measures.
The upshot is that the above evolution amplitude may be written as
a gauge invariant expectation value of a loop operator
$$
M(T,\Gamma)=\langle\tr P\exp\left(ig \oint \!dz\cdot A(z)
\right)\rangle
\,,
\eqn\genloop
$$
where the loop runs over the path $\Gamma$ in the initial time slice,
then over a straight line from $({\bold y},0)$ to $({\bold y},T)$, then
back over the path $\Gamma$ (but now at time $T$) and then directly
down from $({\bold x},T)$ to $({\bold x},0)$, which completes the loop.
For the particular case that $\Gamma$ is taken to be the straight line
joining ${\bold x}$ and ${\bold y}$ one then has the standard, rectangular
Wilson loop.

At lowest order in perturbation theory the potential also displays a
Coulombic $1/R$ fall off\up{\ADM}. Numerical studies, primarily
on the lattice, are widely employed to study the potential between two
static sources at larger separations\up{\LATTPOT}.
It appears that the
results are consistent with a combination of a Coulombic and a linearly
rising term, which should yield confinement.
Variations around the straight line base
can be used to test the accuracy of numerical results for the
potential. It is found in lattice simulations that \lq
smearing\rq\ out the line improves the numerical accuracy\up{\LATTPOT}.
In this context see also Ref.\ \CAHILL.

We may now consider studying the interaction between a dressed quark
and a dressed antiquark in the above spirit. We take a static quark
and a static antiquark at ${\bold x}$ and ${\bold y}$, each dressed as
discussed in Sect.'s 4 and 5. The following overlap
$$
M_{\rm c}(T, {\bold x}-{\bold y})\equiv \langle\bar\psic(0,{\bold x})
\psic(0,{\bold y})\vert\bar\psic(T,{\bold x})\psic(T,{\bold y})\rangle
\,,
\eqn\overlapus
$$
where
$$
\ket{\bar\psic(t,{\bold x})\psic(t,{\bold y})}
\equiv \bar\psic(t,{\bold x})
\psic(t,{\bold y})\ket0
\,,
\eqn\dressket
$$
gives, for large $T$, the energy of the lowest mesonic state, as such
it is analogous to the more traditional Wilson loop.

In analogy to the derivation of the Wilson loop we integrate out
the fermions to obtain two propagators (and drop the terms that
correspond to creation-annihilation effects). In the static
approximation we can, in analogy to Brown and Weisberger\up{\BRWE},
replace the propagators by path ordered exponentials
along straight lines evolving in time. These lines are themselves gauge
invariant except at their ends, but this gauge dependence is annulled
by the four gluonic clouds centred around these ends. The study of such
an object as (\overlapus) should at small distances, where the
perturbative definition of the static dressed quarks is reliable, yield
the usual potential between two static quarks. At larger distances,
however, the potential should become non-perturbative and confining
effects will dominate. Since (\overlapus) is only perturbatively gauge
invariant --- as we saw in Sect.\ 5 no full solution for the dressing
exists --- for large enough separations gauge dependent
effects  eventually
will become visible in the potential as it may be  calculated
from (\overlapus). This explicit breakdown of the quark picture will
correspond to the confinement scale.

The calculation of the potential between dressed quarks involves, as
well as the non-locality inherent in the lines, the non-local gluonic
dressings at the ends of the lines. These last may be removed, however,
by
working in Coulomb gauge. It may be quickly seen that at one loop this
then yields a Coulomb potential between the quarks as usual. This
agreement with the potential from the static Wilson loop is heartening
if not unexpected: the end rungs of the Wilson loop rectangle should be
unimportant
in the large $T$ limit at this level (see, e.g., Ref.\ \ADM). The
question we are faced with is where should this agreement break down?
It has been argued by Fischler\up{\FISCHLER} that the ends of the
static Wilson loop are only of importance for large fields. This would
imply that the gluonic clouds around the ends of the lines are first
important when non-perturbative effects are considered.
A numerical study on the lattice of where the
Gribov problem first raises its head in the potential between the
dressed quarks should be able to measure this.
One difficulty with  such a simulation is raised by the
non-local gluonic clouds present in the two end time slices. By working
in the Coulomb gauge one can remove the non-locality and
also not have to worry about the exact form of the cloud
(for examples of Coulomb gauge lattice investigations see,
e.g., Ref.'s \LATTCOULP\ and \LATTCOULH). The
appearance of the Gribov ambiguity for larger separations
would herald confinement. We would urge that more work be done on
understanding the Gribov ambiguity on the lattice\up{\TASSOS}. We note
also in this context that it seems that many, perhaps most, of these
gauge ambiguities as they are seen in numerical simulations seem to
be lattice artefacts\up{\LATTART} which have no physical meaning. The
breakdown of a quark picture is then to be established from the onset of
non-artefact Gribov ambiguities.

The extension of both the above  approaches
to baryons, glueballs and exotics is more
or less direct. Consider, for example, having three static dressed
quarks. The energy of the lowest lying heavy baryon may be
straightforwardly found, in analogy to  (\elecener) or (\overlapus).
The usual Wilson loop picture for this
has been studied on the lattice\up{\LATTBAR}. Here one considered three
static quarks connected by  strings to some central point. The
results of the
numerical simulations seem to indicate that the lattices only
represent small volumes. (Essentially because the resulting potential
was of a form expressible as a sum over two body forces alone, see
also Ref. \GROMES.)  Long distance physics, and quark confinement
in baryons, would thus appear to require better numerical simulations.

One can also consider extending the above considerations to exotic
particles and glueballs, which are much more poorly experimentally
understood. One can here, e.g., use a globally colourless
configuration of gauge invariant, dressed gluons (such as the solutions
found in Sect.\ 5). We refer to various discussions of constituent
gluons\up{\CGLUE,\MORECGLUE}.

We have now seen that the dressings lead to an efficient method of
calculating the perturbative potential between static charges.
Furthermore, we have suggested various ways of obtaining the
confinement scale within this framework. In the real world we have,
of course, light quarks and we must therefore go beyond the static
approximation. This will be the subject of the rest of this paper.

\vfill\eject

%%%%%%%%%%%%%%%%%%%
%%%             %%%
%%% Section8    %%%
%%%             %%%
%%%%%%%%%%%%%%%%%%%

\secno=8 \meqno=1
\ni
{\bf 8. Lorentz Transformations in Gauge Theories}
\bigskip
\ni
In order to go beyond our previous description of solely static quarks,
we need to address the issue of how Lorentz
transformations must be implemented in gauge theories. We will show that
requiring the Lorentz generators to be gauge invariant modifies their
action on charged states. For simplicity,
this treatment will mainly be carried out in the context of QED --- it
will be clear from our presentation that much
the same conclusions can be reached for any gauge theory. We shall show
that,
even in such an abelian theory, the action of the Lorentz group is
sensitive to whether ghosts are introduced or not.  Important
consequences of our results
for
the construction of non-static dressings will be discussed in the next
section, where we will give explicit solutions for the dressing function
of a moving charge.

In Sect.\ 3 an argument for the breakdown of the  expected action of the
Lorentz group was presented. Till now, this surprising aspect of gauge
theories
has played a minor role in our presentation of charged sectors: its only
implication being as another reason for lowering our
naive expectation of
manifest Lorentz
invariance in the dressing and related gauge fixing.  We have now, though,
reached the
stage in our account of constituent quarks where we want to demonstrate
that the static dressing can be extended to a more dynamical one.
Hence
we need to make clear how Lorentz transformations are to act on such
fields.

In order to make this discussion easier to follow, let us first
summarize what we are going to show. There are two important issues that
we
will address: firstly, how should Lorentz transformations act on the local
fields in a gauge theory (for obvious practical reasons we include here
the gauge
non-invariant ones such as the potential); secondly, how should they act
on
non-local (charged) fields?

By starting with the standard QED Lagrangian we will demonstrate
that  the requirement of gauge
invariance modifies the
action of the Lorentz generators in a profound way.
On general fields this
modification is sufficiently drastic to obstruct them from carrying a
representation of the Lorentz group. However, on local observables the
modified generators will have the same action as the usual ones,
and hence
they will carry the expected representations of the Lorentz group. The
interesting point, though, is that on charged (and hence non-local
but gauge
invariant) fields
the action of these generators is modified. As we will see in Sect.\ 9,
this observation will have important consequences
for our account of such charged states and, in particular, for our
description of the light constituent quarks.

It is also possible, but usually not considered
very useful, to introduce ghosts into QED.
By doing this we will then demonstrate the attractive result
that by insisting on BRST invariance, the standard action of the Lorentz
group on all local fields is recovered --- only in the nonlocal charged
sectors is this not the case.

The QED Lagrangian is
$$
\L_{_{\rm QED}}=-\tfrac14F_{\mu\nu}F^{\mu\nu}+i\bar\psi\gamma^\mu
(\pa_\mu+ieA_\mu)\psi-m\bar\psi\psi\,.\eqn\qedlag
$$
In this
the potential $A_\mu$ is taken to be a four vector and $\psi$ a spinor.
Under the infinitesimal Lorentz transformation
$$
x^\mu\to x'^\mu:=x^\mu-f^\mu\,,\eqn\no
$$
where $f^\mu=-\e^{\mu\nu}x_\nu$, the potential thus  transforms
as\up{\JACKIW,\OHNUKI}
$$
\d_fA_\mu=f^\a\pa_\a A_\mu+(\pa_\mu f^\a)A_\a\,,\eqn\lora
$$
The infinitesimal transformations on
$\psi$ and $\bar\psi$ are then taken to be
$$
\eqalign{
\d_f\psi&=f^\a\pa_\a\psi+\tfrac{i}4\sigma_{\mu\nu}(\pa^\nu f^\mu)\psi\,,\cr
\d_f\bar\psi&=f^\a\pa_\a\bar\psi-\tfrac{i}4(\pa^\nu f^\mu)\bar\psi
\sigma_{\mu\nu}\,,
}
\eqn\lorpsi
$$
where $\sigma_{\mu\nu}=\frac{i}2[\gamma_\mu,\gamma_\nu]$.
Even at this stage, we see an unnaturalness in these transformations as
they
do not reflect the fact that we are dealing with a gauge theory. In
particular,
the
occurrence of derivatives in (\lorpsi) is at odds with the normal
expectation
that by making the gauge symmetry local, derivatives on $\psi$ should be
replaced
by
covariant derivatives. As a consequence, if we have two gauge related
fermions,
$\psi$ and $e^{-ie\Lambda}\psi$, then under the Lorentz transformations
they
are no longer gauge related! Explicitly:
$$
e^{-ie\Lambda}(\psi+\d_f\psi)\ne
e^{-ie\Lambda}\psi+\d_f\bigl(e^{-ie\Lambda}\psi\bigr)\,.\eqn\lorproblem
$$
This should be contrasted with the transformation properties of the
vector potential where two gauge related potentials are still gauge
related after a  Lorentz transformation. This observation highlights
the fact discussed in Sect.\ 3 that it is in the charged sector where more
care is needed when implementing the Lorentz transformations.
To make this unease more precise, and to see how to resolve this problem,
we need to further investigate how the Lorentz transformations are
generated in
QED.

The invariance of the
QED Lagrangian  under the Lorentz transformations (\lora) and (\lorpsi)
allows us to use Noether's theorem to obtain
their generators. Under these transformations the Lagrangian
transforms by a total
derivative
$$
\d_f\L_{_{\rm QED}}=\pa_\mu(f^\mu\L_{_{\rm QED}})\,,\eqn\totaldiv
$$
The associated conserved current is then:
$$
\eqalign
{
M^\mu&=\frac{\d\L_{_{\rm QED}}}{\d(\pa_\mu A_\nu)}\d_f A_\nu
-\frac{\d\L_{_{\rm QED}}}{\d(\pa_\mu\psi)}\d_f\psi-f^\mu\L_{_{\rm QED}}\cr
&=\tfrac12\e_{\a\b}M^{\mu\a\b}
}
\eqn\lorcurrent
$$
where
$$
M^{\mu\a\b}=x^\a T_{\rm c}^{\mu\b}-T_{\rm c}^{\mu\a}x^\b+A^\a
F^{\mu\b}-F^{\mu\a}A^\b-\tfrac12\bar\psi\gamma^\mu\sigma^{\a\b}\psi\,,
\eqn\lorgen
$$
and $T^{\mu\nu}_{\rm c}$ is the canonical stress-energy
tensor\up{\JACKIWOLD}.
$$
T^{\mu\nu}_{\rm c}=i\bar\psi\gamma^\mu\partial^\nu\psi-F^{\mu\rho}
\partial^\nu A_\rho+\tfrac14 g^{\mu\nu}F_{\rho\lambda}
F^{\rho\lambda}\,.\eqn\stress
$$
The generators of the Lorentz transformations  are then identified with
the charges
$$
M^{\a\b}=\int d^3x M^{0\a\b}\,.\eqn\no
$$
Using (\ccrtwo) and (\acrtwo) one finds that
$$
\eqalignno
{
[iM_{\a\b},A_i]&=(x_\a\pa_\b-x_\b\pa_\a)A_i+g_{i\a}A_\b-g_{i\b}A_\a
&\eqnn\no\cr
[iM_{\a\b},\psi]&=(x_\a\pa_\b-x_\b\pa_\a)\psi-\tfrac{i}2
\sigma_{\a\b}\psi  &\eqnn\no\cr
}
$$
which, when multiplied by $\frac12\e^{\a\b}$, recovers
$\d_f A_i$ and $\d_f\psi$.
(Just as we saw for gauge invariance, we do not recover how
$A_0$ transforms as it is not a dynamical field in Eq.\ \qedlag.
We will see
later in this section how to improve upon this.)

The commutator algebra of these generators closes on itself, i.e., it
only
involves linear combinations of the generators:
$$
[M_{\a\b},M_{\rho\sigma}]
=ig_{\a\sigma}M_{\b\rho}-ig_{\a\rho}M_{\b\sigma}+ig_{\b\rho}
M_{\a\sigma}-ig_{\b\sigma}M_{\a\rho}\,.\eqn\LieAlgLor
$$
This tells us that they form a Lie algebra\up{\BARUT} and hence that these
are the generators for a unitary implementation of the Lorentz group on
these fields. That is, under a finite Lorentz transformation described by
the matrix $\Lambda$, where to lowest order
$\Lambda^\a_\b=\d^\a_\b+\e^\a_{\ \b}$, we have
$$
\eqalignno
{
A'_\mu(x)&=U(\Lambda)A_\mu(x)U^{-1}(\Lambda)&\eqnn\no\cr
\psi'(x)&=U(\Lambda)\psi(x)U^{-1}(\Lambda)&\eqnn\no\cr
}
$$
where the unitary operator $U(\Lambda)$ representing the Lorentz
transformation is given by
$$
U(\Lambda)=e^{\frac{i}2M_{\a\b}\e^{\a\b}}\,.\eqn\expgen
$$

Now the difficulties with
this account of Lorentz invariance in such a gauge theory become
clear:
from (\lorgen) and (\stress) we can see that the generators are not
gauge invariant and hence cannot be identified with observables. This
means that we have the former of the alternatives
discussed in Sect.\ 3: the Lorentz
group is implemented unitarily on all the local fields but it is not
generated by
observables,
as such it has no clear action on the physical (gauge invariant)
states or observables.

There is, however, a standard
prescription for improving the stress-energy tensor
(\stress) which, as a consequence, makes it gauge
invariant\up{\JACKIWOLD}.
That is, we consider the gauge invariant stress-energy tensor\note{The
standard prescription actually aims to make the stress-energy tensor
symmetric in $\mu$ and $\nu$, which is a stronger condition than just
making it gauge invariant. To further improve our expression for the
stress-energy tensor so that it is symmetric the term
$\pa_\rho\bigl(F^{\mu\rho}A^\nu-\tfrac14\bar\psi(\gamma^\nu
\sigma^{\rho\mu}+
\gamma^\mu\sigma^{\rho\nu}-\gamma^\rho\sigma^{\mu\nu})\psi\bigr)$
must be added
to the canonical stress-energy tensor. This additional term does not alter
our discussion of the generators, and for simplicity we avoid
this complication. We note, though, that if we wished to couple to gravity
then the symmetric form of the stress-energy tensor must be used.}
$$
T^{\mu\nu}=i\bar\psi\gamma^\mu D^\nu\psi+F^{\mu\rho}F_\rho^{\ \nu}+
\tfrac14 g^{\mu\nu}F_{\rho\lambda}F^{\rho\lambda}\,.\eqn\stressg
$$
This can be viewed as an addition to the  canonical stress-energy
tensor
$$
T^{\mu\nu}=T^{\mu\nu}_{\rm c}+F^{\mu\rho}\pa_\rho A^\nu-
e\bar\psi\gamma^\mu
A^\nu\psi\,.\eqn\no
$$
The argument for why we are free to add the extra terms is that,
after using the equation of motion, $\pa_\rho
F^{\mu\rho}=-e\bar\psi\gamma^\mu\psi$, they can be written as a divergence
$$
T^{\mu\nu}=T^{\mu\nu}_{\rm c}+\pa_\rho(X^{\rho\mu\nu})\,,\eqn\no
$$
where $X^{\rho\mu\nu}=F^{\mu\rho}A^\nu$ is anti-symmetric in
$\rho$ and $\mu$. This last property ensures that the gauge
invariant stress-energy tensor (\stressg)
is conserved as long as the canonical one is.

The same construction can be applied to the Lorentz generators.
We seek a modification
$$
\overline{M}^{\mu\a\b}=M^{\mu\a\b}+\pa_\rho X^{\rho\mu\a\b}\,,\eqn\lorgeng
$$
such that the resulting charge is gauge invariant and conserved.
One finds that if we take
$$
X^{\rho\mu\a\b}=x^\a X^{\rho\mu\b}-X^{\rho\mu\a}x^\b\,,\eqn\no
$$
then the gauge invariant current is given by
$$
\overline{M}^{\mu\a\b}=x^\a T^{\mu\b}-T^{\mu\a}x^\b-\tfrac12
\bar\psi\gamma^\mu
\sigma^{\a\b}\psi\,.\eqn\lorgengauge
$$
It is usually argued that such modifications do not effect the action or
algebra of the charges since it amounts to adding a total
divergence to the generators. However, care is needed since we
have seen that charges imply that surface terms cannot always be
neglected. Also the equations of motion have been used in getting to
the gauge invariant description (\lorgengauge); so again, care is needed.

As it is by construction conserved, this gauge invariant charge will
generate symmetries
of the theory. It is clearly of great interest to know how they are related
to the original Lorentz transformations. One finds for example that
$$
[i\overline{M}_{\a\b},A_i]=x_\a F_{\b i}-x_\b F_{\a i}\,,\eqn\covlora
$$
and
$$
[i\overline{M}_{\a\b},\psi]=(x_\a D_\b-x_\b D_\a)\psi-
\tfrac{i}2\sigma_{\a\b}\psi\,.\eqn\no
$$
Multiplying by $\frac12\e^{\a\b}$ these give the gauge covariant
Lorentz transformations\up{\JACKPAPER} on $A_i$ and $\psi$:
$$
\eqalign{
\bar\d_f A_i&=f^\a F_{\a i}\cr
\bar\d_f\psi&=f^\a D_\a\psi+\tfrac{i}4\sigma_{\a\b}(\pa^\b f^\a)\psi\,.
}
\eqn\gaugecovtrans
$$
These transformations have many interesting properties and
applications\up{\JACKMAN}, and we note that we now have the desired
covariant derivatives on the fermionic fields.
However, these are {\it
not\/} the expected Lorentz transformations. We will now investigate the
physical consequences of this.

It must be emphasised that Lorentz invariance is of central importance
{\it only\/} for physical
states and observables.
As we have repeatedly stressed, gauge theories carry with
them a vast amount of unphysical luggage. Thus the  fact we have been
forced to tinker with the Lorentz generators should not, in itself,
cause too much concern --- as long as we are clear that we have not
altered any physical consequence of the theory. To see this
let us first look at the modification to the stress-energy tensor
(\stress) brought about by our insistence on gauge invariance.

We recall that the canonical stress-energy tensor (\stress) is used to
construct the momentum $P^\nu_{\rm c}=\int d^3x\,T_{\rm c}^{0\nu}$. This
momentum plays a very privileged role in any formulation of quantum field
theory, and one would be loth to see any modification to its role as the
generator of spacetime translations.
The gauge invariant stress-energy tensor (\stressg) gives the momentum
$$
P^\nu=P^\nu_{\rm c}+\int d^3x\bigl(F^{0i}\pa_i A^\nu-eJ^0A^\nu\bigr)
\,.\eqn\no
$$
We note that, by integrating by parts, this can be written as
$$
P^\nu=P^\nu_{\rm c}-\int d^3x\, GA^\nu\,,\eqn\no
$$
where $G=\pa_iF^{0i}+eJ^0$ is Gauss' law, which we recall is the
generator
of the gauge transformations. This means that on gauge invariant
states and observables we recover the action of the normal momentum
{\it as
long as we are free to neglect the surface term.} The potentially
troublesome surface term in this case is
$\lim_{R\to\infty}\int_{S_R}d{\bold s}\cdot{\bold E}A^\nu$. We recall
from our discussion in Sect.\ 3 that the potential is assumed to fall
off as $1/R$ and the electric field as $1/R^2$. These
assumptions on the fall off rates of these fields imply that this surface
term is zero,
so we are free to take the normal momentum action on all gauge invariant
states, i.e., on both charged and uncharged states.
Let us now repeat this analysis for the Lorentz generators.

{}From
(\lorgengauge) we see that
$$
\eqalign{
\overline{M}^{\a\b}&=M^{\a\b}+\int d^3x\,\bigl(F^{0\a}A^\b-F^{0\b}A^\a
+F^{0i}(x^\a\pa_iA^\b-x^\b\pa_iA^\a)\cr
&\hskip6.5cm
-eJ^0(x^\a A^\b-x^\b
A^\a)\bigr)\cr
&=M^{\a\b}-\int d^3x\,(x^\a A^\b-x^\b A^\a)G\,,
}\eqn\no
$$
plus the surface term
$$
\lim_{R\to\infty}\int_{S_R}d{\bold s}\cdot{\bold E}(x^\a A^\b-x^\b
A^\a)\,.
\eqn\surfacetrouble
$$
So again, we see that as long as this surface term is zero, the Lorentz
transformations on gauge invariant objects are recovered. In particular,
on
all local chargeless states this will be the case. But what about charged
states? Here we see that due to the extra
coordinate variable in (\surfacetrouble), the fall off conditions on the
potentials will result in a non-zero surface contribution. To be explicit,
for a static charge
we have $A_i=0$ while $A_0\sim 1/R$,  so for boosts, where the generator
is
$\overline{M}^{0i}$, this surface term will not vanish.

In summary we see that we recover the action of the
Lorentz group on the observables, but on charged states the action is
modified. Given that
these charged states are not observables, this modification does not lead
to
any unphysical consequences. However, we shall now demonstrate that there
is quite a profound technical complication caused by this since it is not
possible to use these infinitesimal transformations to build up finite ones
on any charged state in a unitary way!

The reason why the modified generators do not lead to a representation of
the Lorentz group is because their commutator algebra no longer closes as
in
(\LieAlgLor). Indeed, we can see that
$$
[\overline{M}_{\a\b},\overline{M}_{\rho\sigma}]
=ig_{\a\sigma}\overline{M}_{\b\rho}-ig_{\a\rho}
\overline{M}_{\b\sigma}+ig_{\b\rho}\overline{M}_{\a\sigma}
-ig_{\b\sigma}\overline{M}_{\a\rho}
+iK_{\a\b\rho\sigma}\,,\eqn\centralterm
$$
Before giving an explicit example for the additional term
$K_{\a\b\rho\sigma}$, we note that its addition to the Lorentz algebra
means that we
cannot exponentiate these generators to build up the action of finite
transformations as in (\expgen). For brevity we will call
$K_{\a\b\rho\sigma}$ the central term for the gauge covariant Lorentz
algebra.

The proof of this result is tedious, and we only give a brief sketch here
for
the simpler case when we just deal with that part of the Lorentz group
that
deals with rotations\note{One reason for just looking at rotations is that
they do not mix the spatial and temporal components of the gauge field (as
a
boost would). The reason why this is important is that
we have, as yet, an incomplete description of the modified Lorentz
generators. This is because we are starting with the Lagrangian (\qedlag)
in
which $A_0$ is just a multiplier field, as such we do not know how these
transformations should be extended to $A_0$. This will be improved upon
below.}.
The rotation generators are $\overline{M}_{ij}$. To find the expression
for
the central term  $K_{ijkl}$ in this case we use the Jacobi identity:
$$
[i\overline{M}_{ij},[i\overline{M}_{kl},A_m]]-
[i\overline{M}_{kl},[i\overline{M}_{ij},A_m]]=-[[\overline{M}_{ij},
\overline{M}_{kl}], A_m]\,.
\eqn\Jacobi
$$
{}From (\covlora) and (\centralterm) we find that
$$
[iK_{ijkl},A_m]=\pa_m(x_ix_kF_{lj}-x_ix_lF_{kj}-x_jx_kF_{li}+
x_jx_lF_{ki})\,.\eqn\no
$$
Similarly on $\psi$ we get
$$
[iK_{ijkl},\psi]=ie(x_ix_kF_{lj}-x_ix_lF_{kj}-x_jx_kF_{li}+
x_jx_lF_{ki})\psi\,.\eqn\no
$$
{}From these we deduce that the central term is
$$
\eqalign{
K_{ijkl}=\int d^3x\,\bigl(\pa_m(x_ix_kF_{lj}&-x_ix_lF_{kj}-x_jx_kF_{li}+
x_jx_lF_{ki})E_m\cr
&-(x_ix_kF_{lj}-x_ix_lF_{kj}-x_jx_kF_{li}+
x_jx_lF_{ki})e\psi^\dagger\psi\bigr)
}\eqn\no
$$
Upon integrating by parts this may be rewritten as a sum of two terms:
a part containing Gauss' law
$$
-\int d^3x\,\bigl(x_ix_kF_{lj}-x_ix_lF_{kj}-x_jx_kF_{li}+
x_jx_lF_{ki})(\pa_mE_m+e\psi^\dagger\psi\bigr)\,,\eqn\weakk
$$
and the surface term
$$
\lim_{R\to\infty}\int_{S^2_{_R}} d{\bold s}\cdot{\bold E}\,\bigl(
x_ix_kF_{lj}-x_ix_lF_{kj}-x_jx_kF_{li}+
x_jx_lF_{ki}\bigr)\,.\eqn\no
$$
This surface term can be written as
$$
(\e_{ijk}\d_{ls}-\e_{ijl}\d_{ks})
\lim_{R\to\infty}\int_{S^2_{_R}} d{\bold s}\cdot{\bold E}\,
x_s  x_rB_r\,.\eqn\no
$$
It is now a dynamical problem to determine when this surface term is
zero. Clearly for a static charge it vanishes. This should not come as
too much of a surprise since
the form of the dressing in this case (\coulelec) is naively
rotationally invariant. It is also zero for
charges moving with a constant velocity
since then the magnetic field satisfies $x_rB_r=0$.
This would suggest that in this abelian theory charges can be rotated in
the expected way.  It is intriguing to note that the static quark
does not appear to have this property.

We have seen that gauge invariance dictates the modifications of the
Lorentz generators, a consequence of this is that it is only on the
local physical observables that
the Lorentz group acts in the expected way. On gauge non-invariant and
gauge invariant but charged (and hence non-local) states the action is
modified to such an extent that it cannot be implemented unitarily.
Although this is clearly
physically acceptable, its practical implications are
rather unattractive.
This analysis, though, is incomplete. As we have stressed earlier,
the action
on $A_0$ has not been determined due to the gauge invariance of
(\qedlag). To
remedy this we add a gauge fixing term which breaks the gauge
invariance.
This then reinstates $A_0$ as a dynamical variable, but now the
requirement
that the Lorentz generators are gauge invariant is unnatural
(after all, the gauge fixed theory is no longer gauge invariant).
To proceed
we  have to impose the condition that the
generators must be BRST invariant. Insisting on BRST invariance in this
way puts some restrictions
on
the form of the effective QED Lagrangian. Let us now discuss this
point.

The gauged fixed  QED action based now on the abelian version of the
Lagrangian
(\efftwo) is BRST invariant, but the Lagrangian itself is not: it picks
up a total
divergence. We saw the same thing for the Lorentz generators (\totaldiv)
and
this is quite unobjectionable for a symmetry. (In fact it is mandatory for
a
spacetime symmetry, but optional for an internal one\up{\JACKIWOLD}.)
However, an invariant stress-energy tensor must be built from an invariant
Lagrangian, so we are forced to rewrite (\efftwo) so that it is strictly
invariant under BRST. There are various way to do this, the most
attractive is
to  introduce a
scalar field $B$ and take as our effective Lagrangian
$$
\L_{_{\rm eff}}=\L_{_{\rm QED}}+\tfrac12 B^2+\pa_\mu B A^\mu-i\pa_\mu\bar
c\pa^\mu c\,.\eqn\qedeff
$$
The BRST transformations are then: $\d A_\mu=\pa_\mu c$; $\d c=0$; $\d\bar
c=-iB$; $\d B=0$; $\d\psi=-iec\psi$ and $\d\bar\psi=-ie\bar\psi c$.
We now have, on top of the invariance of (\qedeff), the nilpotency,
$\d^2=0$,
which lies at the heart of the BRST method, without use of an equation
of motion. We see that the new
equations of
motion that follow from (\qedeff) are
$$
\pa_\mu F^{\mu\nu}=e\bar\psi\gamma^\mu\psi-\pa^\mu B\eqn\eqmwithb
$$
and
$$
\pa_\mu A^\mu=\xi B\,.\eqn\no
$$
We note that $A_0$ and $B$ are conjugate variables.

The infinitesimal Lorentz transformations are given by (\lora) and
(\lorpsi)
supplemented by the scalar transformation rule for the ghosts and $B$,
i.e., $\d_f B= f^\mu \pa_\mu B$ and similarly for $c$ and $\bar c$.
These
result in a canonical stress-energy tensor
$$
\eqalign{
T^{\mu\nu}_{\rm c}=i\bar\psi\gamma^\mu\pa^\nu\psi&-F^{\mu\a}\pa^\nu A_\a
+\tfrac14g^{\mu\nu}F_{\lambda\rho}F^{\lambda\rho}\cr
&+A^\mu\pa^\nu B-\tfrac12g^{\mu\nu}\xi B^2-g^{\mu\nu}\pa_\lambda B A^
\lambda\cr
&{}\qquad\quad-i(\pa^\mu\bar c\pa^\nu c+\pa^\nu\bar c\pa^\mu c)
+ig^{\mu\nu}\pa_\lambda\bar c\pa^\lambda c\,.
}\eqn\no
$$
This is evidently
not BRST invariant. To deal with this we improve it by adding
the term
$\pa_\lambda(F^{\mu\nu}A^\nu)$, as we did before. Now after using the
equation
of motion (\eqmwithb) we find the BRST invariant stress-energy tensor
$$
\eqalign{
T^{\mu\nu}=i\bar\psi\gamma^\mu D^\nu\psi&+F^{\mu\lambda}F_\lambda^{\ \nu}
+\tfrac14g^{\mu\nu}F_{\lambda\rho}F^{\lambda\rho}\cr
&+(A^\mu\pa^\nu B+ A^\nu\pa^\mu B)-\tfrac12g^{\mu\nu}\xi B^2-
g^{\mu\nu}\pa_\lambda B A^\lambda\cr
&{}\qquad\quad-i(\pa^\mu\bar c\pa^\nu c+\pa^\nu\bar c\pa^\mu c)
+ig^{\mu\nu}\pa_\lambda\bar c\pa^\lambda c\,.
}\eqn\no
$$
The same procedure goes through for the Lorentz generators and we end
up with
the improved, BRST invariant charge (\lorgengauge); note that we are
now using the above stress-energy tensor.

As before, to investigate the action of this modified generator, we write
$$
\eqalign{
\overline{M}^{\a\b}=M^{\a\b}+&\int d^3x\,\Bigl(F^{0\a}A^\b-F^{0\b}A^\a\cr
&\qquad\qquad+x^\a(F^{0i}\pa_iA^\b-eJ^0A^\b+\dot{B}A^\b)\cr
&\qquad\qquad\qquad-(F^{0i}\pa_iA^\a-eJ^0A^\a+\dot{B}A^\a)x^\b\Bigr)\,.
}\eqn\no
$$
After integrating by parts this gives
$$
\eqalign{
\overline{M}^{\a\b}=M^{\a\b}+\int d^3x\,& (-\pa_iF^{0i}-eJ^0+\dot{B})
(x^\a A^\b-A^\a x^\b)\cr
& +
\lim_{R\to\infty}\int_{S_R}d{\bold s}\cdot{\bold E}(x^\a A^\b-x^\b A^\a)
\,,
}\eqn\brstsurf
$$
where we recognise the surface term (\surfacetrouble) encountered earlier.
Note that $\dot B$ enters these expressions. This velocity can be replaced
using
(\eqmwithb): $-\pa_iF^{0i}-eJ^0+\dot{B}=0$. Hence, in marked
contrast to what happened without ghosts, we now see that on all local
fields, where as we have stated the surface term can be neglected,
$\overline{M}^{\a\b}=M^{\a\b}$, and we recover the normal action
of the Lorentz group.

We are familiar in non-abelian gauge theories with the introduction of
ghosts to maintain unitarity. Our discussion shows that even in the
abelian theory, where unitarity is not an issue in covariant gauges,
ghosts are essential if one wants to preserve the standard form of
the Lorentz generators on local observables and non-observables.
A simple argument to see why this
is the case comes from the example given earlier, (\lorproblem).
Combining the BRST-transformation $\d\psi=-iec\psi$ with the Lorentz
transformations (\lorpsi) and the scalar transformation rule,
$\d_f c=f^\a\pa_\a c$, we see that now
$$
\d(\psi+\d_f\psi)=\d\psi+\d_f(\d\psi)\,.\eqn\no
$$
Hence the reason underlying
our initial unease with the Lorentz transformations may be avoided if
BRST transformations are used instead of the gauge ones. From the above
discussion, though, it is clear that there is still a modification to
the Lorentz generators (even in this BRST-formulation) when charges are
present. This is because the identification of the modified generators
with the standard ones only holds if the surface term in (\brstsurf)
may be neglected.  We have seen for charged states that this is
generally not permitted.
In the next section we will see how this is reflected in the
construction of dynamical dressings for charged particles.

\vfill\eject

%%%%%%%%%%%%%%%%%%%
%%%             %%%
%%% Section9    %%%
%%%             %%%
%%%%%%%%%%%%%%%%%%%

\secno=9 \meqno=1
\ni
{\bf 9. Beyond the Static Quark Approximation}
\bigskip
\ni
In the constituent quark model a proton, say, contains three light quarks,
each of which will be moving. For this reason it is essential to have the
correct dressings for non-static quarks. Indeed even in QED the dressings
for non-static fermions are needed. In this section we will therefore
boost a static charged fermion. This will allow us to find, for the first
time, the dressing for a moving charge. Our discussion in the last
section implies that care might be needed, and we will see that this
is the case. We conclude this section by giving explicit dressings for
relativistic and non-relativistic charges.

Up to now we have analysed in detail how a static dressing can be
applied to the quark fields, and how this procedure allows us to derive
the
essential features of the constituent quark model from QCD. The
existence of heavy quarks means that this static dressing can have
physical implications. Explicitly we have seen in
Sect.'s 6 and 7 how such non-local and
non-covariant fields can be used to cure infra-red problems and to
derive both a running mass
and an inter-quark potential. Such results give us
confidence in the approach we have adopted.
However, we must consider moving quarks.
Thus  our proposed identification of the
constituent quarks with the dressed quarks in QCD can only hope to be
fully successful if we can show how to go beyond this static picture.

{}From our
analysis in Sect.\ 5, we know that for any acceptable gauge fixing there
corresponds a dressing. Thus one possible approach to this problem is to
find an appropriate dressing gauge fixing. For the static dressing we have
seen that the Coulomb gauge is needed. Now even in this simple situation,
there does not seem to be any a priori reason for this connection between
the Coulomb gauge and static charges. In Ref.\ \ADM\ for instance
the Coulomb gauge is
deemed no more than convenient since then the coupling of transverse
gluons  to the heavy quarks may be neglected. In Ref.\ \ADKINS\  it is
again viewed as merely helpful to renormalize the
electron propagator at the static point when in Coulomb gauge. (We have
seen in Sect.\ 6 that there is, in fact,  really no choice here.) This
suggests that, with our present understanding of dressings,
it is not fruitful to approach the form for dynamical dressings solely
in terms of the type of dressing gauge fixing needed.

If we restrict ourselves to dressings that are momentum eigenstates, then
another route to their construction is available --- we can simply boost
the known static dressing. However, as we have seen in some detail in the
previous section, such a proposal comes with its own problems. We have
argued that the addition of ghosts allows us to avoid complications with
the action of the Lorentz group on local fields, but we have seen that
modifications arise when dealing with charged fields. Also, from the
discussion in Sect.\ 3, we cannot expect the Lorentz group to act
unitarily on, for example, a static configuration.

What we propose to do initially is to construct dressings appropriate to
slowly moving charges by just naively boosting. This will allow
us to arrive at a rather natural proposal for such  dressings and to
see at first hand what these general arguments about problems with the
Lorentz group really mean in practice. Of course, we cannot expect this
rather limited class of dynamical dressings to be directly applicable to
complicated bound state problems involving light quarks.
In this context, an extension to, say, dressings that are
angular momentum eigenstates would be very interesting, but this must be
considered future work\note{As we have stated earlier, it seems clear
that the dressings  we are
constructing are related to coherent states, indeed their good infra-red
structure necessitates this connection. These coherent states are based on
creation and annihilation operators, and hence their dynamical
restriction to
dressing with a definite momentum is not so surprising. In this context, we
would expect
that the use of generalized coherent states\up{\COHERENT}
would allow for a more
varied
dynamical dressing.}.

As in the previous section, we will deal in the main with boosting
abelian charges. For the non-abelian case this
corresponds to working only to order $g$. The
extension of this construction to quarks at all orders follows the
algorithm
described for static quarks once the (dynamical) dressing gauge fixing is
identified: we will, though, not enter into these details here.

We recall that the static charge is described by
$$
\psic(x)=\exp\(-\frac{ie}{4\pi}\int d^3 y\frac{\pa_iA_i(x^0,{\bold y})}
{|{\bold x}-{\bold y}|}
\)\psi(x)\,.\eqn\staticnine
$$
A boost by velocity $-{\bold v}$ is generated by the Lorentz
charge, $iv_iM_{0i}$.
Acting on the spinor we have
$$
[iv_iM_{0i},\psi(x)]=(v_ix_0\pa_i-v_ix_i\pa_0)\psi(x)-\tfrac{i}2v_i
\sigma_{0i}\psi(x)\,.\eqn\no
$$
So for small ${\bold v}$ we have
$$
\psi'(x)=S(\Lambda)\psi(x')\,,\eqn\no
$$
where $S(\Lambda)=\bigl(1-\frac{i}2v_j\sigma_{0j}\bigr)$ and
$x'^0=x^0+v_jx^j$, ${\bold x}'={\bold x}-{\bold v}x^0$. Similarly
on the potential we get
$$
A'_i(x^0,{\bold y})=A_i(x^0+v_jy^j,{\bold y}')
+v_iA_0(x^0+v_jy^j,{\bold y}')\,,\eqn\no
$$
where now ${\bold y}'={\bold y}-{\bold v}x^0$.

Naively applying these to the charged field (\staticnine) gives
$$
\psic'(x)=S(\Lambda)\exp\(-\frac{ie}{4\pi}\int d^3 z
\frac{\pa_iA_i(x^0+v_jz^j,{\bold z})+v_iE_i(x^0+v_jz^j,{\bold z})}
{|{\bold x}'-{\bold y}|}
\)\psi(x')\,.\eqn\booste
$$
This suggests that a charged particle slowly moving with velocity
${\bold v}$ can be  described by the field
$$
\psiv(x)= \exp\left(
ie\frac{\partial_i A_i+v_iE_i}{\nabla^2}(x)\right)\psi(x)
+{O}({\bold v}^2)\,,
\eqn\boostv
$$
Note that in this expression we have chosen to neglect the overall
matrix $S(\Lambda)$, and we have identified
the static field $\psic$ with
$\psi_{{\bold 0}}$. In keeping with our previous notation, we can also
write (\boostv) as $\psiv=e^{i\vv}\psi$.

The first thing to note about $\psiv$ is that it is  gauge invariant:
we have in effect only added an electric field term, which is in this
abelian
theory gauge invariant. We can also repeat the  argument
presented in Sect.\ 4, and see that the states created by this field will
have, in addition to the Coulomb electric field, the magnetic field
${\bold B}={\bold v}\times {\bold E}$ expected from such a
non-relativistic
charge, i.e., the correct associated fields up to first
order in the velocity. As it stands, we seem to have been quite
successful in boosting the static configuration ---
one might ask where have the problems with
the implementation of the Lorentz group gone?
Rather than looking at higher orders in ${\bold v}$ (where problems do
arise, but only after much algebra), let us keep to small velocities but
look at how we boost two static quarks.

Starting with two space-like separated static charges:
$$
\psic(x^0-v_ix^i,{\bold x}+{\bold v}x^0)
\psic(x^0-v_iy^i,{\bold y}+{\bold v}x^0)\,,\eqn\twostatic
$$
boosting gives (up to  overall $S(\Lambda)$ factors)
$$
\psiv(x^0,{\bold x})
\psiv(x^0,{\bold y})\,.\eqn\twoboost
$$
Now the initial static fields anti-commute, and that Fermi statistics
apply is an important part of the constituent quark model. However,
the naively boosted charges given in (\twoboost) do not commute,
which is a direct expression of the subtleties associated with
Lorentz transformations of charged states. To see this we
consider the equal time anticommutator of two such fields:
$$
\eqalign{
[\psiv(x),\psiv(y)]_{_+}&=[e^{i\vv(x)},e^{i\vv(y)}]\psi(x)\psi(y)\cr
&=\bigl(1-e^{[\vv(x),\vv(y)]}\bigr)\psiv(x)\psiv(y)\,,
}
\eqn\no
$$
where we have used the Baker-Campbell-Haussdorf relation
$$
e^{A}e^{B}=e^{A+B}e^{\frac12[A,B]} \,,\qquad\hbox{if}\quad
[A,[A,B]]=[B,[A,B]]=0\,,
$$
which holds for the above photonic clouds. Thus, this anticommutator
vanishes
if, and only if, $[\vv(x),\vv(y)]=0$.  However, a direct calculation
shows
that this does not happen. In fact we find that for this specific
motion we get
$$
\eqalign{
[\vv(x),\vv(y)]= -i v_i \int\!d^3z\(
\frac{\pa\ }{\pa z^i}\(\frac1{|{\bold x}-{\bold z}|}\)\frac1{|{\bold y}-
{\bold z}|}-\frac1{|{\bold x}-{\bold z}|}
\frac{\pa\ }{\pa z^i}\(\frac1{|{\bold y}-{\bold z}|}\)
\)
\,.
}\eqn\no
$$
By inspection, this commutator generally
does not vanish. We note, though, that it is, as expected, zero in the
static  limit and further
vanishes for both $x\to y$ and, more specifically,
if ${\bold v}\cdot{\bold x}\to {\bold v}\cdot{\bold y}$.

Here we see a direct manifestation of the fact that the Lorentz
transformations fail to be implemented unitarily on such charged
fields. In deriving (\twoboost) we have assumed that
$$
[iM_{0i},\psi_c(x)\psi_c(y)]=[iM_{0i},\psi_c(x)]\psi_c(y)+\psi_c(x)
[iM_{0i},\psi_c(y)]\,.\eqn\assume
$$
However, in deriving this identity various integrations by parts are
performed
and we are now {\it not} free to discard the resulting surface terms.
On top of this problem, we know from
our discussion in Sect.\ 8 that additional surface terms are
needed to complement the Lorentz transformations in the charged sector.
All in all, we see that the correct implementation of  the boost for the
two charged system (\twostatic) is far from clear.

Rather that attempt to derive the correct form of the boost in this
situation, we assume that they should anti-commute and find a compensating
term\note{A similar problem arises in conformal field theory where a
generalized Klein transformation is needed to ensure the correct
commutation relations for bosonised fields\up{\GODDARD}. In is
interesting to note the
non-locality in that construction as well.}\ that will ensure the
anticommutativity of the boosted fields.

As our ansatz for the two charged system we take
$$
C_{{\rm v}}(x,y)\psiv(x)\psiv(y)\,.\eqn\twoboostguess
$$
We now demand that this satisfies the Fermi statistic relation
$$
C_{{\rm v}}(x,y)\psiv(x)\psiv(y)=-C_{{\rm v}}(y,x)\psiv(y)\psiv(x)
\,.\eqn\fermistat
$$
The solution to this is  found to be
$$
C_{{\rm v}}(x,y)=e^{\(\tfrac12[\vv(x),\vv(y)]\)}\,.\eqn\no
$$
This of course reduces to unity for the cases discussed above where the
dressed fields anticommuted. The need for such phase factors can also be
seen in systems of charges moving with different velocities. In fact even
for a system composed of a static and a moving charge such terms are
present. The non-abelian generalisations of these terms may be readily
seen to be field dependent.

To summarize what we have shown so far in this section: the dressing for
a slowly moving single charged field can be readily written down using
the naive action of the Lorentz boost generators. However, the lack of a
unitary implementation of the Lorentz group means now that additional
terms
are needed to enforce the anti-commutativity of such boosted fields.
We also expect any attempt to go beyond the small velocity approximation
to encounter difficulties when using the naive Lorentz transformations.
Note further that from (\boostv) we
see that the dressing gauge fixing for small velocity is
given by
$$
\pa_iA_i=-v_iE_i\,.\eqn\no
$$

To conclude this discussion of dynamical dressings, we will write down
a description of a relativistically dressed charge. Rather than building
this up from the known static configuration, we will take the form
(\boostv) as a starting point and follow Dirac's original
argument and look now for a gauge invariant field which formally creates
the electric and magnetic fields appropriate for a relativistically moving
charge.

For simplicity, consider the
electron at the point $\bold y$ moving with speed $v$ in the $x^1$
direction, then we recall that its electric field at the point
$\bold x$ is
$$
{\bold E}(x)=\frac{e}{4\pi}\frac1{\sqrt{1-v^2}}\frac{{\bold x}-{\bold
y}}{\vdenom{x}{y}^{\!\!\frac32}}\,.\eqn\no
$$
It also has a  magnetic field given by
$$
{\bold B}(x)={\bold v}\times{\bold E}(x)\,.\eqn\no
$$
The dressing of $\psi$ must reproduce these field configurations in much
the same way as the static dressing did for the Coulomb field. A short
calculation shows that these follow if we take  $\psiv=e^{\vv}\psi$, where
now
$$
\eqalign{
\vv(x)=-\frac{ie}{4\pi}&\frac1{\sqrt{1-v^2}}\cr
&\times\int d^3z\frac{(1-v^2)\pa_1A_1(x^0,{\bold z})+\pa_2A_2(x^0,{\bold
z})+\pa_3A_3(x^0,{\bold z})-vE_1(x^0,{\bold
z})}{\vdenom{x}{z}^{\!\!\frac12}}\,.
}\eqn\vdress
$$
It is straightforward to show that  this
dressed particle is  gauge invariant. Thus we have found
the relativistic dressing for an electron.

In summary, we have seen that it is possible to construct dressings for
moving charges. That they are generally not simple boosts of the static
dressing is a consequence of the way Lorentz transformations must be
implemented on charged states. Similarly we have seen the need for
phase factors when the charged state contains more than one charge.
Such terms need to be better understood. The dressing gauge of a
moving charge is no longer the Coulomb one. An immediate conjecture
is that the propagator for the moving, dressed charges is infra-red
finite only if the appropriate mass shell is used. Similar arguments
apply to the renormalisation of vertices.

\vfil\eject

%%%%%%%%%%%%%%%%%%%
%%%             %%%
%%% Section 10  %%%
%%%             %%%
%%%%%%%%%%%%%%%%%%%

\secno=10 \meqno=1
\ni
{\bf 10. Conclusions}
\bigskip
\ni We have seen that in QCD the colour quantum number is only well
defined upon gauge invariant states. Gauge invariant descriptions of
quarks and gluons have been constructed in perturbation theory. We
understood that the static interquark potential can be found more
efficiently than in other approaches if the dressed quarks are used.
The dressed quark propagator was shown to be free of infra-red
divergences and effects of the QCD vacuum on this propagator were
incorporated. These results support our proposal that constituent
quarks can emerge from QCD as dressed Lagrangian quarks. The fact that
quarks have never been seen in any experiment finds its expression in
a non-perturbative obstruction to constructing the dressing.

Before we review this in more detail and consider
the most obvious tasks at hand in this approach to QCD, we would like
to summarise the implications of these ideas for the rest of the
standard model. We have seen that descriptions of photonic
dressings exist for both static\up{\DIRAC} and moving electrons. Here
we are not faced with a non-perturbative obstruction. In many ways QED
offers us a testing ground for understanding the functions of the
dressings around (colour) charges. It would, for example, be
interesting to study how, from a dressed electron and a dressed positron,
one can form the positronium bound state. We should also stress that
the fact that $U(1)$ theory can confine in three dimensions is in no way
in disagreement with our ability to dress electrons: the lack of
well-defined, gauge invariant, asymptotic fields in a theory
is a sufficient but not a necessary condition for the theory to
confine. The weak interaction poses a
different problem: we observe, e.g., the $W$ and $Z$ bosons and yet
their interactions come from a non-abelian gauge theory. The Gribov
ambiguity can, however, be sidestepped in theories with spontaneous
symmetry breaking by fixing the gauge in the scalar matter (Higgs)
sector of the theory. We refer to Ref.\ \LMSBT\ for further details.
This enables us to construct gauge
invariant physical fields in such theories, see also Ref.\ \THOOFT\ and
\FMS.
Clearly at
high temperatures, when the symmetry is restored, this construction
will, as in QCD, be obstructed by the Gribov ambiguity. The
implications of this, particularly for the early universe, remain to be
explored. We should point out here that renormalisation group analyses
indicate that the electroweak sector is confining at high
temperatures\up{\WETT}. In summary the ideas
presented in this paper are in good agreement with the phenomenology of
the
standard model.

Returning to QCD, we have
seen that a gauge invariant definition of a quark is possible
inside perturbation theory and we have both found explicit
solutions and shown how these solutions may be extended to higher
orders in the coupling constant. Dressings for both static and non-static
quarks were given. These gauge invariant quarks have been
shown to possess various
desirable attributes. For example,
the colour charge, which is usually gauge dependent, was demonstrated to
be gauge invariant on all locally gauge invariant states.
An important consequence of this is that  combinations of such dressed
quarks may be put together in the usual group theoretical fashion
to form hadrons. As well as such essential symmetry properties, the
initial
field theoretical investigations reported here were highly promising:
we have
seen that the infra-red divergence of the one-loop fermion
propagator is removed by the dressing if and only if a physically
motivated mass
shell scheme is employed. Furthermore we have been able to incorporate
non-perturbative vacuum effects, in the shape of the quark condensate,
into the quark propagator in a gauge invariant way. This indicates how
a gauge invariant, non-perturbative constituent quark mass may be
constructed.

An attractive aspect of this approach is that a non-perturbative
construction of constituent quarks is impossible. The  gluonic dressing
necessary to ensure gauge invariance cannot be
fully defined. This is because  the dressing can be used to construct a
global gauge fixing and the Gribov
ambiguity teaches us that such a perfect gauge fixing does not exist in
QCD.
This breakdown would be a catastrophe in an analogous description of an
electron or, say, a $W$ boson since these are observed particles.
However, full descriptions of physical, gauge invariant fields are
available to us in both abelian theories like QED\up{\DIRAC} and in
non-abelian theories with spontaneous symmetry breakdown like the
Salam-Weinberg model\up{\LMSBT, \THOOFT, \FMS}. In QCD on the other
hand we have
quark and gluon confinement and our inability to form a
non-perturbative definition of the dressing fields is a simple
characterization of confinement. The moral of this tale is that
at short
distances, where perturbative physics prevails, it is possible to
describe gauge invariant quarks which, as we have argued, have
properties normally required of constituent quarks. At larger
distances, in the nonperturbative domain, the breakdown tells us that
we can only speak of colourless, gauge invariant hadrons. The quest for
the confinement scale can be understood as finding where this
breakdown takes place. This is, by its very nature a non-perturbative
matter.

The general picture of confinement that emerges from the above
parallels a more familiar one\up{\WEINBERG-\FRITZSCH}: at
one-loop order in perturbation theory the dressing which
removes the infra-red divergence of the quark propagator and allows a
description of an asymptotic coloured field is relatively simple
(essentially that of the one-loop dressed electron propagator), at two
loops the dressing is more complicated, as a result of the non-abelian
nature of the gauge group, and it becomes harder and harder at higher
orders. Non-perturbatively there is no dressing in QCD and the
infra-red problem for asymptotic quarks or gluons at the end of
the day does not permit a cure.

Various  results about charged states emerge naturally in the
approach followed here. Our description of quarks is necessarily both
non-local and non-covariant, which, we once more stress, are expected
characteristics of any charged state. These properties manifest
themselves in distinct ways in both QED and QCD, and the
similarities and differences are instructive. The non-locality is in
each theory a reflection of the
long range nature of the force. An electric charge is surrounded by
an electric field which falls off only as $1/R$. This slow fall off
is known to be responsible\up{\DOLLARD} for the infra-red problem in
QED. Although one can calculate scattering cross-sections in QED
without taking the electromagnetic fields which accompany the
asymptotic fields properly into account, one cannot properly discuss
physical states without them. The situation is
more serious in QCD: here the potential between coloured charges
{\it grows with distance}. This worsens the infra-red behaviour of
Chromodynamics and is
responsible for the strong interaction confining colour charges into
colour singlet hadrons, the residual interactions between which are of
a short range nature.

As well as this non-locality, we have seen that charged states respond
to Lorentz transformations in an unexpected way. For example, when there
is more than one moving charge, phase factors must be introduced to
maintain Fermi statistics. In QED these factors are fields independent.
This does not appear to be the case in QCD.

We have argued in Sect.\ 4 against artificially restoring gauge
invariance by attaching a string to a charged field. This introduces an
unnatural dependence upon the path we give to
the string. Similar criticism can be levelled at constructing hadrons by
attaching strings between quarks. Additionally, it has been shown
by Shabanov\up{\SHABANOV} that if we take two static electric charges and
connect them by a string then this configuration will radiate off until
it eventually approaches the state of a static dressed electron-positron
pair. The QCD equivalent of this is obviously a
very difficult calculation. If the static charges are at short
distances, then presumably they will radiate off gluons in a process
which may be described by perturbation theory. However, as the gluons
get further away from the fermions they will become sensitive to
non-perturbative confining forces, although the vicinity of the fermions
should still be in the realm of perturbation theory. If the fermions
are further apart
then one will presumably see the initial string thicken and form a meson.

The dressed charges allow for a more direct evaluation of the
perturbative static quark potential than has previously been the case.
We have seen two ways of evaluating the confinement scale from such
calculations: one may investigate at which separation of the quarks
the hadronic dressing ceases to factorise into constituent parts or,
alternatively, one can try to directly observe the onset of the Gribov
problem when one moves two quarks away from each other.

There are many other immediate extensions to this work. The perturbative
calculation of the one loop vertices and moving propagators for QED
is one of the most obvious. Here we have various predictions about the
infra-red behaviour of the dressed Green's functions. These are to be
found in Sect.'s  7 and 9. Extensions to higher loops in QCD must also be
carried out. As we have stressed, the constituent quark picture cannot
arise from perturbation theory alone. It has been
possible to include vacuum
effects in the shape of the quark condensate. Further work is required
to understand how gluonic, and in general higher dimensional,
condensates can be included. The main question here is how the
non-locality of the gluonic dressing makes itself felt. In a similar
vein, it would be interesting to see if and how instanton effects can
be incorporated.

Much use is normally made of BRST and anti-BRST invariance in the analysis
of the physical states of gauge theories. Such approaches, though, cannot
distinguish the superselection sectors labelled by the velocity of a
charge (recall our discussion in Sect.~3). In Ref.\ \LMSYM\ we introduced
a new, non-local, non-covariant symmetry for QED which could isolate the
static charged sector of the theory. This symmetry, unlike BRST or
anti-BRST, does not leave the boosted, dressed electrons invariant.
We thus conjecture that our symmetry is but one representative of a
whole class of new symmetries of QED parameterised by the velocities
of the charges. The extension of even the \lq static symmetry\rq\ to
QCD is unknown; however, the work presented here suggests that, within
perturbation theory, it may be possible.

The central role of the Gribov ambiguity in our account of confinement
warrants our recalling that there is a history\up{\HISTORY-\HERSTORY}
of linking this problem with confinement which extends right back to
Gribov's original paper\up{\GRIBOV}. Most recently there have been
attempts\up{\VANB,\TWENTY} to restrict non-abelian gauge theories
to the region unaffected by the Gribov problem, the fundamental modular
domain. This approach aims to construct well defined path integrals, as
such it does not help us to build a dressing (see our discussion of
Lagrangian versus dressing gauge fixings in Sect.\ 5). Such studies
argue that this restriction introduces non-perturbative effects which
could be responsible for confinement. The most difficult aspect of this
idea is that the boundary of the domain contains all the topological
information about the physical configuration space --- this results in
highly non-trivial identifications of various points on the boundary.
The influence of these Gribov copies requires further study.

Finally we wish to list some topics where it seems that it might prove
profitable to apply our dressed quarks and gluons. Two of the most
obvious applications of dressed quarks are jet physics (see articles
in Ref.\ \AACHEN) and the heavy quark effective theory (for a recent
review see Ref.\ \HQET). Multijet events offer us a window upon quarks
and gluons. Our picture is of coloured, dressed quarks and gluons
exiting from the event vertex. Initially the dressings are factorised
but after some time non-perturbative physics tears the dressings apart
and separate colour singlet jets are formed, which hadronise and are
then detected. In the effective theory the heavy quarks are surrounded
by light degrees of freedom, the so-called \lq brown muck\rq\ --- our
approach indicates that some of these light degrees of freedom should
be associated with the heavy quark so as to give it a gauge invariant
definition. It is important to note here that the preeminent role of
the heavy quark's velocity in the effective theory has a natural
counterpart in the boosted dressings that we have constructed. Both
of these applications are highly intriguing.

There is also much interest in QCD at high temperature and
densities\up{\AACHEN}. Here we would expect hadrons to swell
and multiquark structures to form. It also appears that there is
some  phase transition.
The Gribov ambiguity does not vanish at finite temperature. This
means that even at high temperatures individual quarks and gluons
will not be observables.  The implications of this are as yet unclear,
however, we are reminded of the expected confining behaviour in the
high temperature electroweak sector discussed at the start of this
section.

Recent experimental data on that old story the pomeron indicate that it
is composed of one hard gluon surrounded by soft degrees of
freedom\up{\POMMYEXPT}. Assuming that this data holds up, it is
natural to want to describe the hard gluon (where one might hope to
apply perturbation theory) in a gauge invariant way. This would
necessitate use of our formulae for dressed gluons from Sect.\ 5
(or some variant upon them using boosted dressings).

Dressing electric charges correctly can also be used to study
hadronic electromagnetic mass differences.
In QCD sum rules for hadrons with non-vanishing electric charges the
currents, and hence the results of such sum rules, are not invariant under
U(1) gauge transformations\up{\KISS}.
Although this gauge dependence may be neatly removed by connecting the
currents in the correlation function by a string, this is not the most
physical choice (the commutation relations would imply that the
electric field was only non-zero along the string). A carefully
chosen Dirac-type dressing seems more physical. It would be interesting
to see how much this changed the results of Ref.\ \KISS.

In conclusion in this paper we have developed a gauge invariant,
perturbative description of individual quarks and gluons. These we
identify with the constituents of hadrons. We have observed an
obstruction to the non-perturbative construction of these fields. This
we identify with the expected confinement of such coloured particles.
The scale at which hadrons are formed is that where it becomes
energetically favourable for the dressing of a hadronic state to no
longer factorise into a product of the constituent dressings.

\bigskip\bigskip\bigskip
\ni{\bf Acknowledgements} We wish to thank the following for between
them encouraging us, giving constructive criticism, supplying references
and reading drafts of the manuscript: Emili Bagan, David Broadhurst,
Robin Horan, Stephen Huggett, Alex Kalloniatis, Taro Kashiwa,
John Klauder, Peter Landshoff, Peter L\'evay, G.\  M\"unster,
Martin Schaden, Karl Schilcher, Sergei Shabanov, Tom Steele, M.\ Stingl,
John C.\ Taylor, Izumi Tsutsui, Tassos Vladikas and E.\ Werner.
In addition we would like to thank the organisers of the 16th UK
Theory Institute for affording us a chance to finish this paper
together. MJL received support from the research project
CICYT-AEN95-0815. We thank each other's institutes for hospitality.

\bigskip\ni{\bf Note Added} Since the completion of this paper we have 
together with Emili Bagan verified the above conjectured 
infra-red finiteness and multiplicative renormalisability of the 
propagator of the dressed, non-static charge defined by (\vdress) 
at the correct mass 
shell. This has been done both for small\up{\SLOW} and 
relativistic\up{\FAST} velocities. 

\vfill\eject%%%%%%%%%%%%%%%%%%%
%%%             %%%
%%% Appendix    %%%
%%%             %%%
%%%%%%%%%%%%%%%%%%%

\secno=0
\appno=1
\meqno=1

\noindent
{\bf Appendix A. The static quark to order $g^ 3$}
\bigskip
\noindent
In this appendix we shall show how the expression (\psitwo) for the
static quark is derived, and how to extend it to order $g^3$. The
extension to arbitrary order will be clear.

In our description of the physical quarks and gluons, (\psiphysfour) and
(\aphysfour), we expand  $\hc=e^{-\vc}$ in a power series in the coupling
$g$
where
$$
\vc=g\v1+g^2\v2+g^3\v3+\cdots\eqn\no
$$
Then the quarks and gluons are, to these orders,
$$
\eqalign{
\psi^{\hc}=\Bigl(1+&g\v1+g^2\bigl(\tfrac12\v1\v1+
\v2\bigr)\cr
&+g^3\bigl(\tfrac16\v1\v1\v1+\tfrac12\v1\v2
+\tfrac12\v2\v1+\v3\bigr)
\Bigr)\psi
}
\eqn\phihapp
$$
and
$$
\eqalign{
A_i^{\hc}=&A_i-\pa_i\v1\cr
& +g\bigl( [\v1,A_i]+\tfrac12[\pa_i\v1,\v1]-\pa_i\v2\bigr)\cr
&{}\quad +g^2\bigl(
[\v2,A_i]+\tfrac12[\v1,[\v1,A_i]]+\tfrac12[\pa_i\v1,\v2]+
\tfrac12[\pa_i\v2,\v1]\cr
&{}\qquad-\tfrac16[\v1,[\v1,\pa_i\v1]]-\pa_i\v3
\bigr)
}
\eqn\ahapp
$$
Our task now is to find $\v1$, $\v2$ and $\v3$. To this end we
exploit the fact that for the static quark these must be such that
$A^{\hc}_i$ is transverse, {\it i.e.,} $\pa_i A^{\hc}_i=0$. This means that
we must be able to write
$$
A^{\hc}_i=\(\d_{ij}-\frac{\pa_i\pa_j}{\nabla^2}\)\Phi_j\,,\eqn\atrapp
$$
where to this order in $g$ we have $\Phi_i=\Phi_{_{(0)}i}+
g\Phi_{_{(1)}i}+g^2\Phi_{_{(2)}i}$.

To order $g^0$, (\ahapp) and (\atrapp) give
$$
\eqalign{
\(\d_{ij}-\frac{\pa_i\pa_j}{\nabla^2}\)\Phi_{_{(0)}j}&=A_i-\pa_i\v1\cr
&=\(\d_{ij}-\frac{\pa_i\pa_j}{\nabla^2}\)A_j+
\frac{\pa_i\pa_j}{\nabla^2}A_j-\pa_i\v1
}
\eqn\no
$$
from which we deduce that
$$
\v1=\frac{\pa_jA_j}{\nabla^2}\,,\eqn\voneapp
$$
and hence that $\Phi_{_{(0)}i}=A_i$.

To the next order in $g$ we get from (\ahapp) and (\atrapp)
$$
\eqalign{
\(\d_{ij}-\frac{\pa_i\pa_j}{\nabla^2}\)\Phi_{_{(1)}j}&=
\(\d_{ij}-\frac{\pa_i\pa_j}{\nabla^2}\)\([\v1,A_j]+\tfrac12
[\pa_j\v1,\v1]\)\cr
&{}\qquad+\frac{\pa_i\pa_j}{\nabla^2}\([\v1,A_j]+\tfrac12
[\pa_j\v1,\v1]\)-\pa_i\v2
}
\eqn\no
$$
which implies that
$$
\v2=\frac{\pa_j}{\nabla^2}\bigl([\v1,A_j]+\tfrac12
[\pa_j\v1,\v1]\bigr)\,,
\eqn\vtwoapp
$$
and $\Phi_{_{(1)}j}=[\v1,A_j]+\frac12[\pa_j\v1,\v1]$.
Substituting (\voneapp) and (\vtwoapp) into (\phihapp) yields the
expression (\psitwo) for the quark valid to order $g^2$. The expressions
for
$\Phi_{_{(0)}j}$ and $\Phi_{_{(1)}j}$ yield in (\atrapp) the
description (\agfour) for the gluon to order $g$.

This argument can clearly be extended to higher order in $g$. In
particular, repeating this argument to order $g^3$ gives
$$
\eqalign{
\v3=\frac{\pa_j}{\nabla^2}\Bigl([\v2,A_j]+\tfrac12[\v1,[\v1,A_j]]+
&\tfrac12[\pa_j\v1,\v2]+
\tfrac12[\pa_j\v2,\v1]\cr
&-\tfrac16[\v1,[\v1,\pa_j\v1]]\Bigr)
}
\eqn\no
$$
from which the static quark to order $g^3$ can be constructed.

  \vfill\eject\immediate\closeout\reffile%\parindent=20pt
  \centerline{{\bf References}}\bigskip\eightpoint\frenchspacing%
  \input refs.tmp\vfill\eject\nonfrenchspacing

\bye